\documentclass{article}
\usepackage[a4paper, margin=1in]{geometry}
\usepackage{siunitx}
\usepackage{booktabs}
\usepackage{mathtools}
\usepackage{amsmath}
\usepackage{graphicx, color, float}
\usepackage{amsmath,amssymb,amsfonts,amssymb}
\usepackage{subcaption}
\usepackage{bm}
\usepackage{comment}
\usepackage{cite}
\usepackage{xcolor}
\usepackage{array,hyperref}
\usepackage{cancel}
\usepackage{epsfig,epsf}
\usepackage{dcolumn}
\usepackage{float}
\usepackage{bm}
\newcommand{\bse}{\begin{subequations}}
\newcommand{\ese}{\end{subequations}}
\newcommand{\be}{\begin{equation}}
\newcommand{\ee}{\end{equation}}
\newcommand{\bea}{\begin{eqnarray}}
\newcommand{\eea}{\end{eqnarray}}
\newcommand{\ba}{\begin{array}}
\newcommand{\ea}{\end{array}}

\newcommand{\G}{{\tilde{g}}}

\input amssym.def
\input amssym.tex

\title{}

\begin{document}

\begin{titlepage}
	\thispagestyle{empty}

%
	
	\vspace{.4cm}
\begin{center}{\Large \textbf{
Infrared fluctuations and collective entanglement modes in
harmonic lattices}}

\end{center}

\vspace*{15mm}
		\vspace*{1mm}

\begin{center}
{Mahsa Lezgi and  Ali Mollabashi}
\end{center}
\vspace*{1cm}

\begin{center}
{School of Quantum Physics and Matter, Institute for Research in Fundamental Sciences (IPM), P.O.Box 19395-5531, Tehran, Iran}
		
		\vspace*{0.5cm}
		{E-mails: {\tt mahsaLezgi, mollabashi@ipm.ir}}%
\end{center}	
	
\vspace{10mm}

\begin{abstract}

We investigate how infrared fluctuations and spatial separation shape vacuum entanglement between two finite regions in the relativistic harmonic chain and its Lifshitz generalizations. At fixed lattice spacing and finite geometry, we establish that logarithmic negativity remains uniformly bounded as the mass approaches zero, while mutual information diverges. Infrared-enhanced total correlations therefore do not provide an unbounded distillable-entanglement resource. Using local symplectic consolidation, we identify the collective canonical modes carrying the negativity and follow their evolution with separation. Their participation crosses over from localization near the facing boundaries to a collective distribution within each region. Expansion in the adjacent-state canonical basis connects this evolution to progressively weaker, more extended reference modes, including adjacent-halo directions.

Building on the established concentration of long-distance logarithmic negativity into a single canonical pair, we show that its spatial participation is smooth and infrared-dominated despite its ultraviolet canonical structure, a distinction also present in pure-state entanglement modes. These results refine the ultraviolet mode picture by resolving the long-wavelength organization of the oscillators participating in the surviving entanglement. Across the massless dispersions studied, the normalized participation approaches a universal long-distance profile when separation is scaled by the negativity radius. Its detailed form and the entanglement range nevertheless remain sensitive to ultraviolet deformations. Together, these findings distinguish infrared-enhanced correlations from distillable entanglement and reveal the collective spatial organization underlying its persistence at large separation.

\end{abstract}

\end{titlepage}

\newpage

\tableofcontents

\section{Introduction}

How is vacuum entanglement shared between distant regions,
and which spatial scales characterize the degrees of
freedom carrying it?
These questions bring together complementary perspectives
on quantum correlations: as resources for quantum
information processing
\cite{Feynman:1981tf,Bennett:1992zzb,Bennett:1992tv,Bennett:2014rmv}
and as probes of quantum many-body systems
\cite{Amico:2007ag,Laflorencie:2015eck}, quantum field theories
\cite{Casini:2009sr,Casini:2022rlv}, and gravitational physics
\cite{Ryu:2006bv,Solodukhin:2011gn,Almheiri:2020cfm}.
For spatially extended parties, quantifying entanglement
is only part of the problem: one must also identify
the combinations of local degrees of freedom through
which the regions share it.

The vacuum of a field governed by a local Hamiltonian
provides a natural setting for investigating these
questions.
Retaining two finite regions and tracing out their
surroundings produces a generally mixed bipartite state.
Its total correlations need not track its distillable
entanglement, and neither quantity alone determines
the spatial organization of the modes carrying that
entanglement.
In particular, stronger long-wavelength vacuum
fluctuations need not provide a larger entanglement
resource or extend the distance over which it can
be distilled.

A further distinction arises when identifying the
spatial scales of an entanglement-carrying mode.
The canonical observables defining the mode can involve
rapidly oscillating combinations of local fields,
while the participation of the underlying degrees
of freedom extends smoothly across each region.
Understanding long-distance entanglement therefore
requires resolving both its microscopic canonical
structure and its collective spatial participation.

In this work, we investigate these issues in Gaussian
scalar vacua.
We first examine how long-wavelength fluctuations
affect total correlations, distillable entanglement,
and its spatial range.
We then identify the collective canonical modes
carrying the surviving entanglement and follow their
reorganization as the regions are separated.
The relativistic harmonic chain provides our reference
setting, while controlled changes of the dispersion
and Lifshitz generalizations allow us to distinguish
features sensitive to the underlying theory from
structures shared across the models studied.
Our analysis brings together two complementary lines
of research: the infrared sensitivity of vacuum
correlations and the spatial organization of Gaussian
entanglement.
We review these developments in turn.

\paragraph{Vacuum correlations and infrared sensitivity.}

A useful starting point is the distinction between
entanglement with an entire complement and entanglement
shared by two retained regions. For a spatial region
in a pure vacuum state, the reduced entropy measures
its entanglement with the complement. Foundational
calculations established the ultraviolet-divergent
boundary contribution to this entropy
\cite{Bombelli:1986rw,Srednicki:1993im}, while subsequent
work developed area laws for harmonic lattices and
gapped one-dimensional systems
\cite{Plenio:2004he,Hastings:2007iok};
see Ref.~\cite{Eisert:2008ur} for a review.
For two finite regions obtained by tracing out their
surroundings, however, the joint state is generally
mixed, and the entropy of either region no longer
quantifies the entanglement shared between them.

Mutual information and logarithmic negativity address
different aspects of this mixed-state problem.
Mutual information measures total correlations
\cite{Groisman:2005dbo,Wolf:2007tdq}, whereas logarithmic
negativity detects negative-partial-transpose (NPT)
entanglement and provides a computable upper bound
on distillable entanglement \cite{Vidal:2002zz}.

Complementary analytical and numerical approaches have
been developed to study correlations between disjoint
regions. Replica methods provide a field-theoretic
framework for calculating entanglement measures
\cite{Calabrese:2004eu,Calabrese:2009ez}, including
the formulation of logarithmic negativity
\cite{Calabrese:2012ew}.
On the lattice, exact diagonalization enabled early
studies of mutual information between disjoint intervals
\cite{Furukawa:2008uk}, while tensor-network methods
extended the numerical investigation of disjoint-block
entropies and logarithmic negativity
\cite{Alba:2009ek,Calabrese:2013mi}.
For bosonic Gaussian states, these quantities can be
evaluated directly from covariance matrices, making
harmonic lattices a particularly useful setting for
investigating their dependence on geometry and dispersion
\cite{Audenaert:2002xfl, Marcovitch:2008sxc, Anders:2008lus, Calabrese:2014yza, DeNobili:2015dla, Eisler:2016bfd, Zhou:2016ykv, MohammadiMozaffar:2017chk, Angel-Ramelli:2020wfo, Klco:2021biu,Gao:2024nbl}.

We focus on this Gaussian setting to connect the
correlation structure of the retained state with
the spatial organization of its entanglement.
For bipartite Gaussian states, NPT is necessary and
sufficient for distillability under unrestricted local
operations and classical communication \cite{access1}.
Comparing mutual information with logarithmic negativity
therefore distinguishes the growth of total correlations
from the behavior of an upper bound on distillable
entanglement.
A nonzero separation also removes the shared entangling
interface and its associated short-distance divergences,
allowing correlations between the regions to be examined
separately from this boundary contribution.

These studies show that total correlations and
logarithmic negativity can have markedly different
dependence on separation.
In conformal field theories admitting the standard
small-cross-ratio expansion, mutual information has
an algebraic long-distance tail governed by the
lowest-dimension operators, with analogous results
for distant regions in higher dimensions
\cite{Calabrese:2010he,Cardy:2013nua,Agon:2015ftl,
Shiba:2012np}.
Logarithmic negativity is more strongly suppressed:
exponential decay was observed numerically in the
critical harmonic chain \cite{Marcovitch:2008sxc},
while field-theoretic analyses establish decay faster
than any power in the settings considered
\cite{Calabrese:2012nk,Parez:2023xpj}.
An analytic example of exponential suppression at
small cross ratio has recently been obtained for
the free scalar chiral current \cite{Arias:2026bqh}.

This contrast raises a related but distinct question:
at fixed region sizes and separation, do increasingly
strong infrared fluctuations enhance the entanglement
that can be distilled between the regions, or primarily
increase their total correlations?
Addressing this question requires distinguishing the
response of mutual information from that of logarithmic
negativity, which bounds distillable entanglement from
above.
In particular, if logarithmic negativity remains bounded
while mutual information grows without bound, the
increase in total correlations cannot yield an unbounded
distillable-entanglement resource.
The Gaussian setting allows this distinction to be
examined through the different covariance-matrix spectra
that determine the two measures.

\paragraph{Spatial organization of entanglement.}

Beyond determining how much entanglement two regions
share, one can ask which local combinations of degrees
of freedom carry it.
This structural information is not supplied by mutual
information or logarithmic negativity alone.
For pure Gaussian states, the Gaussian Schmidt decomposition
organizes entanglement across a bipartition into independent
pairs of modes through local symplectic transformations
\cite{Botero:2003wov,Botero:2004vpl}.
A general mixed Gaussian state does not admit such
a pairwise decomposition.
The consolidation framework developed in
Refs.~\cite{Klco:2021cxq,Gao:2024esb} nevertheless allows,
when the required conditions hold, the total logarithmic
negativity to be concentrated into pairs of local canonical
modes whose individual negativities sum to that of the
original state.
This organization need not imply a factorization of the
full mixed state into independent pairs.\footnote{A recent application identifies approximate field and
momentum profiles of the most entangled modes between
two disjoint spherical regions in a $(3+1)$-dimensional
massless scalar vacuum and proposes a nonperturbative
mode-swap protocol for transferring their entanglement
to localized probes
\cite{Perche:2026ams}.}
It identifies collective degrees of freedom through which
two regions share NPT entanglement and provides a starting
point for investigating how their spatial participation
changes with separation.
In particular, identifying the negativity-carrying modes
does not by itself explain whether their participation
remains concentrated near the facing boundaries or becomes
distributed across the retained regions.

A complementary approach assigns spatial contributions
to entanglement measures through contour functions
\cite{Vidal:2014aal}, with extensions to logarithmic
negativity developed in Ref.~\cite{Kudler-Flam:2019nhr}.
For bosonic Gaussian states, contour constructions have
been developed for entanglement entropies in harmonic
lattices \cite{Frerot:2015irl,Coser:2017dtb} and, more
recently, for logarithmic negativity and partial-transpose
moments \cite{Zambotti:2026vpm}.
Such constructions connect spectral information to
spatial contributions according to a specified prescription;
the usual consistency requirements do not, in general,
determine a unique contour.
Our analysis follows the individual canonical modes
obtained through local symplectic consolidation and examines
their spatial organization using a canonical participation
function \cite{Botero:2004vpl}.
This function resolves each site's contribution to the
normalization of a mode's canonical commutator.
It can be signed and does not assign a local density
of logarithmic negativity.

The distinction between a mode's canonical components
and its participation profile becomes particularly
relevant at large separation.
For the lattice configurations considered here, the
negativity radius is the threshold separation beyond
which logarithmic negativity vanishes and the retained
state is PPT.
Studies of the relativistic harmonic chain have identified
increasingly UV-dominated partially transposed mode
wavefunctions as this radius is approached
\cite{Klco:2021biu}.
Together with the cutoff dependence of the radius,
this observation highlights the sensitivity of
long-distance negativity to microscopic spatial resolution.
It leaves a complementary question: how is the canonical
participation of the surviving mode distributed across
each region?

Rapid oscillations in a mode's conjugate canonical
components do not preclude a smooth, slowly varying
participation product.
The momentum content of the components and the spatial
distribution of canonical participation therefore
describe distinct aspects of the entanglement-carrying
mode.
This motivates examining whether increasing separation
produces collective participation across the retained
regions, and how such participation relates to the
ultraviolet structure identified in earlier studies.
The consolidation framework allows these questions
to be investigated at the level of individual modes,
beyond determining their contributions to the total
logarithmic negativity.

\paragraph{Scope and main results.}

We address these questions in Gaussian scalar vacua.
For these zero-mean Gaussian states, covariance matrices
completely characterize the state and allow efficient
evaluation of correlation measures
(see e.g. \cite{Adesso:2007tx}).
Their use in free-field entanglement calculations is
reviewed in Ref.~\cite{Casini:2009sr}.
They are also directly relevant to experimental
continuous-variable platforms
\cite{Zhong:2020iql,Zhong:2021dml,Madsen:2022uqz}.
Several of our findings concern the relativistic harmonic
chain itself. Its Lifshitz generalizations, with dynamical
exponent $z$, allow us to compare soft and stiff infrared
dispersions, test the scope of these findings, and distinguish
theory-dependent behavior from structures shared across
different dispersions.
Our main findings are as follows:

\begin{enumerate}

\item \textbf{Infrared fluctuations and distillable entanglement.}
At fixed geometry, logarithmic negativity remains bounded
in the small-mass regimes examined, while mutual
information diverges.
Thus, enhanced long-wavelength correlations do not
produce an unbounded distillable-entanglement resource,
and the ratio of distillable entanglement to mutual
information vanishes.
Our spectral analysis identifies cancellation of the
leading zero-mode divergence in the cases analyzed.

\item \textbf{Infrared and ultraviolet control of the entanglement range.}
Controlled dispersion deformations show that, in the
massless configurations examined, stiffening either
momentum sector extends the negativity range, whereas
softening it shortens the range, with the complementary
sector preserved as specified below.
Along the Lifshitz family, increasing the dynamical
exponent simultaneously softens the infrared dispersion
and stiffens its ultraviolet part.
The observed decrease of the range is consistent with
infrared softening outweighing ultraviolet stiffening
in this regime. Separately, within the undeformed Lifshitz
family, the ordering of the radius with $z$ reverses
in the massive regime.
We also find that the quadratic scaling
$d_{\mathrm{neg}}\simeq\gamma(z)\ell^2/\epsilon$,
previously established for the relativistic chain,
extends numerically to the other exponents examined,
with a dispersion-dependent prefactor.

\item \textbf{Two flows toward collective participation.}
We identify two related changes in the spatial
organization of entanglement.
For adjacent regions, modes contributing less strongly
to logarithmic negativity participate more broadly
across each region.
As the regions are separated, the dominant entangled
mode likewise evolves from participation concentrated
near the facing boundaries to participation spread
across each region.
The boundary-localized regime becomes less prominent
as the mass decreases.

\item \textbf{Connecting the flows through the adjacent basis.}
Expanding the dominant finite-separation mode in the
complete canonical basis of adjacent regions connects
these two observations.
At short separations, its composition remains narrowly
concentrated while shifting toward weaker adjacent
entangled modes.
As participation becomes collective, the expansion
broadens and incorporates directions outside the
adjacent negativity-carrying sector.
Thus, the surviving entanglement involves changing
combinations of local degrees of freedom, including
directions that do not belong to an entangled pair
at adjacency.

\item \textbf{Long-wavelength participation of the surviving pair.}
Building on the established concentration of long-distance
logarithmic negativity into a single canonical pair,
we show that its spatial participation is smooth and
\textit{infrared dominated} despite the pronounced ultraviolet
structure of its conjugate components.
The same distinction appears in the pure-state modes
examined here.
These findings refine earlier high-frequency mode
descriptions \cite{Botero:2004vpl,Klco:2021biu} by
revealing the collective, long-wavelength participation
underlying the microscopic oscillations.
Across the massless (scale-invariant) theories examined, we find
numerical evidence for a universal normalized
participation profile at equal fractions of the
respective entanglement ranges.
Finite mass disrupts this profile collapse.
\end{enumerate}

The rest of the paper is organized as follows.
Section~\ref{sec:IRLN} introduces the Gaussian framework
and the Lifshitz harmonic chain, then examines the
infrared sensitivity of mutual information and
logarithmic negativity and the spatial range of
distillable entanglement.
Section~\ref{sec:Modewise} investigates the spatial
organization of the entanglement-carrying modes.
It follows the two flows toward collective participation,
connects them through the adjacent-state basis expansion,
and examines the long-distance participation profiles
and their relation to the underlying mode structure.
We conclude with a discussion of the implications
and open questions.
Technical details and supporting analyses are
presented in the appendices.

\section{Infrared Structure of Logarithmic Negativity}\label{sec:IRLN}

In this section, we introduce the generalized harmonic lattice model that forms the basis of our analysis and investigate the entanglement structure between two disjoint regions in its vacuum state. Throughout this work, bipartite entanglement in mixed states is quantified by the logarithmic negativity (LN) \cite{Plenio:2005cwa,Vidal:2002zz,Simon:1999lfr,Horodecki:1996nc,Peres:1996dw}. For bipartite Gaussian states, logarithmic negativity provides a computable characterization of distillable entanglement, since every NPT Gaussian state is distillable \cite{access1}. Our analysis is restricted to Gaussian vacuum states within the continuous-variable framework.

It's worth briefly reminding the reader that the entanglement properties of Gaussian states are entirely characterized by the covariance matrix (CM). For a system with $N$ modes, the CM is defined as
\begin{align}
\sigma_{ij}=\frac12\left\langle \{\hat{r}_{i}-\bar{r}_{i},\hat{r}_{j}-\bar{r}_{j}\} \right\rangle,
\label{sec1eq3}
\end{align}
where $\hat{r}=\{q_{1},p_{1},...,q_{N},p_{N}\}$ is a vector of operators, and $\bar{r}$ is the vacuum expectation values. A symplectic form, $\Omega$, encapsulates the commutation relations,
\begin{align}
[\hat{r},\hat{r}^{T}]=i\Omega_{N}~~~~~~~,~~~~~~~~\Omega_{N}=\bigoplus_{j=1}^{N}i\sigma_{y},
\label{sec1eq4}
\end{align}
where $\sigma_{y}$ is the second Pauli matrix \cite{serafini}. 

Negativities characterize the extent to which a CM fails to remain physical under partial transposition (PT) of one subsystem \cite{Horodecki:1996nc}. In the continuous variable formalism, the PT acts as a local time-reversal operation, leaving the field operators unchanged while reversing the sign of the conjugate momenta associated with one part of the bipartition \cite{Simon:1999lfr}. For a bipartite system composed of $N_{A}$ and $N_{B}$ modes, this operation is implemented at the CM, $\sigma^{\Gamma}=\Gamma \sigma \Gamma$, where $\Gamma$ denotes the phase-space representation of the PT
\begin{align}
\Gamma=\left(\displaystyle\bigoplus_{j=1}^{N_{A}}\mathbf{1}_{2}\right)\oplus \left(\displaystyle\bigoplus_{j=N_{A}+1}^{N_{A}+N_{B}}\sigma_{z}\right),
\label{sec1eq5}
\end{align}
where $\sigma_{z}$ is the third Pauli matrix. For entangled bipartitions, performing a PT on one subsystem can drive the resulting operator outside the space of physical states, leading to a violation of positivity in the spectrum of the partially transposed density matrix. At the CM level, this is reflected in the appearance of $\nu_{j}<1/2$, symplectic eigenvalues of $i\Omega \sigma^{\Gamma}$, that drop below 1/2. The logarithmic negativity is entirely encoded in this spectrum and can be expressed as a sum over the PT modes corresponding to eigenvalues below 1/2,
\begin{align}
\mathcal{E}=\sum_{j=1}^{n}{\rm{log}}\;{\rm{max}}\left(1,\frac{1}{2\nu_{j}}\right)
=-\sum_{j=1}^{n_-}\log(2\nu_j),
\label{sec1eq6}
\end{align}
where $\nu_j$ denotes an eigenvalue of $\sqrt{GH^{\Gamma}}$ and $n_{-}$ denotes the number of eigenvalues contributing to negativity.

For comparison, the von Neumann entropy of a Gaussian reduced
state on a region $A$ is determined by the symplectic eigenvalues
$\mu_k^{(A)}\geq1/2$ of its covariance matrix \cite{Adesso:2007tx,serafini}
\begin{align}
S(A)&=\sum_{k=1}^{N_A}
\left[
\left(\mu_k+\frac12\right)\log\left(\mu_k+\frac12\right)
-
\left(\mu_k-\frac12\right)\log\left(\mu_k-\frac12\right)
\right],
\end{align}
where $\mu_k$ denotes an eigenvalue of $\sqrt{GH}$.
The mutual information between the retained regions is then
\begin{align}
    I(A:B)=S(A)+S(B)-S(A\cup B).
\end{align}

\subsection{Lifshitz Harmonic Chain}
The Lifshitz harmonic chain (LHC) provides a versatile framework for investigating quantum correlations in Gaussian field theories with qualitatively different infrared structures, ranging from extremely soft (weakly dispersing) to extremely stiff (strongly dispersing) infrared modes.
The LHC is defined through the following Hamiltonian on a one-dimensional lattice \cite{MohammadiMozaffar:2017nri,He:2017wla}
\begin{align}
H=
\frac{1}{2}\sum_{n=1}^N\left[\frac{p_n^2}{\epsilon}+\epsilon\, m^{2z}q_n^2+\frac{1}{\epsilon^z}\left(\sum_{i=0}^z(-1)^{z+i}\frac{z!}{i!(z-i)!}q_{n+i-1}\right)^2\right],
\label{eq:Hz}
\end{align}
where $[q_r,p_s]=i\delta_{rs}$ and $\epsilon$ is the lattice spacing, and parameter $z$ is called the dynamical exponent. From now on we fix $\epsilon=1$. For positive integer values of $z$, increasing $z$ extends the range of lattice couplings. For $z=1$, the model reduces to the harmonic lattice chain where only nearest-neighbor oscillators interact. For $z=2$, next-to-nearest neighbors are also coupled; for $z=3$, next-to-next-to-nearest neighbors, and so on. The model can be diagonalized as $H=\sum_k \omega_k (n_k+1/2)$ where $n_k$ is the number operator and the dispersion relation is
\be\label{eq:Disp}
\omega_k=\sqrt{\left(2\sin\frac{k}{2}\right)^{2z}+m^{2z}}\,,
\ee
where $k$ denotes the momentum on an infinite lattice and for a periodic lattice with $N$ sites $k\to2\pi k/N$. The non-vanishing blocks of the vacuum state CM are given by
\begin{align}
G_{ij}=\left\langle q_{i}q_{j}\right\rangle=\frac{1}{2}(K^{-1})_{ij}~~~~~~~,~~~~~~~~H_{ij}=\left\langle p_{i}p_{j}\right\rangle=\frac{1}{2}K_{ij}.
\label{sec1eq2}
\end{align}
with
\begin{align}
K_{ij}=\frac{1}{N}\sum_{k=0}^{N-1}{\rm{cos}}\left(\frac{2\pi k}{N}(i-j)\right)\sqrt{m^{2z}+\left(2{\rm{sin}}\frac{\pi k}{N}\right)^{2z}}.
\label{sec1eq1}
\end{align}
In these Gaussian ground states, fluctuations are fully
characterized by the covariance matrix. Decreasing the
mass softens the low-momentum modes and enhances their
position-quadrature fluctuations through the factor
$1/(2\omega_k)$. We refer to these long-wavelength
contributions as infrared fluctuations.

The interaction pattern in \eqref{eq:Hz} is constructed such that in the continuum limit theory retains its canonical quantization with a kinetic term with order-$z$ spatial derivatives given by the Hamiltonian density 
\begin{align}
\mathcal{H}=\frac{1}{2}\left(\pi^2+(\partial^{z}\phi)^{2}+m^{2z}\phi^{2}\right).
\label{eq:Hzcontinuum}
\end{align}
where $\pi(x)$ and $\phi(x)$ are continuous counterparts of $p_n$ and $q_n$ and $m$ denotes the mass parameter of the theory. In the continuum limit, the dispersion relation is
\be\label{eq:ContDisp}
\omega(k)=\sqrt{k^{2z}+m^{2z}}.
\ee
As the $z=1$ case is conformal invariant in the massless limit, the generic $z$ case is invariant under the anisotropic scaling $t\rightarrow \lambda^{z}t$ and $x\rightarrow  \lambda x$ symmetry in the massless limit \cite{Lifshitz1,Lifshitz2,Hartnoll:2009sz,Sachdev}; that is why this model is called the Lifshitz model, and $z$ is the dynamical critical exponent in this context.

The infrared behavior of the lattice dispersion, Eq. \eqref{eq:Disp}, coincides with that of the continuum theory, Eq. \eqref{eq:ContDisp}. 
Throughout this section, it is this infrared structure, rather than the specific lattice realization, that plays the central role.
Although the lattice Hamiltonian in Eq. \eqref{eq:Hz} is naturally defined for positive integer $z$, its diagonal Gaussian representation admits a 
natural analytic continuation to arbitrary positive values of $z$.\footnote{Non-integer values of $z$ correspond to Gaussian theories governed by fractional Laplacians.}
This allows us to continuously interpolate between soft-IR ($z\geq 1$) and stiff-IR ($0<z<1$) theories.
 
Vacuum entanglement in such theories arises because the spatial derivative terms in the Hamiltonian energetically constrain the allowed field configurations. Consequently, the ground-state wavefunctional correlates field amplitudes across space, so that a local measurement partially determines the outcomes of measurements performed elsewhere. For $z=1$ relativistic scalar field, described by the continuum limit of the harmonic chain, these constraints originate from the gradient term, while in Lifshitz theories they are imposed by higher-order spatial derivative terms, giving rise to more intricate spatial correlation patterns. Although the influence of a local measurement generally decreases with distance, it remains nonvanishing throughout the vacuum, reflecting the long-range entanglement of the ground state and though non-trivial contributions from the infrared sector of the theory.

It is worth noting that various aspects of bipartite entanglement in different states of Lifshitz theories, including the vacuum state, and across different spacetime dimensions have been investigated in \cite{Ardonne:2003wa, Fradkin:2006mb, Hsu:2008af, Fradkin:2009dus, Solodukhin:2009sk, Oshikawa:2010kv, Zaletel:2011ir, Rajabpour:2014osa, Zhou:2016ykv, Chen:2016kjp, MohammadiMozaffar:2017nri, Chen:2017txi, He:2017wla, MohammadiMozaffar:2017chk, MohammadiMozaffar:2018vmk, Angel-Ramelli:2019nji, Angel-Ramelli:2020wfo, Angel-Ramelli:2020xvd, Hartmann:2021vrt, Mozaffar:2021nex, Berthiere:2023bwn, Vasli:2024mrf, MohammadiMozaffar:2026osc}.

\subsection{Infrared Structure of the Negativity Sphere}
\label{subsec:IRNS}

In this section, we investigate the entanglement structure
between spatially separated, non-complementary regions in
the vacuum state. Such configurations provide a direct probe
of long-range quantum correlations. In lattice scalar-field
vacua, logarithmic negativity can vanish beyond a finite
separation even though cross-region correlations persist.
This behavior was observed for separated groups of oscillators
in early harmonic-chain studies \cite{Audenaert:2002xfl}.
Subsequent work characterized the associated
\textit{negativity sphere} in latticized massless and massive
scalar fields, including its dependence on the region size
and spatial resolution \cite{Klco:2021biu,Gao:2024nbl}.
Beyond this threshold, the reduced state is PPT, although
vanishing logarithmic negativity alone does not establish
separability for a general multimode Gaussian state.
For the relativistic scalar-field configurations studied
in these works, the negativity radius diverges as the lattice
spacing is taken to zero at fixed physical region sizes
\cite{Klco:2021biu,Gao:2024nbl}.

In the present work, we investigate how the infrared structure of Gaussian field theories influences the negativity sphere in the LHC vacuum.
To distinguish the loss of NPT entanglement from separability, we also apply the Gaussian separability algorithm of Ref.~\cite{PhysRevLett.87.167904}, which provides a necessary and sufficient criterion for separability in arbitrary bipartite multimode Gaussian systems.
For the configurations tested, the algorithm identifies the states beyond the negativity threshold as separable. Within this numerical scope, the negativity sphere therefore coincides with the entanglement sphere.
This conclusion follows from the separate separability test, rather than from vanishing logarithmic negativity alone.

Throughout this work, numerical calculations were performed
in Mathematica using arbitrary-precision arithmetic.
The numerical settings were adjusted separately for
different parameter regimes, with
\texttt{WorkingPrecision} ranging from 200
to 1200,
\texttt{MaxRecursion} from 150
to 1100, and
\texttt{\$MaxExtraPrecision} from 200
to 1200.
Starting from comparatively inexpensive settings, we
increased the precision and integration controls until
further refinement left the quantities of interest
stable to the accuracy reported.
The production calculations used the lowest settings
that passed these convergence checks.

\begin{figure}[t!]
\begin{center}
\includegraphics[scale=0.35]{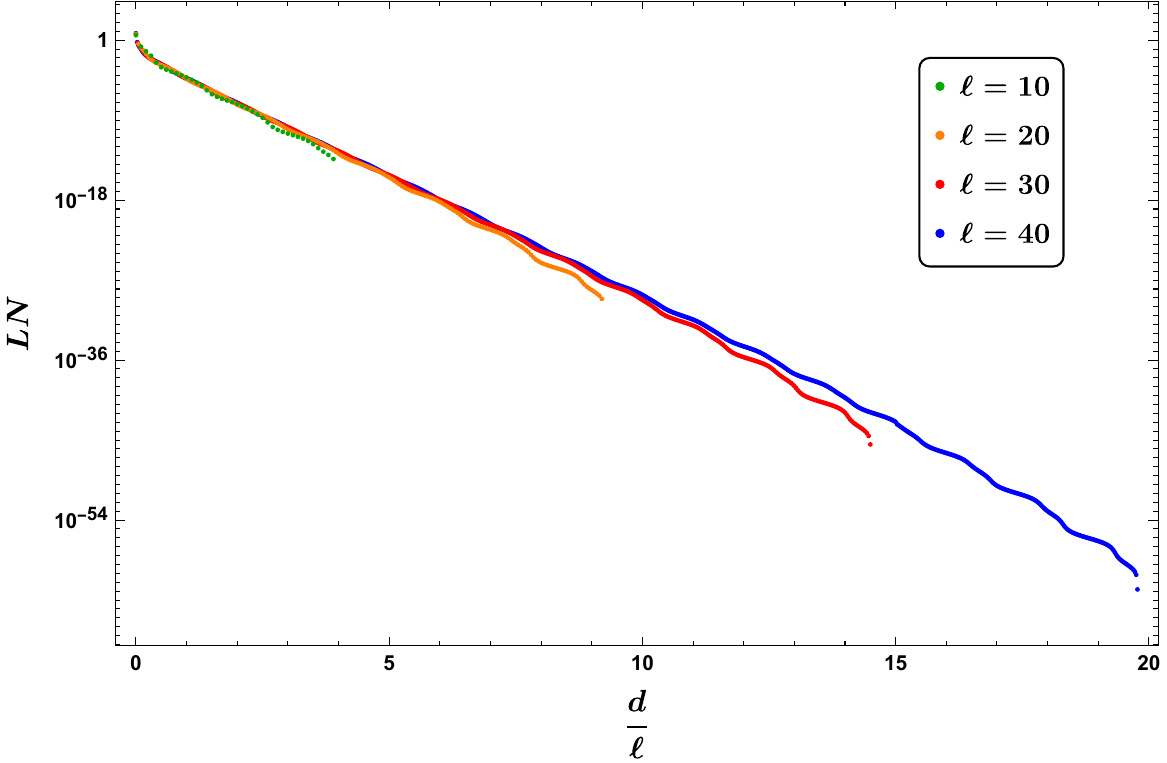}
\end{center}
\caption{Logarithmic negativity for $z=3$, different values of $\ell$ and $m=\frac{10^{-2}}{\ell}$, as a function of $\frac{d}{\ell}$.}
\label{fig:scaling}
\end{figure}

An appealing physical interpretation of the fact that the existence of the negativity sphere is a lattice artifact is that extracting spacelike entanglement at sufficiently large separations requires interaction profiles with support at frequencies higher than those allowed by the lattice cutoff \cite{Gao:2024nbl}. In other words, the lattice fails to resolve the high-frequency 
modes necessary to access the long-distance entanglement.

In the following, we distinguish between two physically relevant regimes. The continuum massless regime is characterized by $\epsilon\ll d,\ell\ll\xi$ where $\epsilon$ is the lattice spacing, $d$ is the separation between the intervals, $\ell$ is the interval length, and $\xi=1/m$ denotes the correlation length. In this regime, the infrared divergence dominates the field correlations while continuum physics is preserved. By contrast, the (highly) massive regime corresponds to $d,\ell\gtrsim(\gg) \xi$ where correlations decay over distances (much) shorter than the subsystem size and separation.

Since several of our results involve comparing different subsystem sizes, it is important first to establish the appropriate continuum scaling. When comparing subsystems of different physical sizes in the continuum, the ratio between the subsystem size and the correlation length must be kept fixed. Otherwise, the subsystems probe different physical regimes, making a direct comparison meaningless. While this is well-known for the harmonic chain (see for instance \cite{Klco:2021cxq}), Figure \ref{fig:scaling} demonstrates that, after fixing $m\ell$, the LN exhibits a universal scaling behavior also for $z\neq1$. This confirms that the observed behavior is governed by the dimensionless ratio $m\ell$ rather than by the subsystem size itself.

\begin{figure}[t!]
\begin{center}
\includegraphics[scale=0.35]{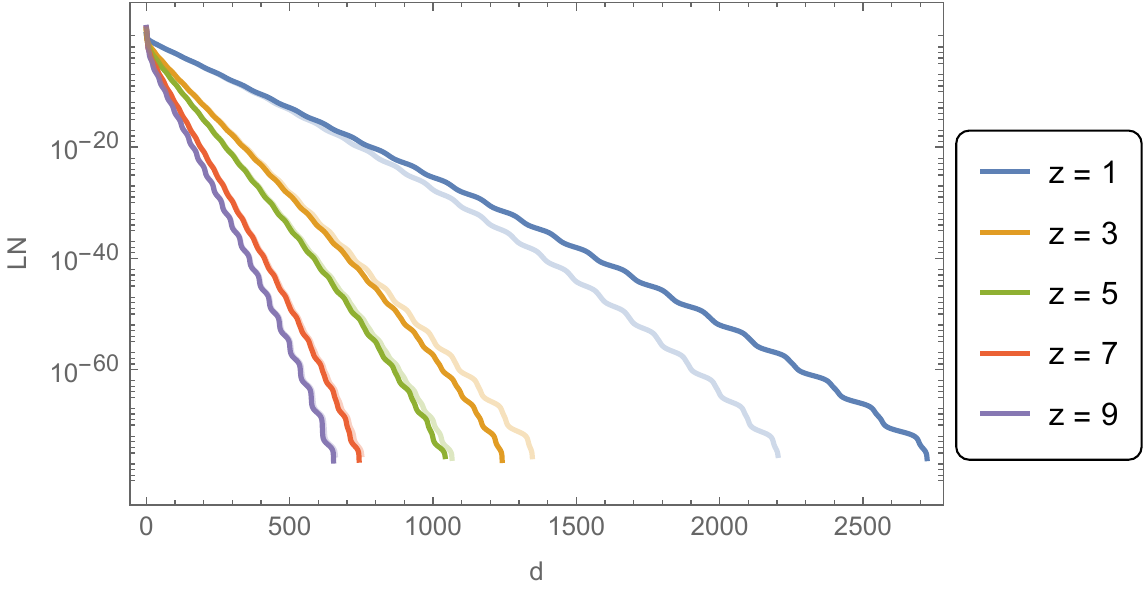}
\hspace{4mm}
\includegraphics[scale=0.35]{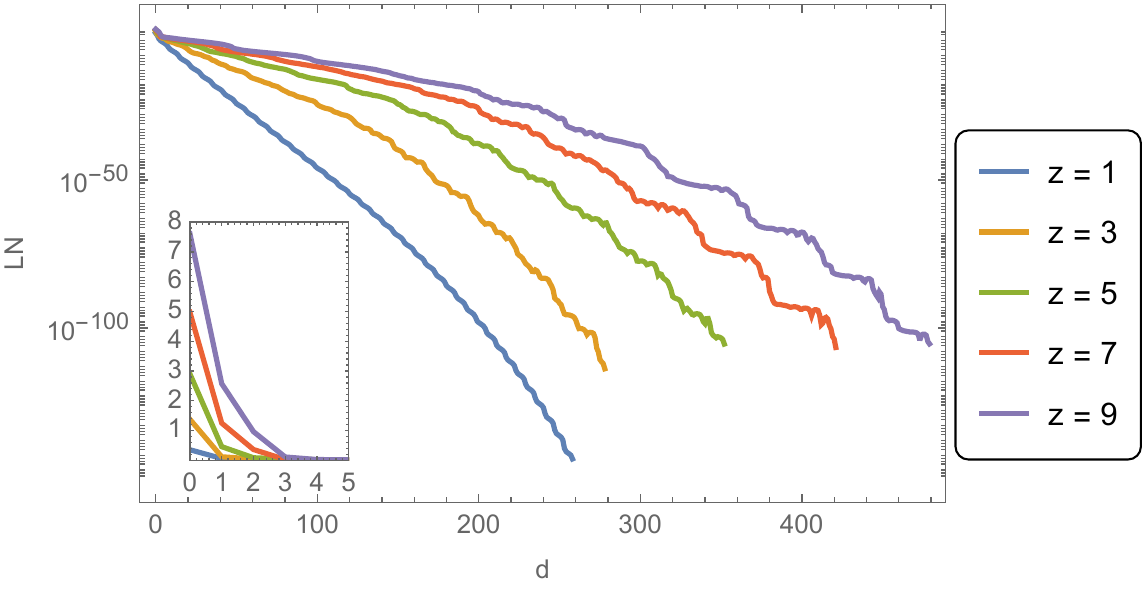}
\end{center}
\caption{LN for massless and massive regimes as a function of $d$. Left: massless regime with $m\ell=0.00005$ (pale curves correspond to $m\ell=0.05$). We have used $\ell=50$. Here the best fit finds that $f_z$ increases with $z$ as $z^{0.65}$. Right: massive regime with $m\ell=40$. We have used $\ell=80$. Here the best fit find $g_z\sim z^{-0.69}$.
The insets show the adjacent limit where the ultraviolet modes are dominant. 
}
\label{fig:LN}
\end{figure}

The numerical results presented in Figure \ref{fig:LN} indicate that, in the massless (scale-invariant) regime, increasing $z$—which corresponds to a softer infrared sector—makes the logarithmic negativity more spatially localized, resulting in a smaller negativity sphere radius $d_{\cancel{\mathcal{E}}}$. At first sight, this behavior appears counterintuitive if one focuses solely on the interaction pattern of the LHC Hamiltonian, Eq. \eqref{eq:Hz}, where the interaction range increases with $z$. As we demonstrate in the following subsection, this behavior originates from the interplay between the field and momentum correlation channels, together with the cancellation of the leading infrared contribution in logarithmic negativity. In other words, in the massless (scale-invariant) regime, increasing $z$ concentrates quantum correlations into a shorter spatial range, making them stronger at short distances but weaker at large separations, despite the extended interaction range of the underlying Hamiltonian. By contrast, when mass deformations are introduced, the negativity sphere radius increases with $z$, closely following the increase in the interaction range of the Hamiltonian. This trend is consistent with the behavior previously observed when comparing nearest-neighbor interactions with interactions between sites separated by up to three lattice spacings \cite{Klco:2019yrb}.

The numerical data further reveal distinct scaling laws in the massless and massive regimes. In the massless regime, LN decays exponentially with the normalized separation, $\mathcal{E}\sim e^{-f_z\frac{d}{\ell}}$, where the coefficient $f_z$ is an increasing function of $z$, extending the $z=1$ result reported in \cite{Marcovitch:2008sxc,Klco:2021biu}. In the massive regime, logarithmic negativity initially exhibits an exponential decay, $\mathcal{E}\sim e^{-g_z\,md}$, where $g_z$ is a decreasing function of $z$. At larger separations, however, the decay crosses over to a quadratic exponential form, $\mathcal{E}\sim e^{-h_z\,(md)^2}$, thereby generalizing the $z=1$ behavior reported in \cite{Gao:2024nbl,Marcovitch:2008sxc}.

Table \ref{tab:overall} summarizes the behavior of $d_{\cancel{\mathcal{E}}}$ in both the massless and massive regimes. The first two rows, corresponding to the massive regime, show that $d_{\cancel{\mathcal{E}}}$ increases with $z$. In contrast, the last two rows, corresponding to the massless regime, show the opposite trend: $d_{\cancel{\mathcal{E}}}$ decreases as $z$ increases.

\begin{table}[h]
        \centering
        \begin{tabular}{c||ccccc}
            $m$ & $z=1$ & $z=3$ & $z=5$ & $z=7$ & $z=9$ \\
            \hline\hline
            $0.5$ & 158 & 170 & 218 & 248 & 282 \\
            $0.4$ & 178 & 199 & 239 & 299 & 319 \\
            $0.02$ & 787 & 856 & 852 & 1043 & 1258 \\  
            $10^{-3}$ & 2203 & 1346 & 1066 & 751 & 659 \\
            $10^{-6}$ & 2723 & 1241 & 1043 & 742 & 653 \\   
            $10^{-9}$ & 2723 & 1241 & 1043 & 742 & 653 \\   
        \end{tabular}  
\caption{The $z$ and $m$ dependence of $d_{\cancel{\mathcal{E}}}$ for odd $z$ with fixed $\ell=50$. The first two lines correspond to the massive regime where $m\ell=25,20$, and the third line with $m\ell=1$ marginally corresponds to the same regime. The last two lines correspond to the massless regime with $m\ell=5\times 10^{-5},5\times 10^{-8}$ while the $m=10^{-3}$ line with $m\ell=5\times 10^{-2}$ and larger $md$ behaves more like the massless regime.}
        \label{tab:overall}
\end{table}

An additional feature of our results is the non-monotonic dependence of the negativity sphere radius on $z$ in the intermediate regime, $m\ell\sim1$. In this crossover regime, where the correlation length becomes comparable to the subsystem size, neither the massless nor the massive behavior dominates. Instead, the negativity sphere smoothly interpolates between the two limiting regimes, giving rise to a non-monotonic dependence on $z$. This crossover further supports the conclusion that the negativity sphere is governed by the interplay between the infrared structure and the physical length scales of the problem.

These opposite trends demonstrate that \textit{the negativity sphere provides a sensitive probe of the infrared structure of the vacuum rather than merely reflecting the interaction range of the underlying lattice Hamiltonian. }

\begin{figure}[h]
\begin{center}
\includegraphics[scale=0.4]{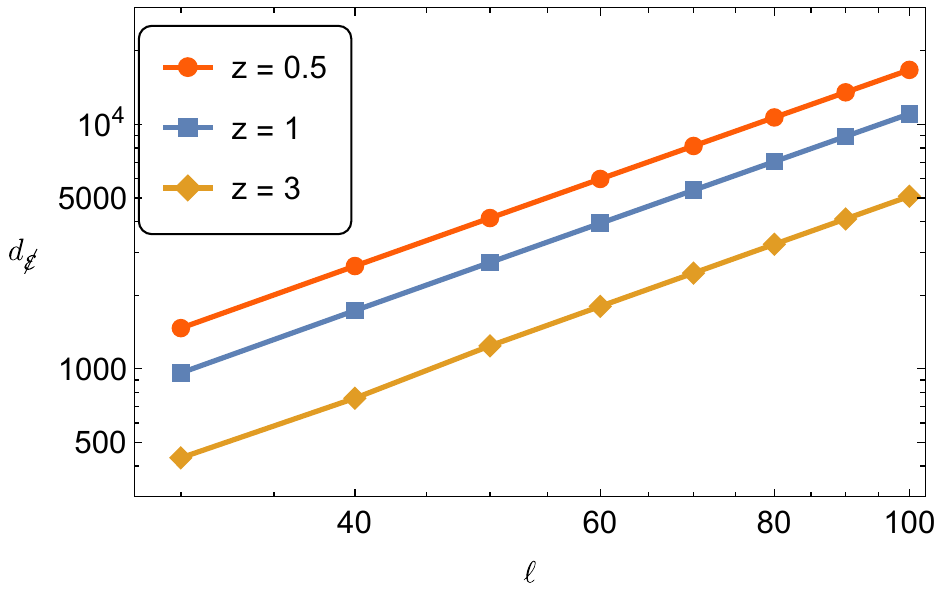}
\end{center}
\caption {The scaling of $d_{\cancel{\mathcal{E}}}$ with $\ell$. All data sets are fitted by $\ell^\alpha$ where $\alpha\approx2$ with $< \%1$ error.}
\label{fig:dneg}
\end{figure}

\paragraph{Dependence on interval size.}
We next examine how the negativity-sphere radius depends on the interval size in the massless regime. At fixed lattice spacing $\epsilon=1$ and mass $m=10^{-6}$, our
numerical results for $\ell=30,40,\ldots,100$ are consistent with
\begin{align}
d_{\cancel{\mathcal{E}}}\simeq\gamma(z)\ell^2,
\end{align}
for all three exponents examined, $z=1/2,1,3$, as shown in
Fig.~\ref{fig:dneg}.  The fitted coefficients are
$\gamma(1/2)\simeq1.69$, $\gamma(1)\simeq1.12$, and
$\gamma(3)\simeq0.52$. Thus, the dispersion exponent changes the
amplitude of the negativity range while leaving its observed
size-scaling exponent unchanged. The $z=1$ result agrees with the
relativistic scalar-field scaling reported in
Ref.~\cite{Klco:2021biu}.

Restoring the lattice spacing, this scaling takes the form
\begin{align}
\frac{d_{\cancel{\mathcal{E}}}}{\ell}
\simeq \gamma(z)\frac{\ell}{\epsilon}.
\end{align}
Within the massless regime studied, the negativity range
in units of the interval size grows linearly with the spatial
resolution $\ell/\epsilon$, or equivalently with
$\Lambda_{\rm UV}\ell$ for a momentum cutoff
$\Lambda_{\rm UV}\propto\epsilon^{-1}$.
These results extend the observed UV-IR scaling beyond the
relativistic case to the Lifshitz dispersions examined here:
finer spatial resolution supports NPT entanglement over larger
relative separations, while $\gamma(z)$ captures the dispersion
dependence, including the influence of the infrared sector.
The following analysis clarifies how infrared fluctuations
affect $d_{\cancel{\mathcal{E}}}$.
In Section~\ref{sec:Modewise}, we further examine the
long-wavelength spatial participation of the mode pair
carrying the surviving entanglement near the negativity sphere.

\subsection{Correlation Functions and Infrared Cancellation in Logarithmic Negativity}
The numerical results above indicate that the infrared structure of the theory governs the negativity sphere in a highly nontrivial manner. To provide a physical interpretation for the origin of this behavior, we now examine the field and momentum correlation functions and identify the specific combinations of field and momentum correlations that determine logarithmic negativity. In this subsection, we discuss the \textit{integer} $z\geq 1$ case, and the next subsection is devoted to positive $z<1$.

\subsubsection{Massless Regime}
In the massless limit, the momentum channel correlators for two spatial points at distance $r$ and non-even $z>0$ in the continuum are given by
\be\label{eq:H}
H_z(r;m)\equiv\langle\pi(0)\pi(r)\rangle= \int\,\frac{dk}{2\pi}\, |k^z| e^{ikr}=-\frac{1}{4\pi r^{1+z}}\sin \left(\frac{\pi z }{2}\right) \Gamma (1+z),\qquad m\to0\,,
\ee
while the field channel suffers from an infrared divergence for $z>1$. In this case, the exact behavior of the field channel is given by Meijer G-functions, where not only the elements but also the length of the vector of parameters depends on $z$, and we avoid reporting its exact form since it is not instructive for our discussion.

For odd Lifshitz exponents $z=2n+1>1$, the continuum
field correlator has the following small-mass expansion
at fixed nonzero separation $r$:
\begin{equation}\label{eq:Gz>1}
G_z(r;m)\equiv\langle\phi(0)\phi(r)\rangle
=
\sum_{p=0}^{n-1}
A_{z,p}\,r^{2p}m^{1-z+2p}
+
B_z\,r^{z-1}\log(mr)
+
\cdots,
\qquad m\to0.
\end{equation}
Here $A_{z,p}$ and $B_z$ are independent of $m$ and $r$,
and the remainder is bounded as $m\to0$ at fixed $r$.\footnote{
For the continuum dispersion
$\omega_k=\sqrt{k^{2z}+m^{2z}}$, with the normalization
$G_z(r;m)=\int_{-\infty}^{\infty}
\frac{dk}{2\pi}\frac{e^{ikr}}{2\omega_k}$, the coefficients
for $z=2n+1>1$ are
$$
B_z=\frac{(-1)^{n+1}}{2\pi(2n)!}
\qquad,\qquad
A_{z,p}
=
\frac{(-1)^p
\Gamma\!\left(\frac{2p+1}{2z}\right)
\Gamma\!\left(\frac12-\frac{2p+1}{2z}\right)}
{4\pi z(2p)!\sqrt{\pi}},
\qquad p=0,\ldots,n-1\,.
$$
}
The leading divergence, $A_{z,0}m^{1-z}$, is spatially
uniform. For $z\geq5$, additional power-law divergences
occur with polynomial dependence on the separation.
Their coefficients are nonzero, so these terms must
also be retained when analyzing infrared cancellation.
The case $z=1$ is marginal: its leading infrared
divergence is logarithmic.

\begin{figure}[t]
\begin{center}
\includegraphics[scale=0.23]{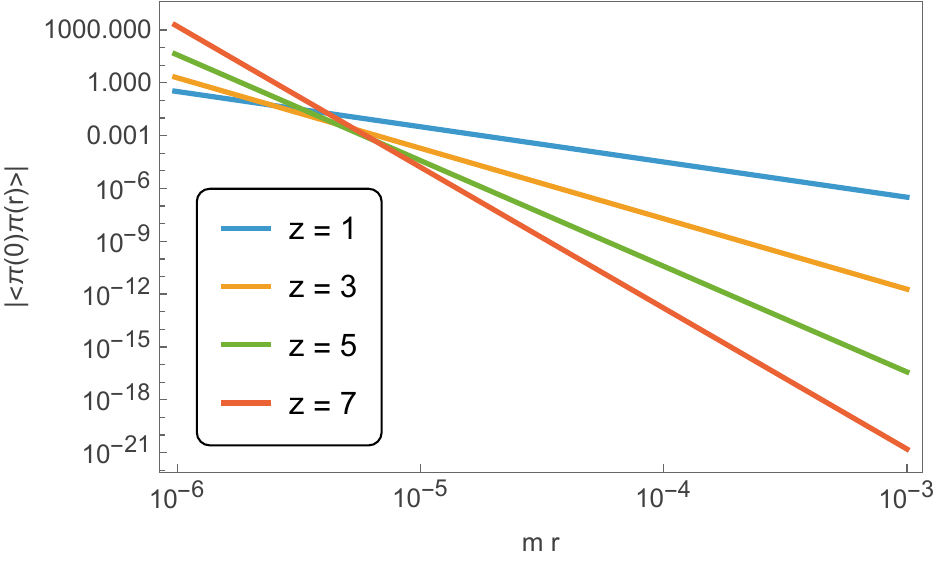}
\hspace{2mm}
\includegraphics[scale=0.23]{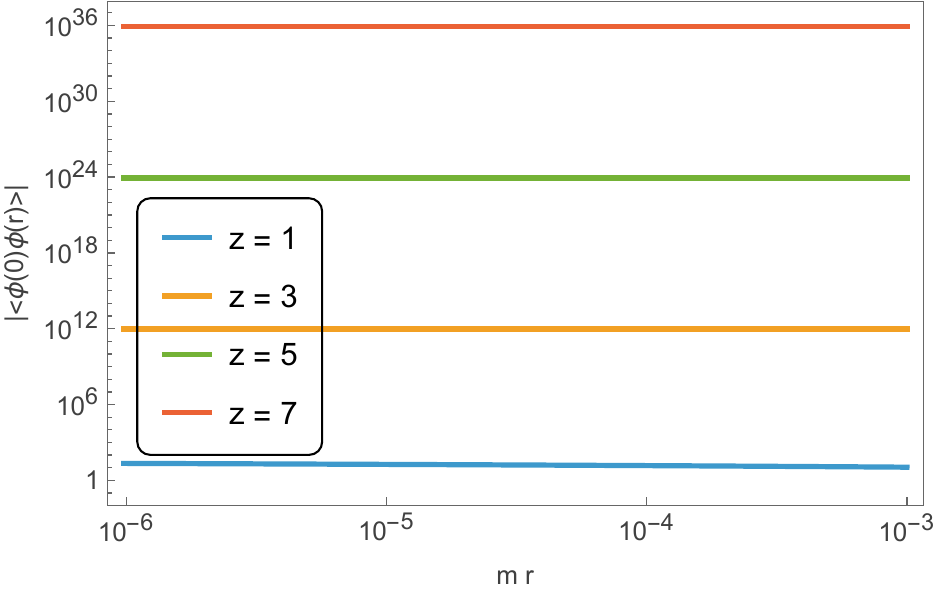}
\hspace{2mm}
\includegraphics[scale=0.23]{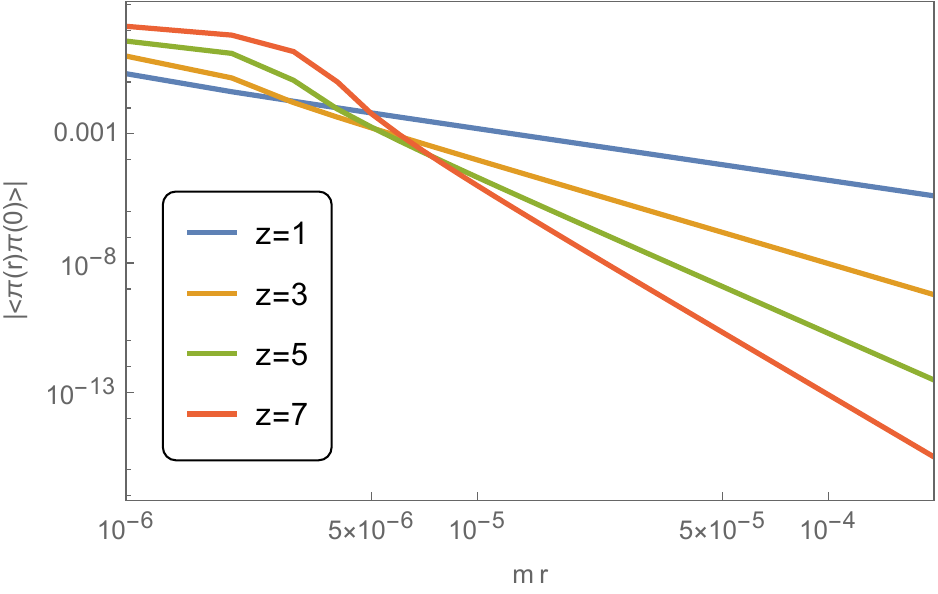}
\hspace{2mm}
\includegraphics[scale=0.23]{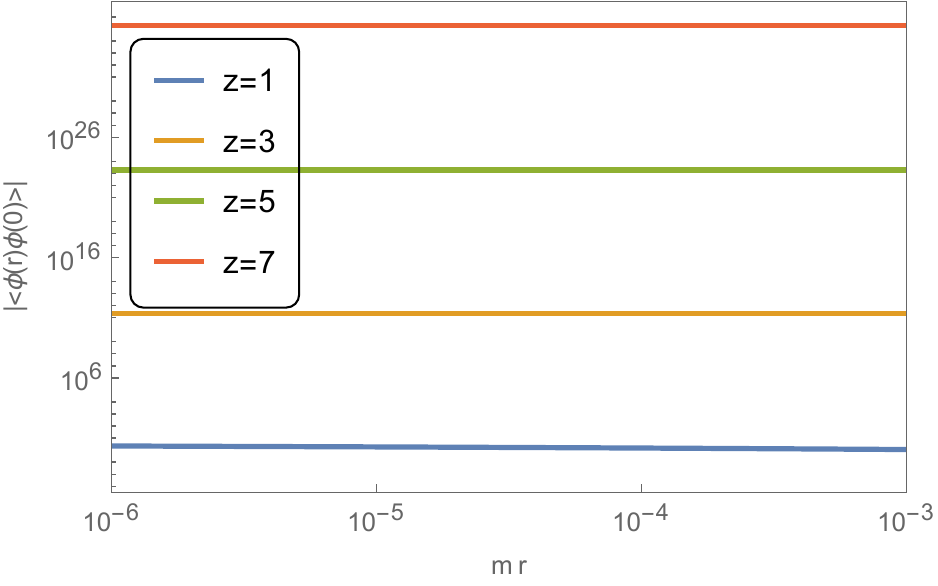}
\end{center}
\caption {Correlation functions for the field and momentum channels for $z\geq1$ in the massless regime. The two left panels belong to the continuous theory where the field-channel computed numerically agrees with the estimated function given in Eq. \eqref{eq:Gz>1}.
The two right panels correspond to the same correlation functions calculated on the lattice, which qualitatively agree with the continuum results. In all plots, we have set $m=10^{-6}$.}
\label{fig:zg1CorrMassless}
\end{figure}

In Fig.~\ref{fig:zg1CorrMassless} we illustrate the behavior of the massless correlation functions given in Eqs.~\eqref{eq:H} and \eqref{eq:Gz>1}.
Due to the power-law behavior given in Eq.~\eqref{eq:H}, momentum correlations are stronger for larger $z$ in the $r\to0$ limit, while this ordering is reversed as the separation increases.
Note that the momentum channel (upper left-most panel) does not depend on an infrared cutoff; nevertheless, to obtain uniform plots, we have scaled the $r$-axis of the left panel.
The field-channel correlations being dominated by the infrared divergence given in Eg.\eqref{eq:Gz>1}, behave differently; larger values of $z$ yield stronger correlations. 

Before examining the role of infrared modes in quantum correlations, we briefly discuss the contribution of ultraviolet modes. Regardless of the mass parameter—which is irrelevant in the UV—the
dispersion relation assigns higher energy to larger values of $z$ at fixed momentum. Consequently, in the adjacent limit of the subregions, the correlations become stronger for higher $z$. The correlation functions exhibit the same physics as that illustrated in Fig.~\ref{fig:zg1CorrMassless} for the massless regime. In the logarithmic-negativity pattern, these ultraviolet modes are responsible for the larger LN values observed in the adjacent limit in both regimes (see the insets of Fig.~\ref{fig:LN}).
   
In the massless regime, increasing $  z  $ suppresses the dispersion of low-momentum modes, $  |k|^z  $, rendering the long-wavelength sector increasingly soft. In the momentum channel, this alters the weighting of low-momentum modes and thereby weakens the long-distance correlations. By contrast, the field-channel correlations exhibit the opposite trend: the leading infrared divergence dominates such that it yields stronger field correlations at larger $z$ at arbitrary separation. This tension between the momentum- and field-channel behaviours seems to preclude a clear interpretation of the observed LN pattern.

Let us first clarify what Fig.~\ref{fig:LN} and the data in Table~\ref{tab:overall} reveal about the $m\to0$ limit. A comparison of the last two rows of the table clearly indicates that $d_{\cancel{\mathcal{E}}}$ is independent of the infrared divergence due to the existence of the zero-mode. Even more strikingly, a comparison of the pale and colored curves in the left panel of Fig.~\ref{fig:LN} shows that, as the infrared sector of the theory becomes softer, the logarithmic negativity becomes mass-independent for arbitrary separations between the subregions. In particular, in the adjacent limit—where the momentum-channel correlation pattern agrees with the diverging field-channel pattern in assigning larger values to larger $z$—the LN exhibits an $m$-independent behavior.

\begin{figure}[t]
\begin{center}
\includegraphics[scale=0.25]{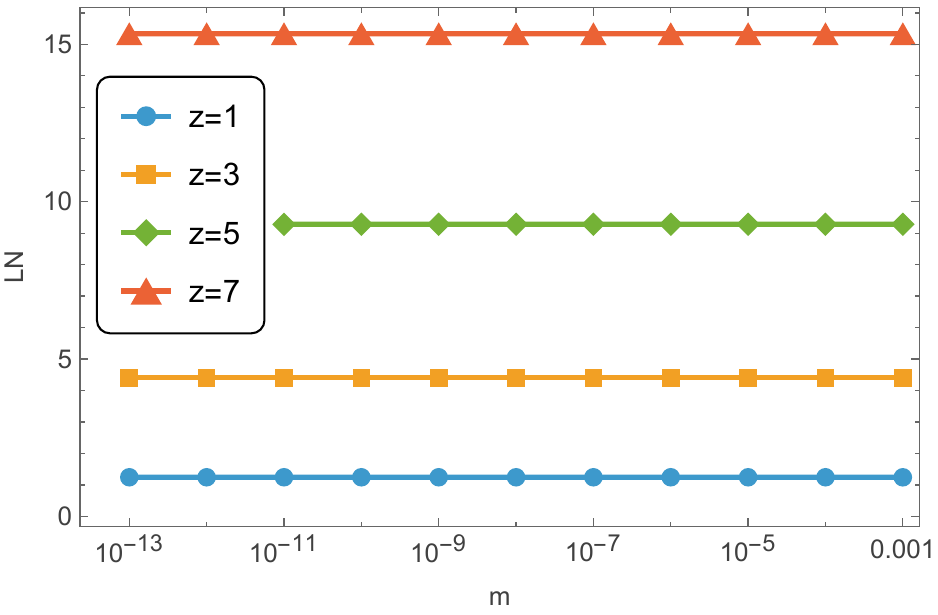}
\includegraphics[scale=0.25]{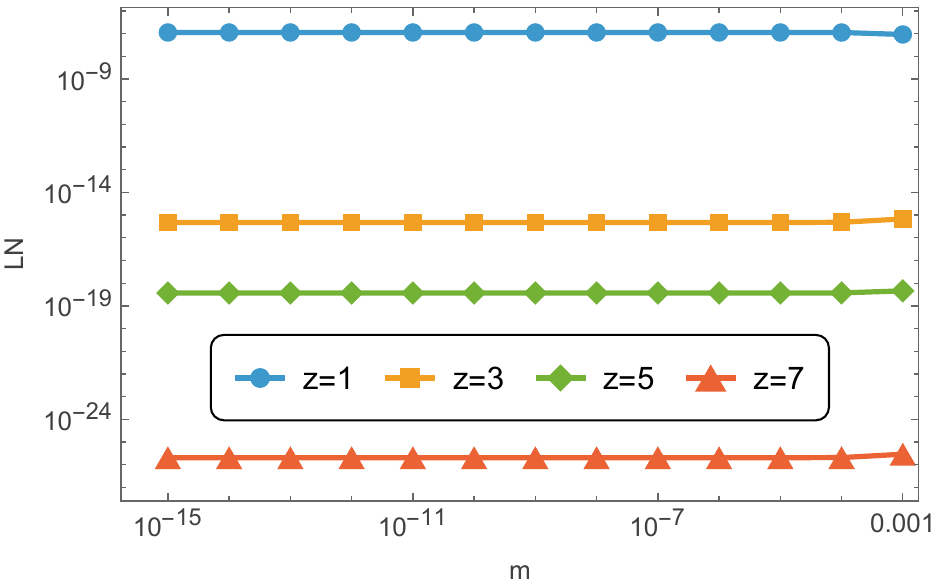}
\includegraphics[scale=0.25]{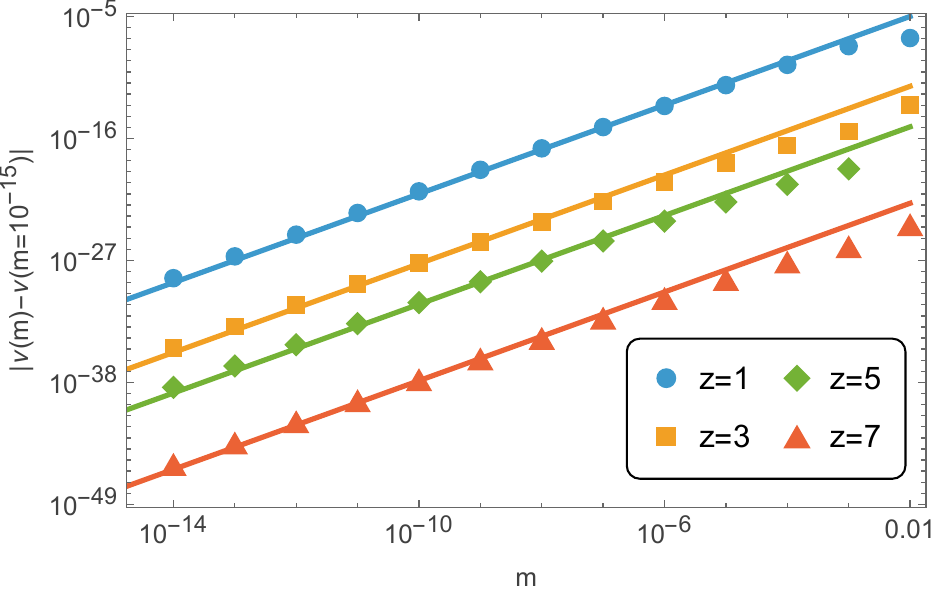}
\includegraphics[scale=0.22]{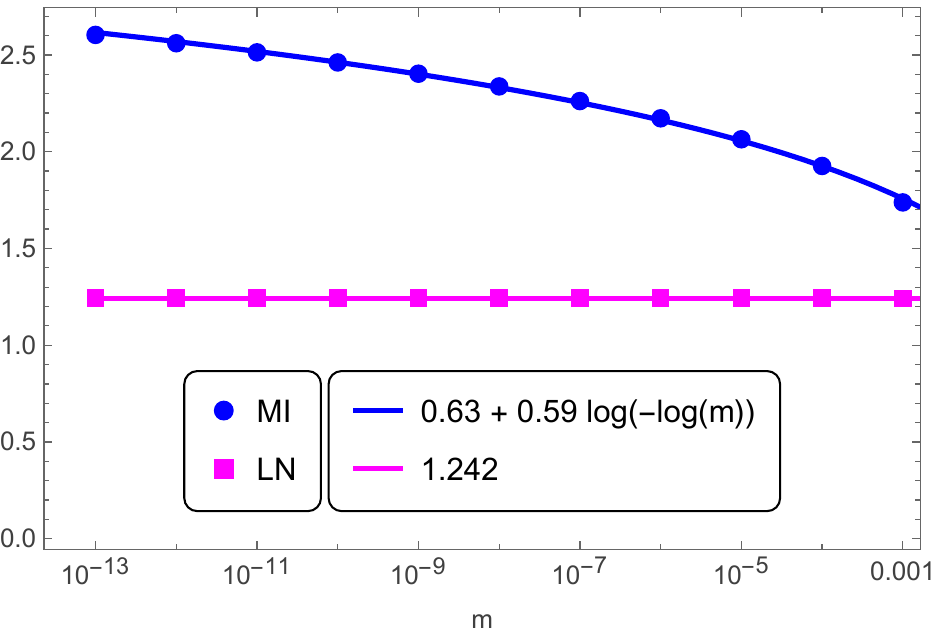}
\includegraphics[scale=0.25]{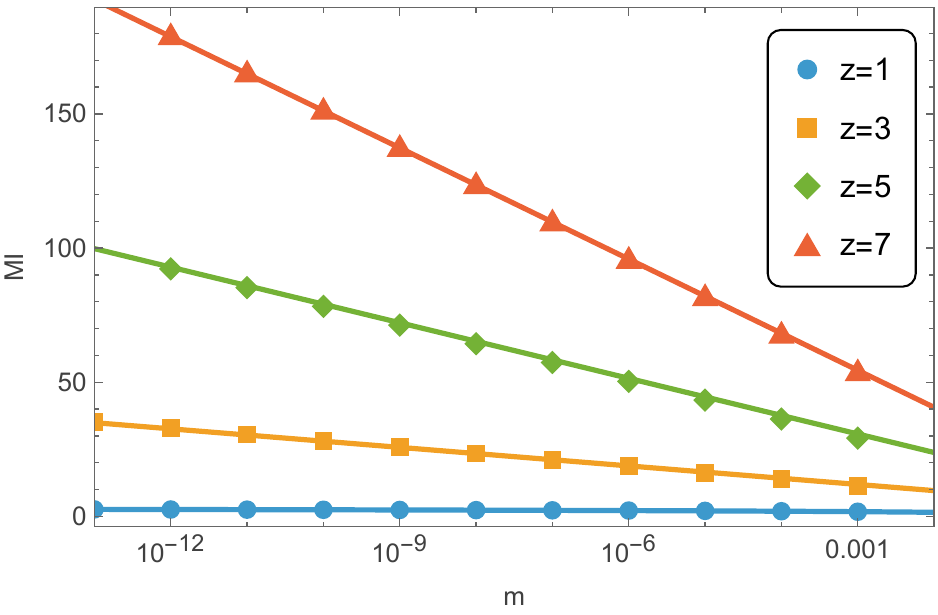}
\includegraphics[scale=0.25]{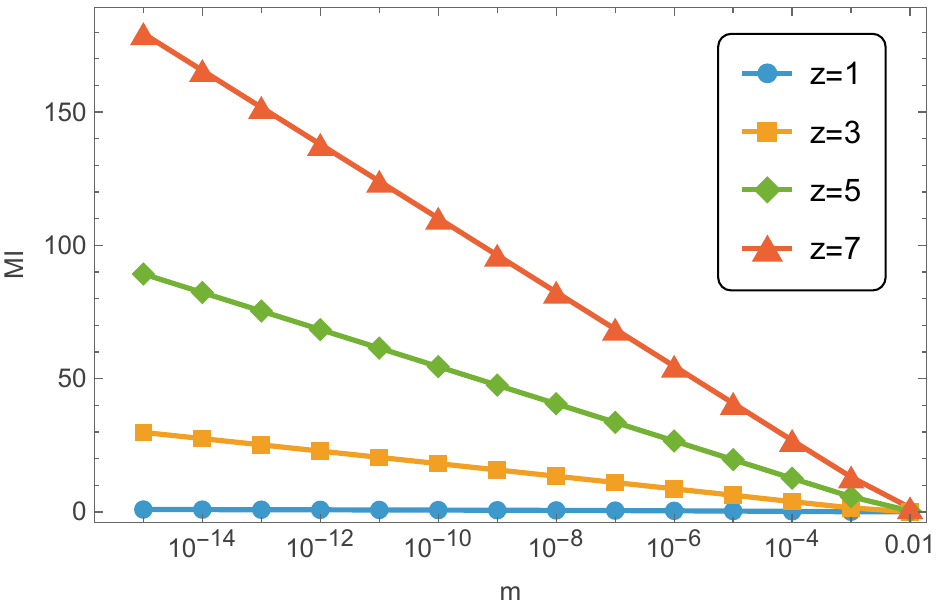}
\includegraphics[scale=0.25]{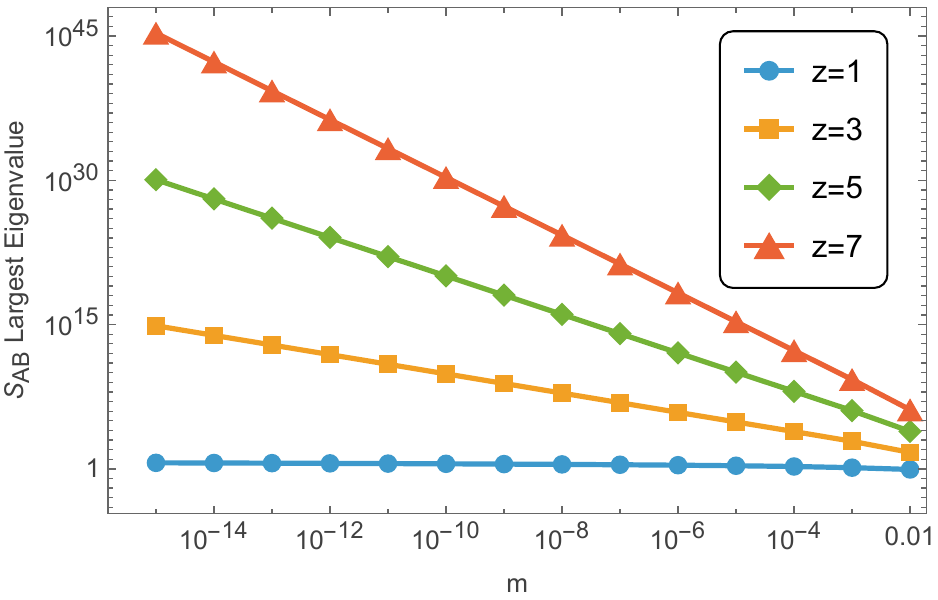}
\includegraphics[scale=0.22]{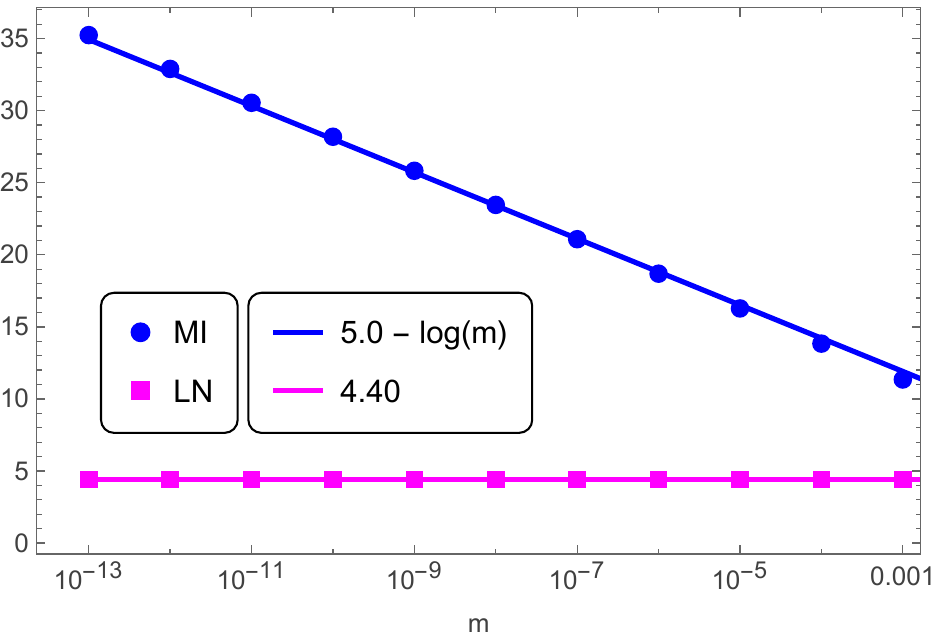}
\end{center}
\caption{
The leading mass dependence of LN and MI in the massless regime. The first column from the left corresponds to the values of LN and MI in the adjacent limit. The second column from the left shows LN and MI in the large-distance limit, where $d=5\ell=250$. Note the reversed pattern of LN at large distance compared to the adjacent case, compatible with the correlation pattern.
The third column from the left shows the largest contributing eigenvalues to LN and MI in the large-distance limit. In the top panel, the solid curves indicate the $m^2$ scaling of the subtracted largest contributing eigenvalue. The bottom panel shows the largest eigenvalue of $S_{AB}$ where the solid lines show $\sqrt{-\log(m)}$ for $z=1$ and $m^{-n}$ for $z=2n+1>1$. The largest eigenvalue of $S_A$ exhibits the same scaling. For a clearer comparison, the right column shows LN versus MI for $z=1$ in the top panel and for $z=3$ in the bottom panel in the adjacent limit. Note the scales of the vertical axis on the different panels. 
}
\label{fig:LNm0MIm0}
\end{figure}

In Fig.~\ref{fig:LNm0MIm0} we show that the largest value of the LN is unaffected by the leading infrared divergence. Note that the eigenvalues contributing to LN and von Neumann entropy form an exponentially decaying hierarchy, which makes it sufficient to examine the largest contributing eigenvalues \cite{Klco:2020rga}. In the other panel, we confirm this $m$-independence of LN in the adjacent limit. To emphasize that this behavior is a distinctive feature of quantum correlations, we also show that the mutual information depends clearly on the infrared divergence.

This behavior shows a specific cancellation of (at least) the leading infrared divergence in LN. Such a mechanism should be tractable through a careful analysis of the structure of the PT symplectic eigenvalues. Suppose there is a single mode in each subregion. Then we parametrize the elements of $G$ by $g_0$ and $g_r$ and similarly for $H$, we find that the PT symplectic eigenvalues are 
\be
\nu_1=\sqrt{(g_0+g_r)(h_0-h_r)}
\;\;\;\;,\;\;\;\;
\nu_2=\sqrt{(g_0-g_r)(h_0+h_r)}\,.
\ee
In the massless regime, due to the separation-independent power-law divergence of the elements of $G$, $\nu_2$ is the eigenvalue that may contribute to LN, where the $(g_0-g_r)$ factor results in the cancellation of the leading infrared divergence in LN.

Note that the symplectic eigenvalues in this case are given by
\be
\lambda_1=\sqrt{(g_0-g_r)(h_0-h_r)}
\;\;\;\;,\;\;\;\;
\lambda_2=\sqrt{(g_0+g_r)(h_0+h_r)}
\ee
where this structure clearly shows that mutual information, which is known as a quantity that is ultraviolet divergent-free, does not have such a structure (due to the contribution of $\lambda_2$), and its value does depend on the infrared cutoff.

This simple example illustrates the cancellation of the
leading spatially uniform infrared contribution.
For higher odd $z$, differences of field correlators can
still contain subleading power-law divergences, so this
observation alone does not establish the boundedness of
logarithmic negativity for extended regions.
Nevertheless, excluding the $m=0$ point, boundedness can be established without
resolving these divergent terms individually.

For fixed finite regions $R=A\cup B$ on the infinite
lattice, the restricted covariance blocks satisfy
\begin{equation}
G_R(m)\geq g_*\,\mathbb I,
\qquad
H_R(m)\geq h_*\,\mathbb I,
\qquad 0<m\leq m_0,
\end{equation}
with positive constants $g_*$ and $h_*$ independent of
$m$.
The first bound follows from the finite upper bound
on the lattice dispersion.
The second follows from
$H_R(m)\geq H_R(0)$ and the strict positivity of the
massless momentum covariance restricted to a finite
set of sites.
Writing $\Gamma$ for the diagonal matrix that reverses
the momenta in $B$, the squared PT symplectic
eigenvalues are the eigenvalues of
\begin{equation}
G_R^{1/2}\Gamma H_R \Gamma G_R^{1/2}
\geq h_*G_R
\geq g_*h_*\,\mathbb{I}.
\end{equation}
Consequently,
\begin{equation}
\nu_j(m)\geq\sqrt{g_*h_*}>0.
\label{eq:PT_uniform_lower_bound}
\end{equation}
No PT symplectic eigenvalue can therefore approach
zero as the mass vanishes.
Since logarithmic negativity is a finite sum of
nonnegative contributions that diverge only when a
PT symplectic eigenvalue approaches zero, it remains
uniformly bounded in this limit.
The detailed proof is presented in
Appendix~\ref{app:LNbound}.
This statement holds at fixed lattice spacing and
fixed finite regions; the bound need not remain
uniform as the lattice spacing vanishes or the
regions grow.

The bound excludes a divergence of logarithmic
negativity but does not determine its limiting value
or its leading mass corrections.
Our numerical results support convergence to a finite
massless limit $\mathcal E_0$.
As summarized in Table~\ref{tab:massless_scaling},
the difference
$\Delta\mathcal E(m)\equiv\mathcal E(m)-\mathcal E_0$
vanishes with a positive power of $m$ in the regimes
studied.
These results characterize the approach to the
massless limit beyond the general boundedness
statement.

\begin{table}[h]
\centering
\begin{tabular}{lcccc}
\toprule
& \multicolumn{2}{c}{Adjacent}
& \multicolumn{2}{c}{Large separation $d\gg\ell$} \\
\cmidrule(lr){2-3}\cmidrule(lr){4-5}
Quantity
& $z=1$ & $z>1$
& $z=1$ & $z>1$ \\
\midrule
$\delta\nu_{\max}$
& $m^2$ & $m^{\alpha_\nu}$
& $m^2$ & $m^2$ \\

$\lambda_{\max}$
& $\sqrt{-\log m}$ & $m^{-n}$
& $\sqrt{-\log m}$ & $m^{-n}$ \\

$\delta\mathcal E$
& $m^2$ & $m^{\alpha_{\mathcal E}}$
& $m^2$ & $m^2$ \\

$I(A:B)$
& $\log(-\log m)$ & $-\log m$
& $\log(-\log m)$ & $-\log m$ \\
\bottomrule
\end{tabular}
\caption{
Leading mass dependence in the massless regime
at fixed geometry, with $m$ expressed in lattice units.
Here $\delta\nu_{\max}$ and $\delta\mathcal E$ denote
corrections relative to the corresponding finite
$m\to0$ limits, whereas the entries for $\lambda_{\max}$
and $I(A:B)$ describe their leading divergences.
Multiplicative coefficients, including their signs,
are suppressed.
The fractional exponents satisfy
$0<\alpha_\nu,\alpha_{\mathcal E}<1$ and need not coincide.
The $z>1$ columns refer to the odd integer exponents
examined, with $z=2n+1$.
}
\label{tab:massless_scaling}
\end{table}

More specifically, for well-separated intervals, the
largest eigenvalue contributing to logarithmic negativity
is numerically consistent with
\begin{equation}
\nu_{\max}(m)\simeq A_z+B_zm^2,
\end{equation}
where the coefficients depend on the geometry as well
as on $z$. We find $B_1<0$, whereas
$B_z>0$ for $z=3,5,7,9$.
The logarithmic negativity likewise exhibits a quadratic
mass correction,
\begin{equation}
\mathcal E(m)\simeq \mathcal E_0+C_zm^2.
\end{equation}
Here $\mathcal E_0$ is its finite massless limit.
The coefficient $C_z$ incorporates the logarithmic
dependence on the contributing eigenvalues and should
be distinguished from $B_z$.

In contrast, the largest eigenvalue entering the
von Neumann entropies behaves as
$\lambda_{\max}\sim\sqrt{-\log m}$ for $z=1$ and
$\lambda_{\max}\sim m^{-n}$ for odd
$z=2n+1>1$, at fixed geometry.
The resulting mutual information retains an infrared
divergence: its leading mass dependence is
$I(A:B)\sim\log(-\log m)$ for $z=1$ and
$I(A:B)\sim-\log m$ for the odd exponents $z>1$
examined here.%
\footnote{
For $z=1$, the infrared scaling of the largest
eigenvalue and the entanglement entropy has been
studied in Ref.~\cite{Yazdi:2016cxn}, while the
double-logarithmic contribution to the entanglement
entropy was discussed in Ref.~\cite{Casini:2005zv};
see also Refs.~\cite{Pye:2015tta,Jain:2021ppx}.
}

For higher odd Lifshitz exponents, the field correlator contains subleading power-law infrared divergences in addition to its spatially uniform leading term. Their coefficients depend polynomially on the separation, so cancellation of the uniform contribution alone does not establish infrared finiteness. The uniform bound established above excludes a divergence
of logarithmic negativity from the full hierarchy of
infrared-divergent correlation terms at fixed lattice
spacing and fixed finite regions.
Our numerical results further support convergence to a
finite massless limit with positive-power corrections.
The detailed spectral mechanism beyond the leading-order
cancellation remains to be understood; boundedness does
not exclude finite infrared effects on the limiting
negativity or its mass corrections.

The infrared finiteness observed above distinguishes the
entanglement between two finite regions from that of a
pure bipartition.
Let $C$ denote the complement of $A\cup B$, so that
the global vacuum is pure on $ABC$.
Across the bipartition $A:(BC)$, the entropy $S(A)$
measures entanglement with the entire complement and
retains the infrared-divergent contribution.
Tracing out $C$ leaves $S(A)$ unchanged, but this
entropy no longer measures entanglement between
$A$ and $B$.
Its infrared enhancement therefore cannot be
identified directly with an enhancement of the
entanglement shared by the retained regions.

The same distinction arises when logarithmic
negativity is used for both bipartitions.
For the pure global state,
$\mathcal E(A:BC)=S_{1/2}(A)$, where $S_{1/2}$
is the R\'enyi entropy of order $1/2$.
This quantity inherits the infrared divergence
associated with a divergent reduced symplectic
eigenvalue.
For the mixed state on $AB$, however, our
small-subsystem analysis and numerical results
for extended intervals show that the leading
divergence cancels from the negativity-carrying
sector in the cases studied.
Mutual information remains sensitive to this
divergence.
The contrast thus reflects the different ways
infrared correlations enter the two bipartitions,
and cannot be attributed solely to the use of
different entanglement measures.

This cancellation concerns the leading divergent
contribution and does not eliminate all infrared
dependence.
Nevertheless, because logarithmic negativity
bounds distillable entanglement from above,
its finite massless limit excludes an unbounded
distillation rate between the retained intervals
as the infrared regulator is removed.
It does not establish that finite infrared
contributions to distillable entanglement vanish.

Finally, even integer values of $z$ require
separate consideration.
In the strictly massless continuum theory,
the momentum-channel correlator is proportional
to $\delta^{(z)}(r)$ and therefore has support
only at coincident points.
The field-channel correlator remains infrared
sensitive and requires a regulator, but the
momentum cross-correlations vanish between
regions separated by a nonzero distance.
For these Gaussian vacua, this structure implies
vanishing logarithmic negativity between
disjoint regions in the continuum.
It does not imply the absence of all correlations
or of entanglement across other bipartitions.
Related behavior has been established for the
$z=2$ quantum Lifshitz model
\cite{Chen:2017txi}.
On the lattice, nonzero logarithmic negativity
can persist below a finite separation threshold
for even $z$.
This threshold reflects the spatial regulator
and disappears in the continuum limit
\cite{Angel-Ramelli:2020wfo,MohammadiMozaffar:2017chk}.

\subsubsection{Massive Regime}
\begin{figure}[t]
\begin{center}
\includegraphics[scale=0.23]{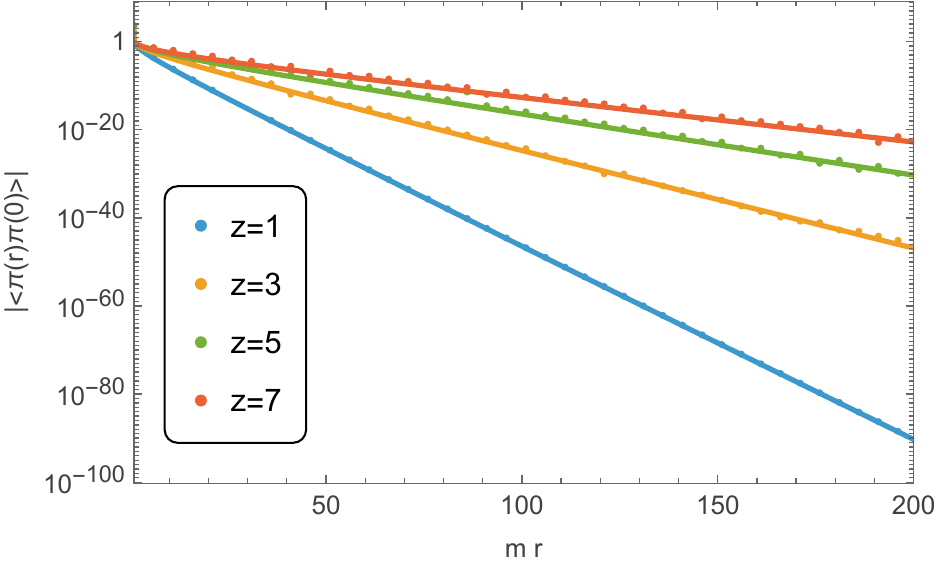}
\hspace{2mm}
\includegraphics[scale=0.23]{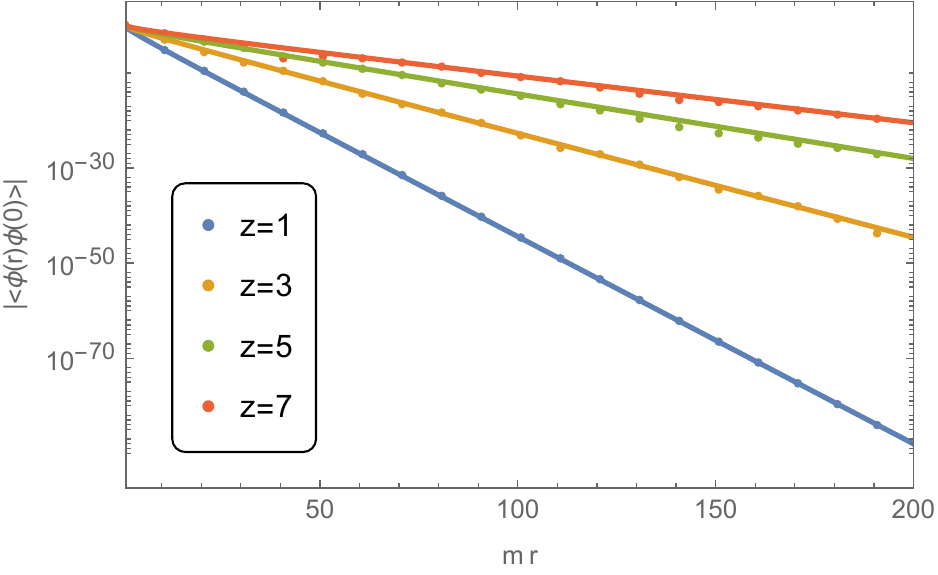}
\hspace{2mm}
\includegraphics[scale=0.23]{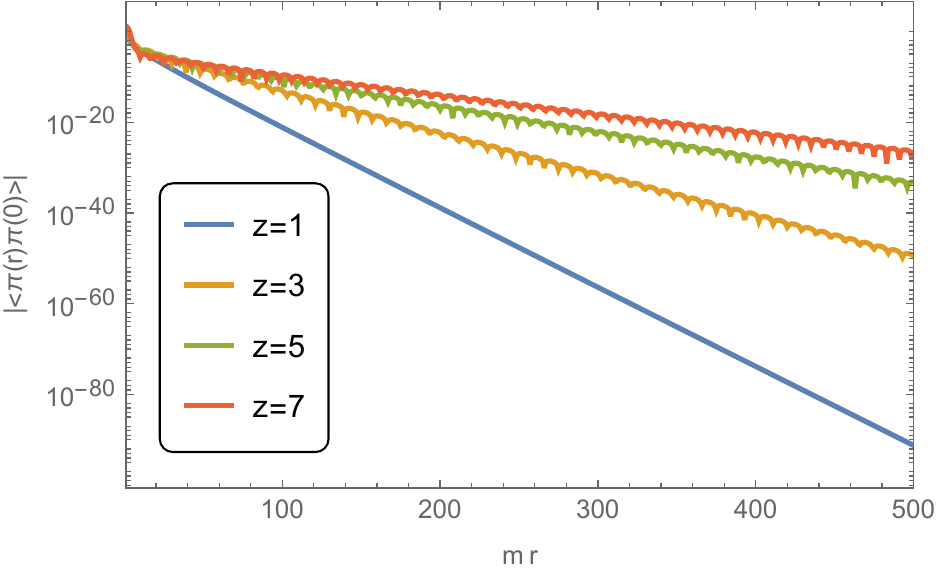}
\hspace{2mm}
\includegraphics[scale=0.23]{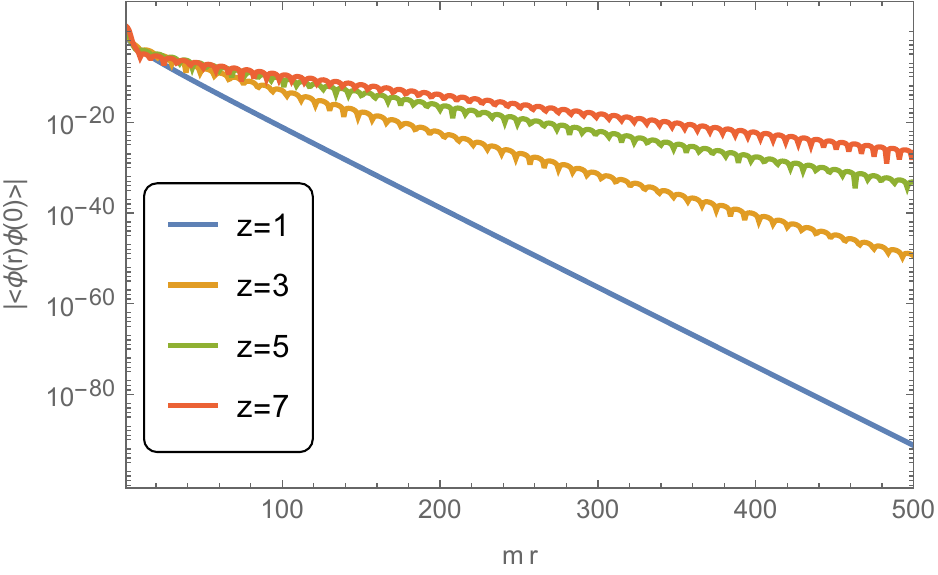}
\end{center}
\caption {Correlation functions for the field and momentum channels for $z\geq1$ in the massive regime.
The left panels are from the continuum theory. The solid curves show the estimated function given in Eq. \eqref{eq:GHmassive}, and the dots show the results from numerical calculation.
The two right panels show the same correlations calculated on the lattice. We have set $m=1$ in the continuum and $m=0.4$ on the lattice (lattice spacing is set to unity). The lattice-regularized results qualitatively agree with the continuum theory results.}
\label{fig:zg1CorrMassive}
\end{figure}

So far we have discussed the massless regime where the softness of the infrared sector played a key role in shortening the negativity sphere radius. In this part, we analyze the massive regime, where the behavior of negativity is again consistent with the correlation pattern, though the correlation pattern is drastically different from the massless regime in the sense that in the large-separation limit both of the momentum and field channels show stronger correlations for larger values of $z$.

For integer $z\geq1$, the large-distance behavior of the continuum correlators is controlled by the complex branch points of the dispersion.
The branch points closest to the real momentum axis have imaginary part $\kappa_z=m\sin(\pi/2z)$. Consequently, for $mr\gg1$, the field and momentum correlators have decay envelopes
\begin{align}
\begin{split}
\operatorname{Env}\!\left[
\langle\phi(0)\phi(r)\rangle
\right]
&\sim
\mathcal A_z(m)\,
\frac{e^{-\kappa_z r}}{r^{1/2}},
\\
\operatorname{Env}\!\left[
\langle\pi(0)\pi(r)\rangle
\right]
&\sim
\mathcal B_z(m)\,
\frac{e^{-\kappa_z r}}{r^{3/2}},
\end{split}
\label{eq:GHmassive}
\end{align}
where $\kappa_z=m\sin\!\left(\frac{\pi}{2z}\right)$ and $\mathcal A_z(m)$ and $\mathcal B_z(m)$ are independent of $r$.
For $z>1$, these envelopes multiply oscillatory factors with wave number $q_z=m\cos(\pi/2z)$ and generally different phases in the two channels.
For $z=1$, the leading decay is nonoscillatory. Thus Eq.~\eqref{eq:GHmassive} describes the large-distance envelopes rather than the full signed correlators.

The behavior of these correlation functions is illustrated in the right-hand panels of Fig.~\ref{fig:zg1CorrMassive}.
The left panels compare the numerical continuum correlators with the asymptotic decay envelopes in Eq.~\eqref{eq:GHmassive}. The right panels present the same correlation functions computed on an infinite lattice, showing that the lattice results closely reproduce the continuum behavior.

The behavior of the negativity-sphere radius is fully consistent with this correlation-based picture. As shown in the right panel of Fig.~\ref{fig:LN} and in the first two rows of Table~\ref{tab:overall}, the negativity-sphere radius increases with $z$, in agreement with the trend exhibited by the correlation functions.

\begin{figure}[h]
\begin{center}
\includegraphics[scale=0.23]{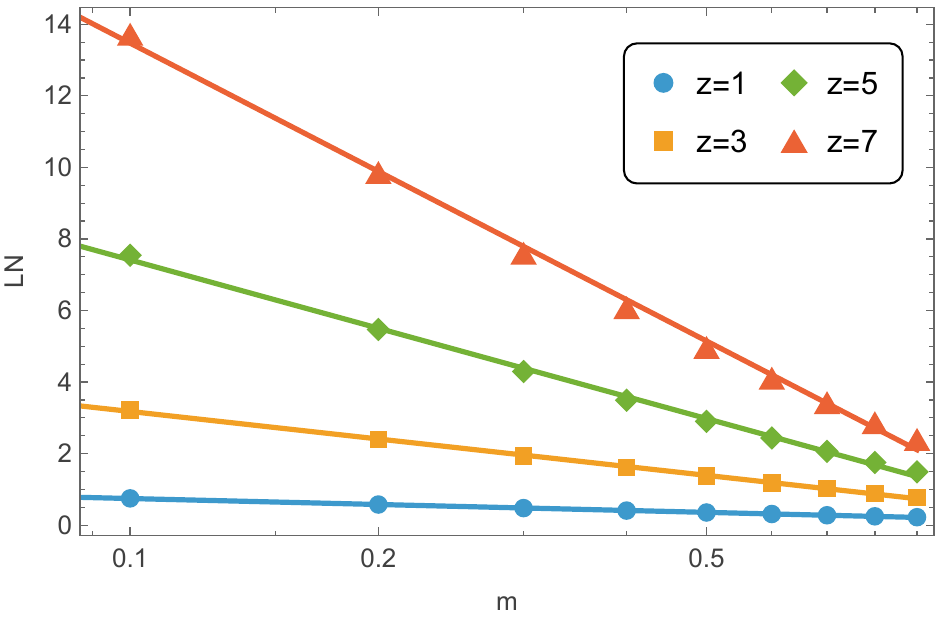}
\hspace{2mm}
\includegraphics[scale=0.23]{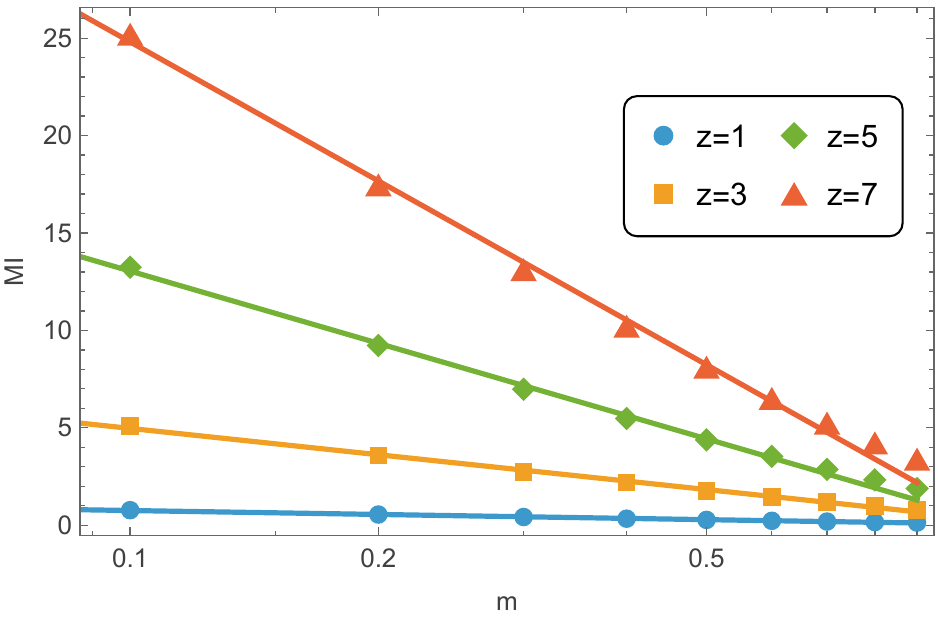}
\hspace{2mm}
\includegraphics[scale=0.23]{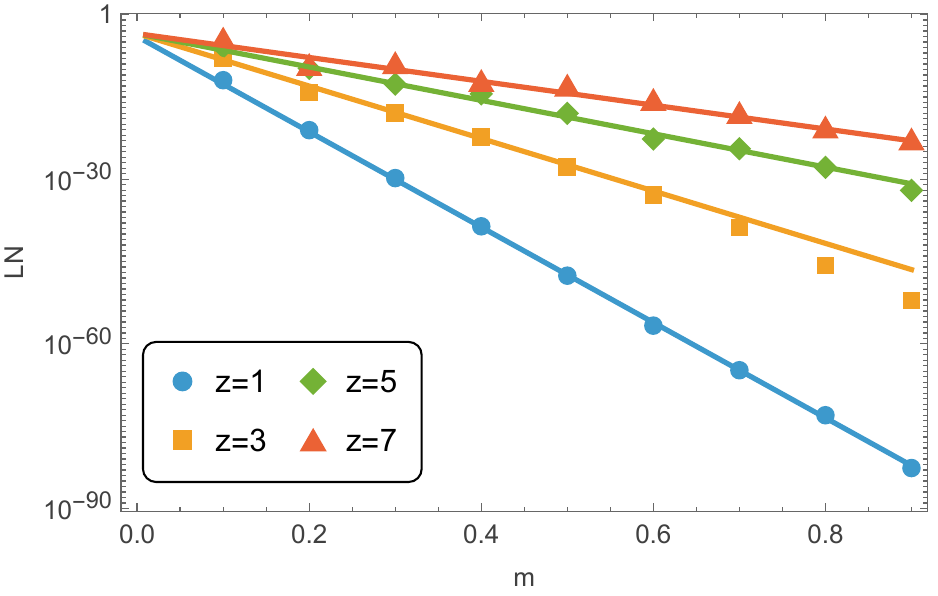}
\hspace{2mm}
\includegraphics[scale=0.23]{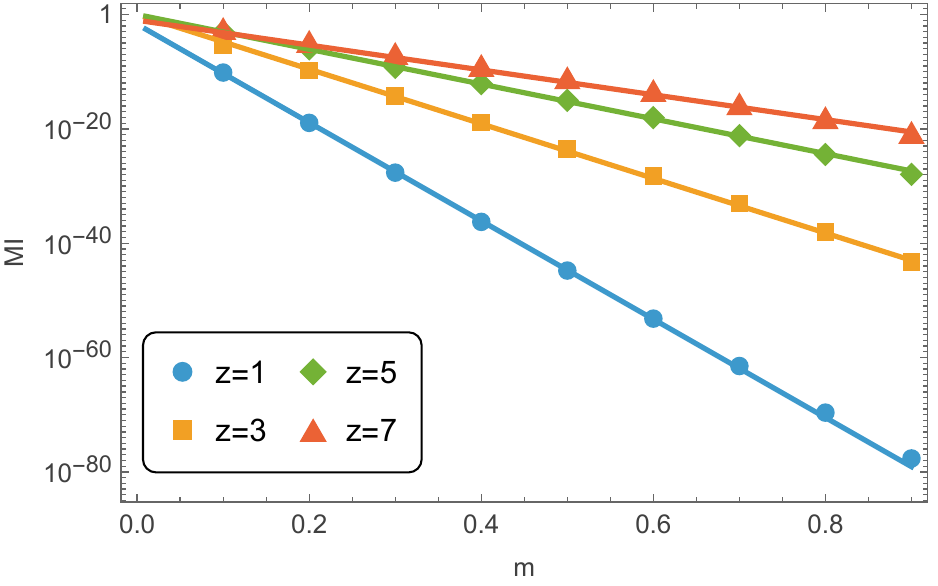}
\end{center}
\caption{
Logarithmic negativity and mutual information in the massive regime. The two left panels correspond to the adjacent limit. The solid lines in the adjacent LN correspond to $a-b\,\log m$ where $b=0.24, 1.11, 2.76, 5.17$ for $z=1,3,5,7$ respectively. For MI, the same functionality holds with $b=0.31, 2.00, 5.35, 10.30$. The two right panels correspond to disjoint subregions with $d=2\ell=100$. For the disjoint case, the solid curves for both LN and MI are $\sim e^{-a m}$ where $a=200,110,70,50$ for $z=1,3,5,7$.
}
\label{fig:LNMImassive}
\end{figure}

In the massive regime, as shown in Fig. \ref{fig:LNMImassive}, we demonstrate that in the adjacent limit, LN scales as $\log m$ for $z=1,3,5,7$. This behavior, including the value of its prefactor, expected to be 1/4 for $z=1$, is consistent with the scaling arguments in Ref.~\cite{Calabrese:2012nk}.\footnote{Subsequent analytical work in massive
relativistic quantum field theories established the
leading exponential approach to saturation for a
finite interval adjacent to a semi-infinite region,
and the exponential decay of logarithmic negativity
between two separated semi-infinite regions
\cite{Blondeau-Fournier:2015yoa}.
These results concern geometries and scaling limits.
distinct from the fixed finite intervals considered here.} The scaling of MI is also logarithmic in this case. For disjoint intervals with $d\gtrsim\ell$, we find an exponential decay for both LN and MI in the massive regime, as illustrated in the two right panels of Fig. \ref{fig:LNMImassive}.

\subsection{Stiffer Dispersion $z<1$}
When the theory is defined through the dispersion relation in Eq.~\eqref{eq:ContDisp} with $z<1$, as for any non-integer value of $z$, the corresponding spatial operator in the Hamiltonian is the fractional Laplacian. Consequently, the theory is intrinsically spatially nonlocal, with algebraically decaying interactions in real space. This nonlocality makes the infrared sector particularly prominent, significantly influencing both the dispersion relation and the resulting structure of quantum correlations.

The behavior of the momentum channel given in Eq.\eqref{eq:H} is valid for the massless regime in the case of $z<1$ as well. While the field channel suffered from an infrared divergence for $z\geq1$, it is IR-finite for $z<1$ given by
\be\label{eq:Gz<1}
\langle\phi(0)\phi(r)\rangle=\frac12 \int\,\frac{dk}{2\pi}\,\frac{e^{ikr}}{|k^z|}
=
\frac{1}{4\pi r^{1-z}}\sin \left(\frac{\pi z }{2}\right) \Gamma (1-z)\,.
\ee

\begin{figure}[t]
\begin{center}
\includegraphics[scale=0.31]{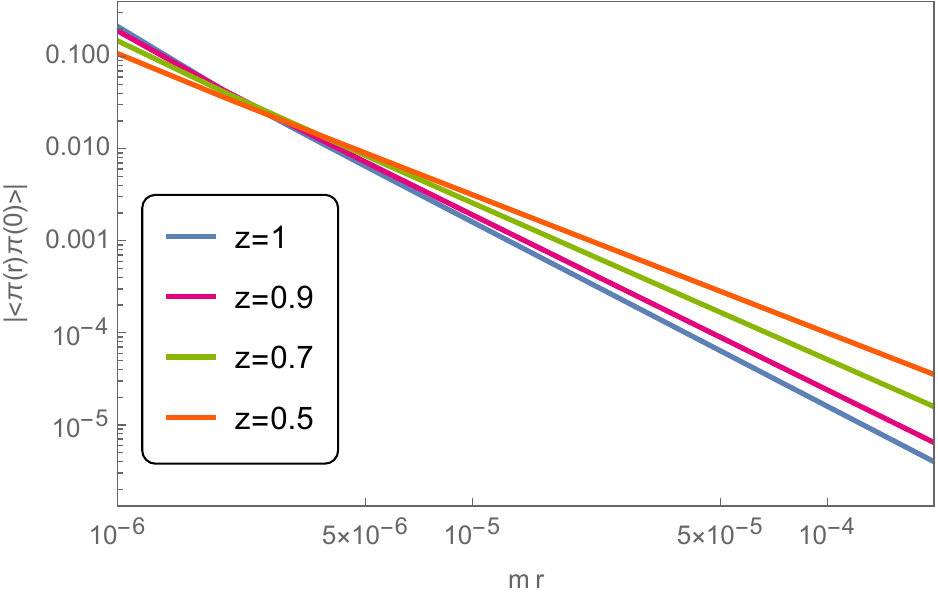}
\hspace{2mm}
\includegraphics[scale=0.31]{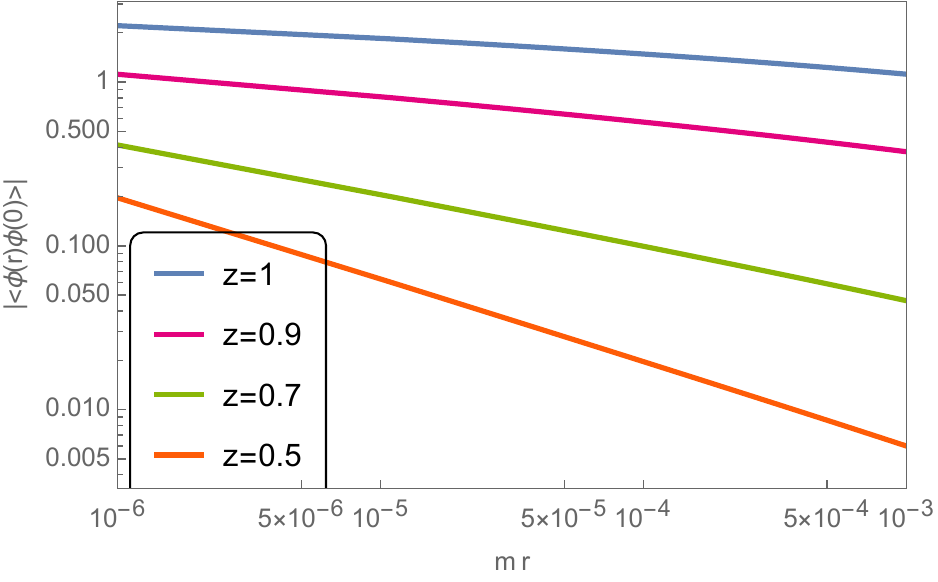}
\hspace{2mm}
\includegraphics[scale=0.31]{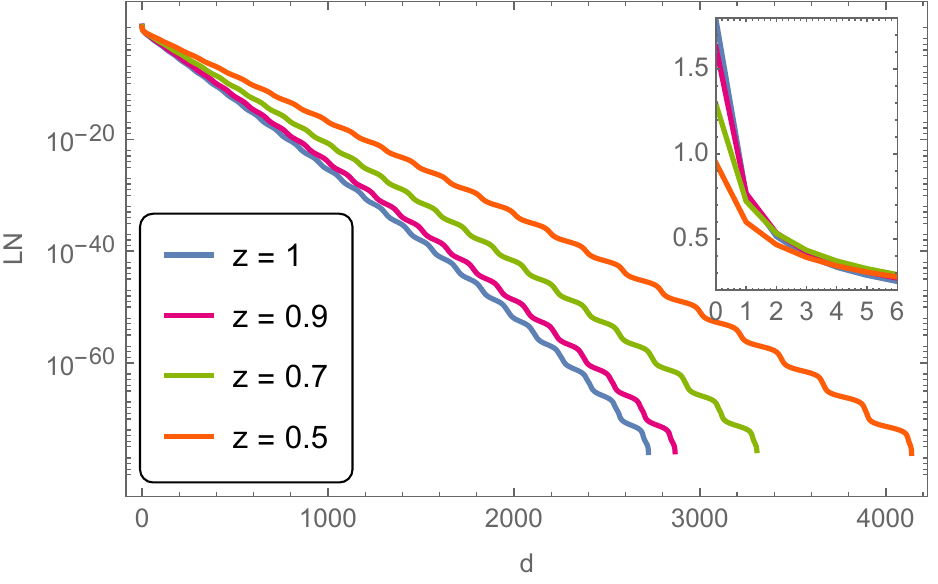}
\end{center}
\caption {Correlation functions for the field and momentum channels and LN for $z<1$. The left and middle panels belong to the correlation functions in the massless regime. Here we present lattice results for $m=10^{-6}$ which qualitatively agrees with what we find from analytically from the continuous theory. The right panel shows the behavior of LN in this regime where we have included the $z=1$ in the plot for comparison. LN is plotted for $\ell=50$ and $m=10^{-6}$. The inset of the right panel shows the adjacent limit where the ultraviolet modes are dominant.}
\label{fig:zl1}
\end{figure}

These correlations are shown in Fig.\ref{fig:zl1}. Similar to the previous cases, the structure of the momentum-channel correlation at small and large distances provides a physical interpretation of what we find in the bahavior of LN which is presented in the same Figure. The behviour of LN is similar to what we reported in the previous section for integer $z\geq 1$ in the sense that for smaller values of $z$ we find a weaker decay of LN with separation between the subregions and a larger radius for the negativity sphere. The negativity sphere radius is found to 4138, 3306, 2866, and 2723 for $z=0.5, 0.7, 0.9, 1$ respectively.

In the massive regime, one can verify that 
\begin{align}\label{eq:GHmassivezl1}
\begin{split}
\langle\phi(0)\phi(r)\rangle
&=\frac12 \int\,\frac{dk}{2\pi}\,\frac{e^{ikr}}{\sqrt{k^{2z}+m^{2z}}}
\sim
\frac{1}{r^{2z+1}}\,,
\\
\langle\pi(0)\pi(r)\rangle
&=\frac12 \int\,\frac{dk}{2\pi}\,\sqrt{k^{2z}+m^{2z}} e^{ikr}
\sim
\frac{1}{r^{2z+1}}\,,
\end{split}
\end{align}
where the mass affects the amplitudes and crossover scales of the correlations, while their asymptotic algebraic decay exponent remains unchanged at fixed $0<z<1$%
\footnote{In a different context, the field-channel correlation has been reported in \cite{Rajabpour:2014osa}.}.
Although the mass opens a spectral gap, it does not remove the nonanalyticity of $|k|^{2z}$ at $k=0$. For integer $z$, the massive dispersion is analytic near the origin, and its complex branch points produce exponentially decaying correlation envelopes. By contrast, for $0<z<1$, the nonanalyticity at the origin yields algebraic tails even at nonzero mass. Thus, the scale $m^{-1}$ characterizes a crossover rather than a finite exponential correlation length.

For subregions of size $\ell=20$, our numerical results in the massive regime show that the negativity-sphere radius increases with $z$ over the investigated range $0<z<1$. For larger subregions, such as $\ell=80$, logarithmic negativity remains nonzero up to the largest separations accessible to our computations, so we can determine only lower bounds on the negativity-sphere radius. The persistence of NPT entanglement over these distances is consistent with the algebraic correlation tails, although those tails alone do not determine the NPT threshold. We leave a systematic investigation of this regime to future work.

\subsection{Separate ultraviolet and infrared deformations}
\label{sec:UVIRdeformations}

Varying the Lifshitz exponent changes both the infrared
and ultraviolet sectors of the dispersion. The decrease
of the negativity-sphere radius with increasing $z$ in
the massless regime therefore motivates a more
controlled comparison in which these sectors are
modified separately. At fixed lattice spacing and
interval size, we consider two families that allow
either stiffening or softening. Ultraviolet deformations
preserve the leading infrared dispersion, whereas
infrared deformations leave a finite ultraviolet
momentum band exactly unchanged. We restrict this comparison to the massless regime, to clarify the decrease of the negativity-sphere radius
observed along the Lifshitz family. The response to these
deformations in the massive regime is not addressed here. Throughout this
subsection, momenta and masses are expressed in lattice
units, and we write $\widehat{k}=2\sin(k/2)$.
The formulas below specify the nonnegative-momentum
branch; the dispersions are extended evenly to
negative momenta.

The two deformation families are
\begin{align}
\omega_{\mathrm{UV}}^2(k;a)
&=
m^{2z}+\widehat{k}^{2z}
\bigl(1+a\widehat{k}^{2}\bigr),
\qquad a>-\frac14,
\label{eq:UVdeformed_dispersion}
\\
\omega_{\mathrm{IR}}^2(k;b)
&=
m^{2z}+\widehat{k}^{2z}
\left(\frac{\widehat{k}}{k_c}\right)^{
-2b\,\chi(\widehat{k}/k_c)},
\qquad b<z,
\label{eq:IRdeformed_dispersion}
\end{align}
where $0<k_c<2$ specifies the cutoff in
$\widehat{k}$, and
\begin{equation}
\chi(t)=
\begin{cases}
\displaystyle
\exp\!\left[-\frac{t^2}{1-t^2}\right],
& 0\leq t<1,\\[2mm]
0,
& t\geq1.
\end{cases}
\label{eq:IRdeformation_window}
\end{equation}
Both families recover the undeformed dispersion
$\omega^2(k)=m^{2z}+\widehat{k}^{2z}$ when their
deformation parameter vanishes. Positive $a$ or $b$
stiffens the corresponding sector, while negative
values soften it.

For the ultraviolet family, the leading
small-momentum term remains $\widehat{k}^{2z}$,
and the modification enters at order
$\widehat{k}^{2z+2}$. The restriction $a>-1/4$
keeps the spatial factor positive throughout the
Brillouin zone and avoids an additional zero at
its edge. This deformation changes intermediate-
and high-momentum frequencies while preserving
the leading infrared behavior.

For the infrared family,
\begin{equation}
\omega_{\mathrm{IR}}^2(k;b)
\simeq
m^{2z}
+k_c^{2b}\widehat{k}^{2(z-b)},
\qquad \widehat{k}\ll k_c.
\end{equation}
Positive $b$ lowers the infrared exponent and
increases the low-momentum frequencies, whereas
negative $b$ raises the exponent and decreases them.
The condition $b<z$ ensures that the spatial term
vanishes continuously at $\widehat{k}=0$, where
we define $\omega_{\mathrm{IR}}(0;b)=m^z$.
For $\widehat{k}\geq k_c$, the dispersion coincides
exactly with the undeformed one. The switching
function and all its derivatives vanish at
$\widehat{k}=k_c$, so the matching introduces no
discontinuity or derivative kink. This exact
restriction of the deformation to a momentum band
generally introduces spatially nonlocal couplings.

In the massless configurations examined,
stiffening either the ultraviolet or infrared sector
increases the negativity-sphere radius, whereas
softening either sector decreases it.
For example,
at $z=1$, $m=10^{-6}$, and $\ell=50$, the undeformed radius is $d_{\cancel{\mathcal{E}}}=2723$. With $k_c=1$ for the infrared deformation, ultraviolet softening with $a=-1/8$ and infrared softening with $b=-1$ reduce the radius to $2347$ and $862$, respectively. Conversely, ultraviolet stiffening with $a=1$ and infrared stiffening with $b=0.9$ increase it to $4319$ and $4591$, respectively.
The ultraviolet deformation preserves the leading
infrared dispersion, while the infrared deformation
leaves the ultraviolet band exactly unchanged.
Thus, relative to the undeformed vacuum, raising
the excitation energies in either sector extends
the entanglement range in these examples, while
lowering them shortens it.
The state remains the vacuum of the deformed
Hamiltonian: these changes concern excitation
energies and vacuum fluctuations, rather than
populations of excited modes.

The two-site example discussed above illustrates
how such changes enter the entanglement criterion.
For one oscillator in each region, define
\begin{equation}
V_q=g_0-g_r
=\frac12\langle(q_A-q_B)^2\rangle,
\qquad
V_p=h_0+h_r
=\frac12\langle(p_A+p_B)^2\rangle.
\end{equation}
The corresponding PT symplectic eigenvalue satisfies
$\nu_2^{\,2}=V_qV_p$ and contributes
to logarithmic negativity when
$\nu_{2}<1/2$. At fixed positive mass,
\begin{align}
V_q
&=\int_0^\pi\frac{dk}{2\pi}\,
\frac{1-\cos(kr)}{\omega(k)},
\\
V_p
&=\int_0^\pi\frac{dk}{2\pi}\,
\omega(k)\bigl[1+\cos(kr)\bigr].
\end{align}
Stiffening a momentum sector decreases $V_q$ but
increases $V_p$, while softening produces the
opposite changes. The response of entanglement
therefore depends on the balance between the
fractional changes in these two conjugate
fluctuation channels.

More explicitly, consider a small deformation
$\omega_s(k)=\omega(k)e^{s f(k)}$. Its effect is
\begin{equation}
\left.
\frac{d\log\nu_{2}^{\,2}}{ds}
\right|_{s=0}
=
\langle f\rangle_p-\langle f\rangle_q,
\label{eq:PT_dispersion_response}
\end{equation}
where the normalized averages are
\begin{align}
\langle f\rangle_q
&=
\frac{1}{V_q}
\int_0^\pi\frac{dk}{2\pi}\,
\frac{1-\cos(kr)}{\omega(k)}\,f(k),
\\
\langle f\rangle_p
&=
\frac{1}{V_p}
\int_0^\pi\frac{dk}{2\pi}\,
\omega(k)\bigl[1+\cos(kr)\bigr]\,f(k).
\end{align}
For stiffening to enhance the negativity associated
with this eigenvalue, the fractional reduction in
$V_q$ must exceed the fractional increase in $V_p$.
This example explains why changes in excitation
energies can affect the entanglement range without
assigning a predetermined sign to the response.
It illustrates the competition between correlation
channels; it does not by itself determine the
threshold for extended intervals.

These comparisons show that preserving the leading infrared
dispersion does not fix the negativity-sphere radius, and
that the radius can also change when a finite ultraviolet
band is held exactly fixed. In the near-massless examples
examined, stiffening either sector extends the range,
whereas softening it shortens the range. These opposite
responses provide a qualitative interpretation of the
Lifshitz-family results: the decrease of the radius with
increasing $z$ is consistent with infrared softening
outweighing the accompanying ultraviolet stiffening.
The controlled deformations do not, however, quantitatively
decompose the change along the Lifshitz family into
separate sector contributions. Indeed, a uniform rescaling
$\omega(k)\mapsto c\,\omega(k)$, with $c>0$, sends
$G\mapsto G/c$ and $H\mapsto cH$, leaving the
PT symplectic spectrum unchanged. The entanglement
range therefore responds to the momentum dependence
of the excitation energies rather than their
overall scale.

These results concern the distance over which NPT
entanglement survives. A distinct question is how
the deformations affect the spatial organization
of the surviving entanglement. We return to the
same ultraviolet dispersion in Section~\ref{sec:Modewise},
where we compare the canonical participation
profiles and partially transposed mode wavefunctions.

\vspace{3mm}

The analysis above establishes how infrared structure
affects the magnitude and spatial range of logarithmic
negativity, and how its response differs from that of
mutual information. These correlation measures do not,
however, identify the spatial degrees of freedom carrying
the surviving entanglement. In particular, the cancellation
of the leading infrared divergence from the
negativity-carrying spectrum does not determine the
spatial or momentum-space structure of the corresponding
modes. To resolve this structure, we now turn to a
mode-wise description based on local canonical
transformations within each region. This allows us to
follow how the entanglement-carrying modes reorganize
with separation and to examine how their spatial
participation relates to the ultraviolet and infrared
content of their canonical components.

\section{Mode-Wise Entanglement Structure and UV--IR Connection}\label{sec:Modewise}

The organization of quantum correlations in Gaussian continuous-variable systems, and in particular the identification of the elementary carriers of entanglement, has motivated extensive studies of entanglement reorganization. A hierarchy of entangled mode pairs can provide an interface between a many-body field state and low-dimensional quantum detectors, offering a possible route toward the simulation, extraction, and characterization of quantum-field-theoretic entanglement using quantum-information protocols \cite{Marshall:2015mna,Bauer:2021sgi}. A review of earlier work can be found in Sec.~III A of Ref.~\cite{Klco:2021cxq} and the references therein.

One particularly appealing form of entanglement reorganization is a reduction of the many-body entanglement structure to an effective one-to-one pairing. In such a representation, the correlations between two parties $A$ and $B$ are concentrated into a set of $\bigl(1_A\times1_B\bigr)$ mode pairs. For pure Gaussian states this structure follows directly from the Gaussian Schmidt decomposition, while for mixed states it is more restrictive and requires a separate consolidation procedure. This distinction will be important below.

The mode-wise viewpoint also provides a natural setting in which to revisit the UV--IR structure of vacuum entanglement. From an effective-field-theory perspective, long-distance observables are usually expected to be governed by infrared physics and to become insensitive to ultraviolet modifications such as momentum truncations or the integration out of high-energy degrees of freedom. Vacuum entanglement between finite spatial regions is more subtle. Early studies showed that the persistence of two-region vacuum entanglement at large separation requires probes with increasingly large energy gaps \cite{Reznik:2002fz,Reznik:2003mnx}. Ref.~\cite{Botero:2004vpl} subsequently related this persistence to the localization of high-frequency inner entanglement modes in the harmonic chain. This mode-wise picture later motivated the analysis of mixed states in lattice scalar field theory, where the maximal range of two-region vacuum entanglement was shown to retain a nontrivial sensitivity to UV degrees of freedom \cite{Klco:2021biu}.

Appendix~\ref{app:correlation_channels}.

In this section, we examine how the spatial participation of entanglement-carrying modes relates to their ultraviolet canonical structure. Using the relativistic harmonic chain as our reference, we study pure bipartitions and the mixed state of two finite subregions. Our central finding concerns the pair known to carry the full logarithmic negativity at large separation. Although its conjugate canonical components retain pronounced UV oscillations, the participation of the original oscillators is smooth and IR-dominated. The same distinction appears in the pure-state modes examined here. This provides a more complete physical picture of long-distance vacuum entanglement by \textit{distinguishing its microscopic canonical realization from its collective spatial participation}. We trace the emergence of this collective structure through two related flows -along the adjacent-mode hierarchy and with increasing separation- and connect
them through the adjacent-state canonical basis. Lifshitz generalizations test the scope of this picture and reveal a common normalized long-distance participation profile across the massless dispersions studied. A controlled UV deformation preserves IR dominance while changing the detailed profile and entanglement range.  A complementary singular-value analysis of the cross-region covariance blocks is presented in Appendix~\ref{app:correlation_channels}. It reveals an increasingly pronounced hierarchy of spatial correlation channels with separation, without identifying these channels with the negativity-carrying canonical modes.

\subsection*{Pure states}

For a pure Gaussian state, the mode-wise structure is particularly transparent. Under a bipartition into subsystems $A$ and $B$, a multimode pure Gaussian state can always be transformed by local symplectic operations into a product of two-mode squeezed states, together with local single-mode vacua whenever the two subsystems have unequal numbers of modes \cite{Botero:2003wov}. Thus each entangled mode in $A$ has a unique partner in $B$, and the number of entangled pairs is at most $\min(N_A,N_B)$. The same construction extends to the special class of isotropic mixed Gaussian states, whose covariance matrices have completely degenerate symplectic spectra \cite{Botero:2003wov}.

For oscillator chains without $q$--$p$ correlations, Ref.~\cite{Botero:2004vpl} related the required local symplectic transformation to the eigenvectors of the restricted covariance blocks. For subsystem $A$ one may write
\begin{align}
	\widetilde q_j=\sum_i u_i^{(j)}q_i,
	\qquad
	\widetilde p_j=\sum_i v_i^{(j)}p_i,
\end{align}
where $u^{(j)}$ and $v^{(j)}$ are right eigenvectors of $H_A G_A$ and $G_A H_A$, respectively. Since the transformation from the local variables $(q_i,p_i)$ to the collective canonical variables $(\widetilde q_j,\widetilde p_j)$ is symplectic rather than orthogonal, the position and momentum components of a single mode need not have the same spatial profile. A useful quantity that combines them is obtained by expressing the collective annihilation operator in terms of the local site operators \cite{Botero:2004vpl},
\begin{align}
	a_j=\sum_i\sqrt{P_i^{(j)}}\left(\cosh \tau_i^{(j)}a_i
	+\sinh \tau_i^{(j)}a_i^\dagger\right),
	\label{modewise1}
\end{align}
with
\begin{align}
	P_i^{(j)}\equiv u_i^{(j)}v_i^{(j)},
	\label{modewise2}
\end{align}
and
\begin{align}
	\tau_i^{(j)}\equiv \tanh^{-1}
	\left(\frac{u_i^{(j)}-\zeta_i^2 v_i^{(j)}}
	{u_i^{(j)}+\zeta_i^2 v_i^{(j)}}\right),
\end{align}
where $\zeta$ is an arbitrary dimensional parameter. With the biorthogonal normalization $u^{(j)}\!\cdot v^{(j)}=1$, the participation function obeys $\sum_i P_i^{(j)}=1$, though $[\widetilde q_j,\widetilde p_k]= i\delta_{jk}$.
The hyperbolic parametrization is understood with consistent branch choices and, where necessary, by continuity. In particular, $\tau_i^{(j)}$ need not be real when $P_i^{(j)}<0$. The participation function itself is real and can have either sign.
This interpretation will be especially useful in the mixed-state problem below:
Once a negativity-carrying $(1\times1)_{\rm mixed}$ pair has been consolidated, the corresponding $P_i=u_i v_i$ characterizes the spatial canonical participation of its constituent local mode.

Canonical participation should be distinguished from the participation weights used to construct entanglement contours \cite{Vidal:2014aal,Coser:2017dtb,Zambotti:2026vpm}.
In the bosonic negativity-contour construction of Ref.~\cite{Zambotti:2026vpm}, nonnegative spatial weights obtained from a Williamson decomposition of the partially
transposed covariance matrix distribute the spectral
contributions to the total logarithmic negativity over sites.
That construction does not require the local pairwise
consolidation employed here.
Our $P_i^{(j)}$ instead describes an individual local canonical
mode, is normalized independently of its entanglement
contribution, and can be signed.
The two constructions therefore provide different spatial
diagnostics of the same Gaussian state.

\begin{figure}[htbp]
	\centering
	\includegraphics[width=0.3\textwidth]{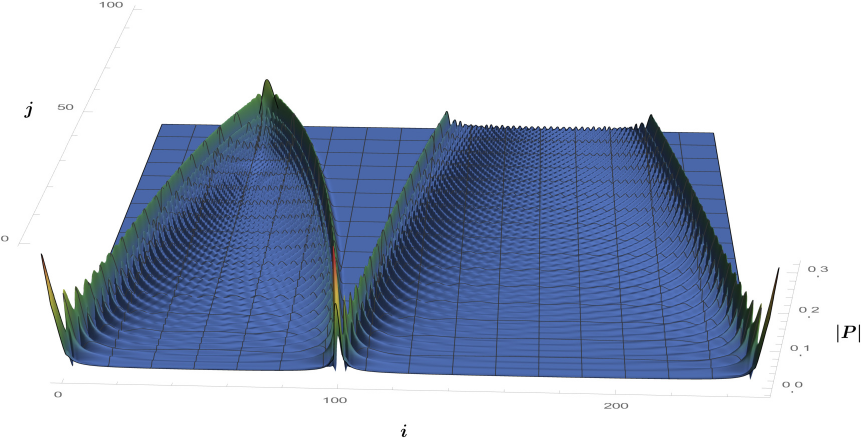}
	\hspace{0.02\textwidth}
	\includegraphics[width=0.3\textwidth]{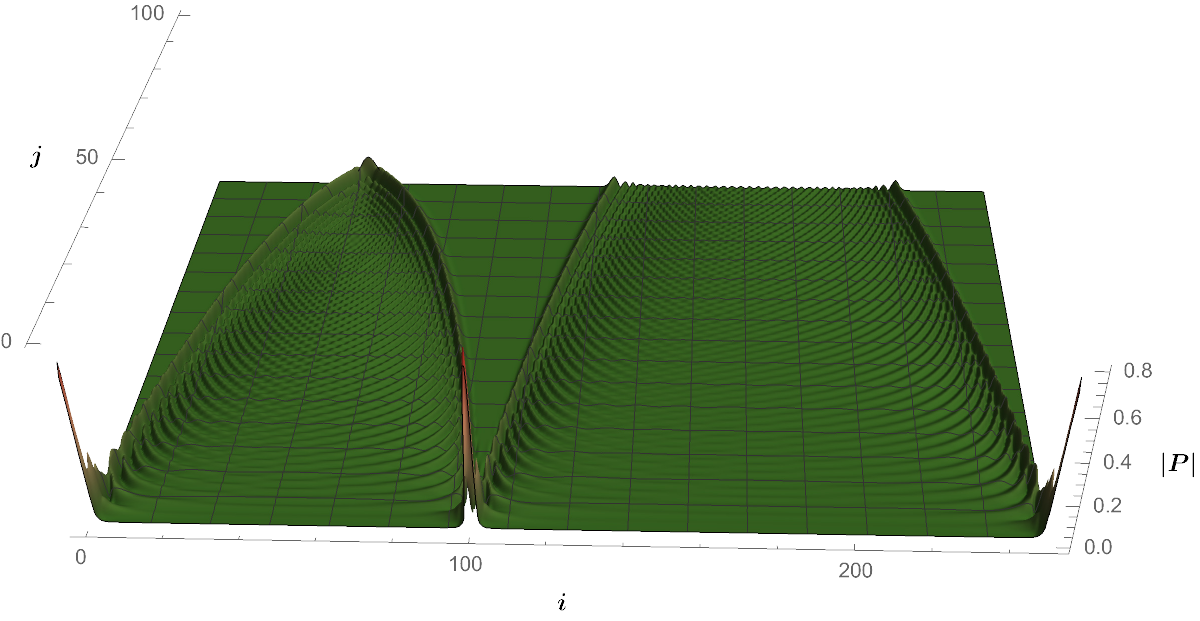}
	\hspace{0.02\textwidth}
	\includegraphics[width=0.3\textwidth]{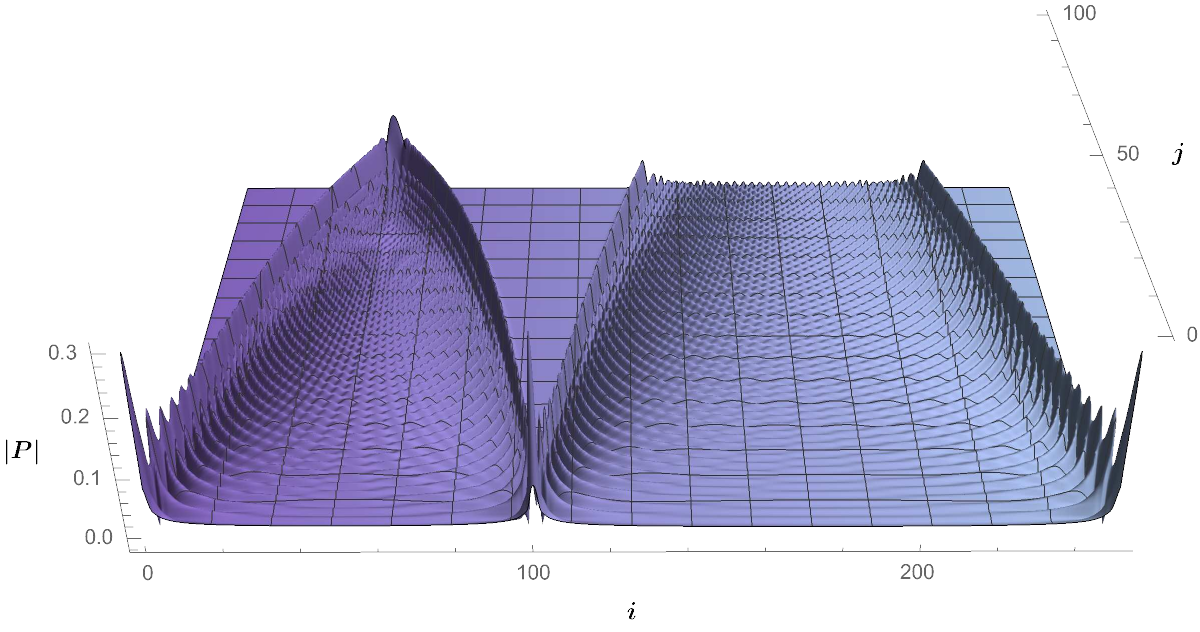}
	\caption{Absolute value of the site participation function, $|P|$, for the entangled canonical modes in a chain composed of $N=100+150$ contiguous oscillators, for $m=10^{-3}$ and, from left to right, $z=1$, $z=3$, and $z=1/2$. The axes $i$ and $j$ denote the lattice site and the mode number, respectively. The modes are ordered by their entanglement contribution, with the dominant mode in the foreground.}
	\label{figmode}
\end{figure}

Figure~\ref{figmode} shows $|P_i^{(j)}|$ for two complementary unequal subregions in the LHC vacuum. The modes are ordered according to their entanglement contribution, with the dominant mode in the foreground. The leading modes are concentrated near the entangling interface, whereas modes with progressively smaller contributions move into the interior of the subregions and develop a more structured spatial profile. For the larger subsystem only $\min(N_A,N_B)$ modes participate in entangled pairs; the remaining canonical modes are locally pure. The comparison among $z=1$, $z=3$, and $z=1/2$ also shows that the detailed spatial extent of the participation profiles depends on the Lifshitz exponent, even though the overall boundary-to-interior hierarchy of the entangled modes persists.

\subsection*{Mixed states}

Beyond pure states, a one-to-one Gaussian mode decomposition is not generic. The construction of Refs.~\cite{Botero:2004vpl,Botero:2003wov} extends directly only to isotropic mixed states. Ref.~\cite{Klco:2021cxq} developed a different procedure for the mixed state obtained by tracing out the complement of two finite subregions in the vacuum of the harmonic chain. The resulting \textit{consolidation procedure} reorganizes the negativity-carrying subspace into effective $\bigl(1\times1\bigr)_{\rm mixed}$ mode pairs whenever a suitable local symplectic structure exists. The relevant directions are obtained from the eigenvectors associated with the partially transposed (PT) symplectic eigenvalues that contribute to the logarithmic negativity. Subsequent developments and applications of this framework can be found in Refs.~\cite{Gao:2024nbl,Klco:2023ojt,Gao:2024esb}.

It is important to distinguish these mixed-state negativity modes from the Gaussian Schmidt modes of the pure-state discussion above. For a generic mixed state, the negativity-carrying directions are determined by the PT covariance matrix of the full bipartite state, rather than by a local mode decomposition of either reduced state alone. When the one-to-one consolidation condition is satisfied, these directions can be reorganized into mixed two-mode core pairs. Even when this condition fails, the PT-negative subspace itself remains well defined and can still be used to study the spatial organization of the degrees of freedom responsible for logarithmic negativity.

We first summarize the consolidation protocol needed to identify the local canonical modes carrying logarithmic negativity, then examine their spatial organization.
Throughout, we distinguish the participation function $P=u\,v$, which characterizes how a consolidated canonical mode is assembled from the original oscillators, from the PT-mode wavefunction $\Psi_{\rm SK}$, which resolves its canonical structure.

\subsection{Consolidation protocol}\label{s1}

Using the two-point correlation functions in Eq.~\eqref{sec1eq2}, evaluated in the infinite-volume limit, the left and right eigenvectors relevant for partial transpose satisfy
\begin{align}\label{eq:uv}
	H^{\Gamma}G \bar{u}_{j}=\nu_j^2 \bar{u}_{j},
	\qquad
	GH^{\Gamma}\bar{v}_{j}=\nu_j^2 \bar{v}_{j},
\end{align}
where $\bullet^{\Gamma}$ denotes partial transpose and $\nu_j$ are the PT symplectic eigenvalues. Imposing the normalizations $\bar u_{j}^{T}G \bar u_{j}=\nu_j$, $\bar v_{j}^{T}H^{\Gamma}\bar v_{j}=\nu_j$, and $\bar u \bar v^{T}=\mathbb{I}$, one obtains local phase-space directions in $A$ and $B$. After restricting these vectors to the two subregions and applying a local symplectic Gram--Schmidt procedure, one constructs local symplectic transformations $S_A$ and $S_B$. Their direct sum,
\begin{align}
	S=S_A\oplus S_B,
\end{align}
acts on the covariance matrix as $\sigma'=S\sigma S^T$.

When the state is consolidable, the first $n_-$ transformed mode pairs---where $n_-$ is the number of PT symplectic eigenvalues contributing to the logarithmic negativity---capture the negativity in independent $\bigl(1\times1\bigr)_{\rm mixed}$ pairs. These pairs constitute the \textit{negativity core}. In this representation the logarithmic negativity is additive over the core pairs.

Logarithmic negativity provides a computable characterization of NPT entanglement and, more generally, an upper bound on distillable entanglement \cite{access1}. In addition, entangled two-mode Gaussian states are distillable \cite{Giedke:2000ruw}. The consolidated core therefore isolates a sector whose entanglement is operationally accessible through the corresponding mixed two-mode pairs. For symmetric two-mode states, the entanglement of formation also admits the special relation discussed in Ref.~\cite{Giedke:2003wns}. The remaining modes form the so-called halo sector. In the terminology of Ref.~\cite{Klco:2021cxq}, this sector can contain entanglement resources that are obscured by mixing and are not directly accessible through the isolated core-pair structure. This is closely related to the notion of separability-obscured entanglement, in which an apparently separable mixed state may contain underlying entanglement that becomes identifiable only when additional classical information about the preparation is available \cite{Klco:2021cxq,cohen}.

An equivalent construction begins directly from the symplectic eigenproblem $i\Omega\sigma^{\Gamma} V_j=\nu_j V_j$, where the eigenvalues appear in $\pm$ pairs. Separating the real and imaginary parts and then applying symplectic Gram--Schmidt again yields the local transformations; details are given in Appendix B2 of Ref.~\cite{Klco:2021cxq}. The more general formulation of Ref.~\cite{Gao:2024esb} constructs a symplectic transformation $\widetilde S$ that provides a normal-mode decomposition of $\sigma^{\Gamma}$ and identifies the negativity-contributing subspace $\mathcal V_{\mathcal N}$. The local components of the corresponding rows or columns of $\widetilde S$ are then used to build $S_A$ and $S_B$. All three constructions agree when consolidation is possible, and for isotropic states they reduce to the mode-wise transformation of Ref.~\cite{Botero:2004vpl}.\footnote{It's worth noting that, recently, analytical expressions have been obtained for the negativity and its core-mode functions between arbitrary disjoint intervals of the massless relativistic scalar field \cite{Pye:2026siw,Pye:2026ehc}. The continuum construction restricts the observables to avoid coupling to the zero mode.}

\paragraph{Consolidation condition.}
Ref.~\cite{Gao:2024esb} identified a necessary and sufficient condition for the above one-to-one consolidation. Consolidable Gaussian states belong to the class of states that are Symplectically Orthogonal Locally in the negativity-contributing subspace, denoted $\mathcal N$-SOL. In the notation used here,
\begin{align}
	\langle \widetilde\nu_{i,A}|\Omega|\widetilde\nu_{j,A}\rangle
	=\langle \widetilde\nu_{i,B}|\Omega|\widetilde\nu_{j,B}\rangle
	=\Omega_{ij}/2,
	\label{c2}
\end{align}
where $|\widetilde\nu_{j,A}\rangle$ denotes the local $A$-space component of the normal-mode basis of $\sigma^{\Gamma}$ before consolidation. The condition requires each negativity-contributing PT mode to define an independent two-dimensional symplectic subspace in both $A$ and $B$, so that the local symplectic form remains non-degenerate and different core pairs do not overlap symplectically.

\paragraph{Separability of the consolidated state.}
The post-consolidation state can be analyzed using the Gaussian separability algorithm of Ref.~\cite{PhysRevLett.87.167904}. Applying the algorithm to the core--halo partition of
\begin{align}
	\sigma'=\begin{pmatrix}
		\sigma'_{cc} & \sigma'_{ch}\\
		(\sigma'_{ch})^T & \sigma'_{hh}
	\end{pmatrix}
\end{align}
returns, whenever the cut is separable, a decomposition
\begin{align}
	\sigma'=\sigma_c\oplus\sigma_h+Y,
	\label{c3}
\end{align}
where $Y=\left(\begin{smallmatrix}\alpha&\beta\\ \beta^T&\delta\end{smallmatrix}\right)$ contains Gaussian classical correlations,
\begin{align}
	\sigma_c&=\sigma'_{cc}-\mathrm{Re}(X)-\|\mathrm{Im}(X)\|_{\rm op}\,\mathbb I,\\
	X&=\sigma'_{ch}(\sigma'_{hh}-i\Omega/2)^{-1}(\sigma'_{ch})^T,\\
	\sigma_h&=\sigma'_{hh}-(\sigma'_{ch})^T(\sigma'_{cc}-\sigma_c)^{-1}\sigma'_{ch}.
\end{align}
Here $\|O\|_{\rm op}$ is the operator norm. The explicit steps for this problem are reviewed in Appendix C of Ref.~\cite{Klco:2021cxq}. The halo can then be tested separately across the $A$--$B$ cut,
\begin{align}
	\mathrm{Tr}_c(\sigma')=\sigma_h+\delta
	=\sigma_{h,A}\oplus\sigma_{h,B}+Y_h.
	\label{c4}
\end{align}
As emphasized in Ref.~\cite{Klco:2021cxq}, Eqs.~\eqref{c3} and \eqref{c4} need not be simultaneously realizable because the classical correlations encoded in $Y$ and $Y_h$ may be incompatible.

\subsection{Core-halo structure of the reduced vacuum}
For two finite regions, consolidation provides the canonical organization on which our spatial analysis is based.
We consider equal intervals of length $\ell$ separated by $d$, using the relativistic harmonic chain as the
reference case.
The construction follows Ref.~\cite{Klco:2021cxq}; its Lifshitz generalizations allow us to examine how dispersion affects the
core-halo structure.
The $\bigl(1_A\times1_B\bigr)_{\rm mixed}$ description is used where the consolidation conditions hold.

As a result of the consolidation protocol, logarithmic negativity is captured by independent $\bigl(1_A\times1_B\bigr)_{\rm mixed}$ core pairs. The separability analysis of the post-consolidation state yields
\begin{align}
	\sigma'=\sigma_{c,1}\oplus\sigma_{c,2}\oplus\ldots\oplus\sigma_{c,n_c}\oplus\sigma_h+Y,
	\label{q1}
\end{align}
where each $\sigma_{c,i}$ is an entangled two-mode core covariance matrix that can be isolated by sequential application of the algorithm of Ref.~\cite{PhysRevLett.87.167904}. In the LHC vacuum, the core--halo decomposition and the separate $A$--$B$ separability condition for the isolated halo are not generally compatible. The preparation covariance $\sigma_h$ can have nonzero $A$--$B$ logarithmic negativity. This should be distinguished from the negativity of the physical halo marginal, $\sigma'_{hh}=\sigma_h+\delta$, which includes the added noise.

\begin{figure}[t]
	\begin{center}
		\includegraphics[scale=0.26]{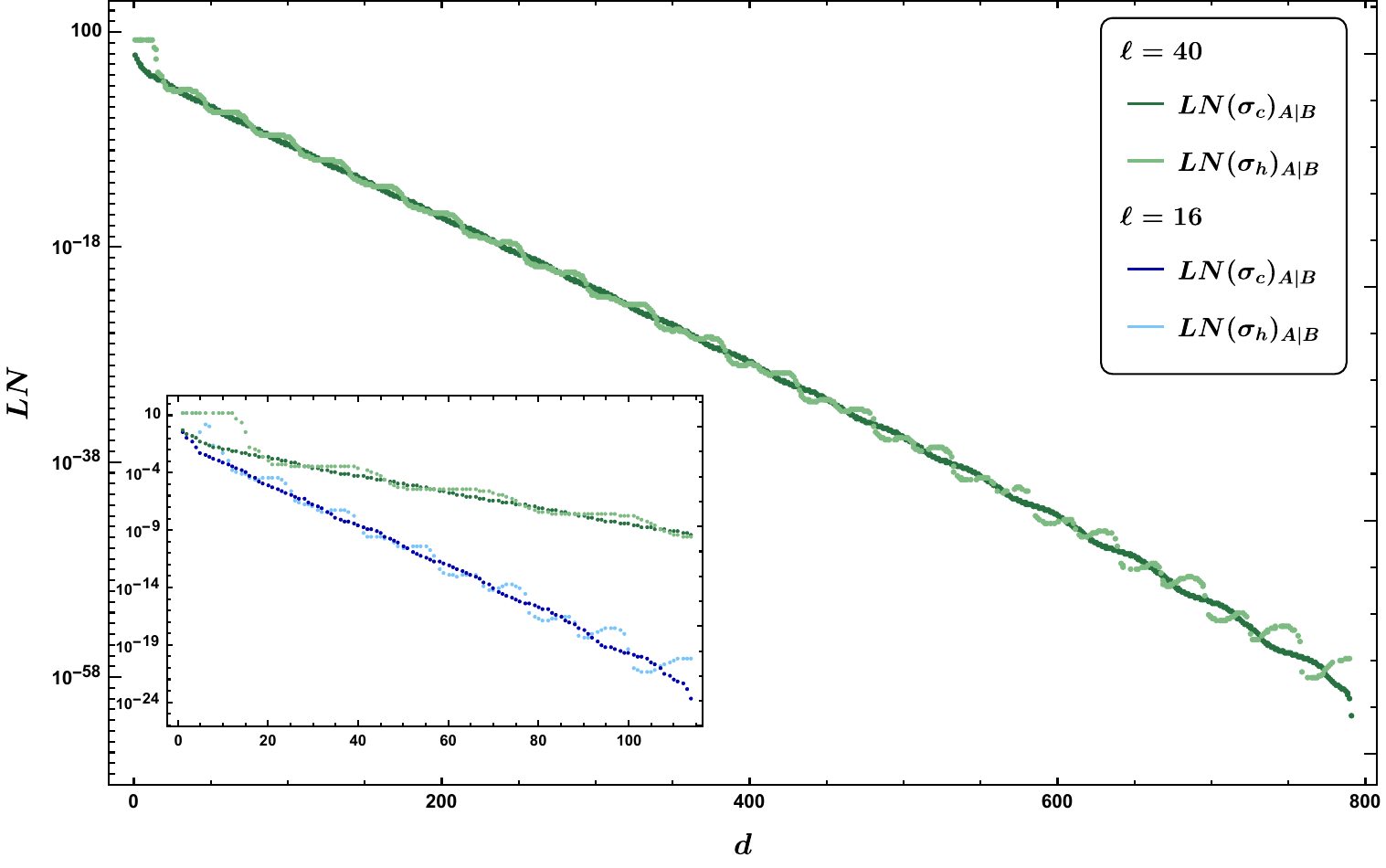}
	\end{center}
	\caption{Logarithmic negativity of the core and halo sectors for two disjoint regions in the vacuum of a latticized massive Lifshitz scalar field with $m=0.00025$, plotted against the separation $d$ for regions of size $\ell=40$ and $z=3$.
    The inset shows the corresponding result for $\ell=16$ and illustrates how lattice-scale halo oscillations decrease toward the continuum limit.
    }
	\label{fig1}
\end{figure}

Figure~\ref{fig1} shows the logarithmic negativities associated with the core and halo sectors as functions of the separation. The halo contribution disappears as the total state approaches its separability threshold.
Both core and halo curves display lattice-scale fluctuations whose amplitudes decrease toward the continuum limit. In contrast with the relativistic harmonic chain, where core and halo contributions have comparable separation dependence \cite{Klco:2021cxq}, the LHC exhibits a pronounced hierarchy at small separation: the halo contribution can be substantially larger than the core contribution and decreases rapidly as $d$ grows. For $z=3,5,7$ and $\ell=64$, the corresponding numerical values are collected in Table~\ref{tableL64} of Appendix~\ref{Appendix:halo}. Increasing $z$ enhances both the halo contribution and its dominance over the core at small $d$, indicating a stronger reorganization of the mixed-state entanglement outside the directly accessible two-mode core sector. The decomposition in Eq.~\eqref{q1}, constructed using the techniques of Ref.~\cite{PhysRevLett.87.167904}, is not unique.

The enhanced halo entanglement at larger $z$ also has implications for state preparation (for numerical data see Appendix~\ref{Appendix:halo}). In constructions that prepare the reduced vacuum through the core-halo decomposition used here, it indicates a larger
entanglement resource assigned to the halo \cite{Klco:2021cxq}. This is a property of the chosen preparation construction, rather than a determination of the minimum resource cost over all possible protocols.

For $z<1$, the consolidation condition is satisfied exactly across the entire range of separations considered. The absence of a pronounced halo dominance at small separations for $z<1$ contrasts with the hierarchy observed for odd-integer $z>1$. Instead, the two contributions remain comparable in this regime, with the halo contribution generally remaining smaller than the core contribution. Moreover, as $z$ increases, both the halo and core contributions increase. The corresponding numerical results for $z=0.5$, $0.7$, and $0.9$ are presented in Table~\ref{tableL64fractionalz} of Appendix~\ref{Appendix:halo}.

\subsection{Spatial organization of the negativity-carrying modes}

\begin{figure}[t]
\begin{center}
\includegraphics[scale=0.45]{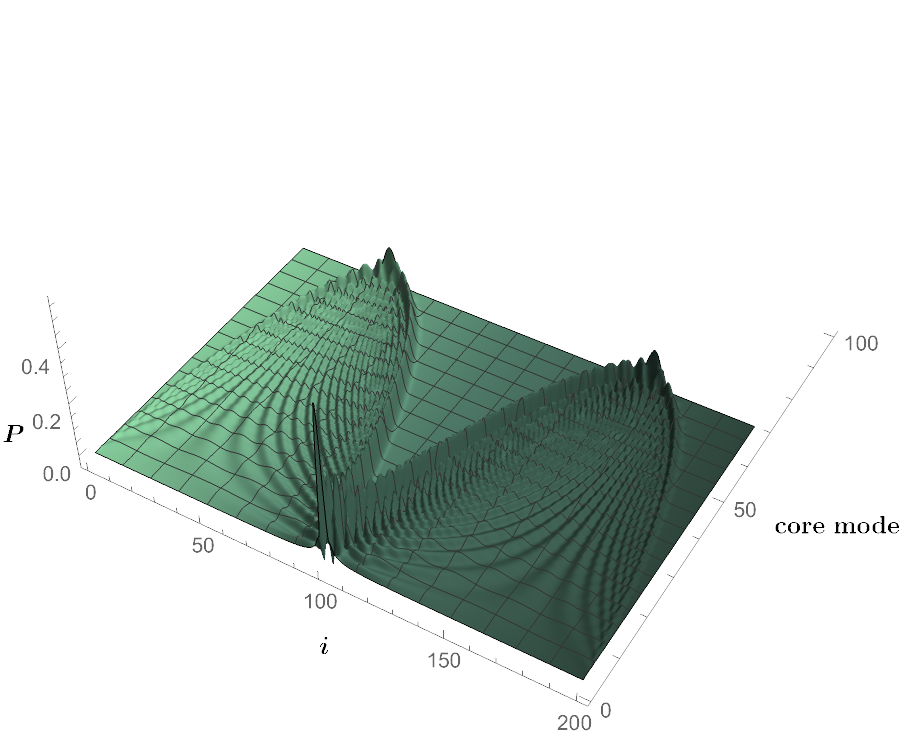}
\hspace{4mm}
\includegraphics[scale=0.45]{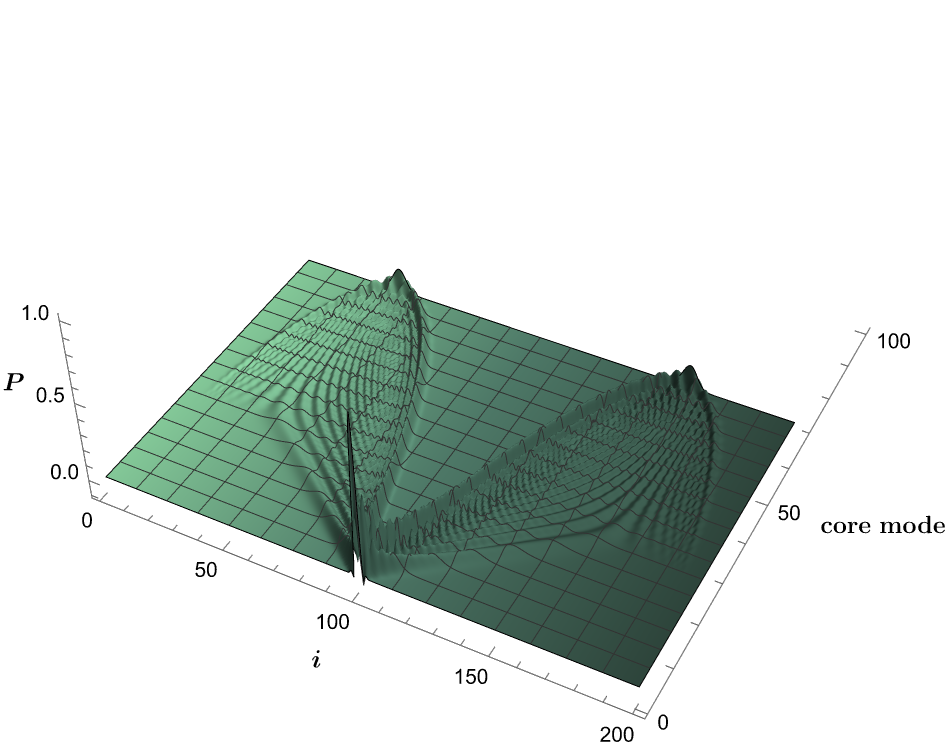}
\end{center}
\caption{Site participation function, $P$, for the core modes in an infinite chain consisting of two subregions of length $\ell=100$ each, with zero separation between them for $m=10^{-3}$ (left), $m=0.4$ (right) for $z=1$ and $n_{c}=67$. From the first to the last mode, the contribution to the logarithmic negativity decreases.}
\label{fig1Cores}
\end{figure}

Having identified the canonical pairs that reproduce the logarithmic negativity, we now examine how the original oscillators participate in those pairs. This spatial information distinguishes their collective organization from the magnitude of their entanglement. Here we use ``collective'' for profiles that become approximately reflection symmetric about the subregion center, including profiles with two separated peaks. At larger mode index or separation, these can develop a standing-wave-like shape. This description refers to their spatial form; the modes are not energy eigenmodes of the LHC Hamiltonian.

\paragraph{Adjacent regions and the mode-index flow.}
For two adjacent equal subregions, the modes can be ordered by decreasing contribution to the logarithmic negativity. The leading modes are localized near the shared boundary, with negligible weight near the opposite boundary. As the mode index $j$ increases, a second, reflected peak develops near the far boundary. Beyond an empirically observed crossover scale	$j_0\sim m\epsilon$,
the profiles become approximately reflection symmetric about the center of each subregion. For still larger $j$, the two peaks move toward one another and the modes acquire an increasingly collective, standing-wave-like character, eventually becoming concentrated near the subregion center and suppressed near both boundaries. See Fig. \ref{fig1Cores}. For details of the behavior of $u$ and $v$ see Appendix \ref{Appendix:uvConfig}.

The progression from an interface-localized profile to a reflection-symmetric two-peak profile, and eventually to a center-peaked profile, shows how the modes increasingly resolve the full subregion. The leading modes are concentrated at the entangling interface, whereas the subleading modes are sensitive to both boundaries. 

\begin{figure}[h!]
	\begin{center}
		\includegraphics[scale=0.38]{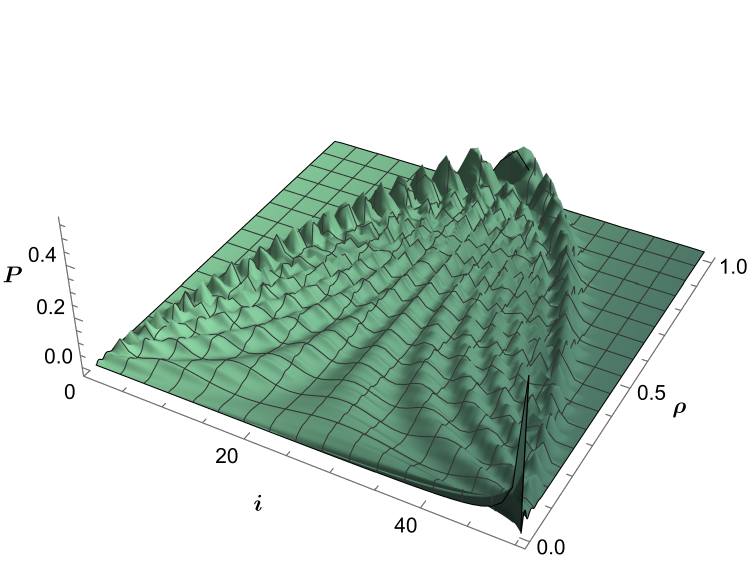}
		\hspace{2mm}
		\includegraphics[scale=0.38]{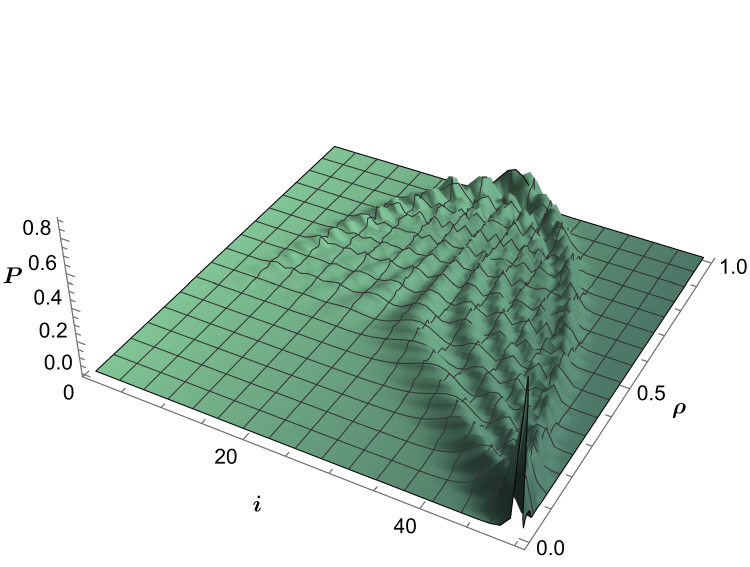}
		\hspace{2mm}
		\includegraphics[scale=0.38]{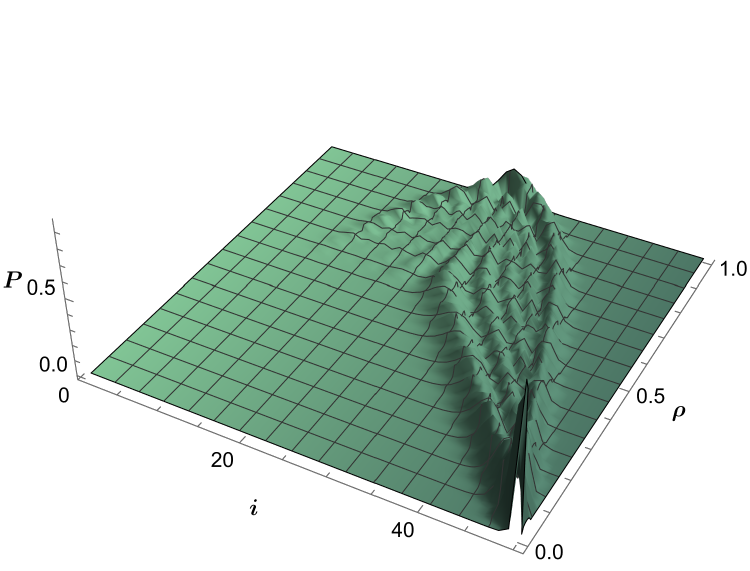}
	\end{center}
	\caption{Site participation function of the dominant negativity-carrying mode as the separation $d$ is varied, for $\ell=50$, $z=1$, and masses $m=10^{-3}$ (left), $m=0.4$ (middle), and $m=0.8$ (right), shown in subregion $A$. Increasing separation drives the mode from an interface-localized profile toward a collective profile that resolves the full subregion.}
	\label{fig:dominant_core_vs_d}
\end{figure}

\paragraph{Separation flow of the dominant mode.}
A similar evolution occurs when the separation is increased while following the dominant negativity-carrying mode. The hierarchy of contributions associated with the PT symplectic eigenvalues motivates this focus: the second-largest contribution to the logarithmic negativity is exponentially suppressed relative to the leading one. At $d=0$ the dominant mode is localized near the shared boundary. As $d$ grows, a reflected peak appears near the far boundary, and the two peaks move inward. At sufficiently large separation, the profile becomes centered within the subregion and standing-wave-like, as illustrated in Fig.~\ref{fig:dominant_core_vs_d} where $\rho=d/d_{{\cancel{\mathcal E}}}(z)$. The scaling analysis below shows that this shape is already essentially established before the negativity sphere is reached: the last surviving mode subsequently loses its NPT character as the separation increases further.\footnote{A related separation dependence has recently been
observed for the most entangled modes between two
spherical regions in a $(3+1)$-dimensional massless
scalar vacuum: their field and momentum profiles
broaden and shift away from the facing boundaries
while becoming more oscillatory
\cite{Perche:2026ams}.
}

The similarity between the mode-index and separation flows motivates a direct comparison in a fixed basis. We therefore take a more precise look at this structure by expanding the dominant finite-separation mode in the complete adjacent-state basis before turning to its momentum-space structure.

\subsection{Expansion in the adjacent-state canonical basis}
\label{subsec:adjacent_basis_expansion}

Expanding the finite-separation modes in the adjacent-state
canonical basis identifies which reference directions
participate as the regions are moved apart. For fixed mass
$m$ and interval size $\ell$, let
$S_0=S_A(0)\oplus S_B(0)$ denote the full local symplectic
transformation obtained from the consolidation construction
at $d=0$. Its core and halo modes together form a complete
canonical basis for the retained phase space. Completeness
follows from the invertibility of the full symplectic
transformation and does not require the halo state to
decompose into independent
$(1_A\times1_B)_{\rm mixed}$ pairs.

We identify the sites within each interval using the same
internal coordinates at all separations. If $S_d$ denotes
the corresponding transformation at separation $d$, the
canonical coordinates obey
\begin{align}
r^{(d)}=S_dS_0^{-1}r^{(0)}.
\end{align}
Thus the adjacent basis provides an exact reference for
the finite-separation modes, even though the reduced state
and its entanglement structure change with $d$. The pairwise
interpretation of the adjacent core applies where the
consolidation condition is satisfied.

We describe the construction in region $A$; the same
analysis applies to region $B$. For the transformations
considered here, the position and momentum sectors can
be treated separately. The adjacent canonical
modes are
\begin{align}
Q_j^{(0)}&=(u_j^{(0)})^Tq_A,
&
\Pi_j^{(0)}&=(v_j^{(0)})^Tp_A,
\qquad j=1,\ldots,\ell,
\end{align}
with
\begin{align}
(u_j^{(0)})^Tv_k^{(0)}=\delta_{jk}.
\label{eq:adjacent_biorthogonality}
\end{align}
The core modes are ordered by decreasing contribution
to the adjacent logarithmic negativity. Their index set
is denoted by $C_0=\{1,\ldots,n_-(0)\}$, while the remaining
indices form the adjacent halo set $H_0$. These labels
refer throughout to the reference state at $d=0$.

A local canonical pair at separation $d$,
\begin{align}
Q_r(d)&=u_r(d)^Tq_A,
&
\Pi_r(d)&=v_r(d)^Tp_A,
\end{align}
admits the expansions
\begin{align}
u_r(d)&=\sum_{j=1}^{\ell}a_{jr}(d)u_j^{(0)},
&
v_r(d)&=\sum_{j=1}^{\ell}b_{jr}(d)v_j^{(0)}.
\label{eq:adjacent_mode_expansion}
\end{align}
Because the reference vectors are biorthogonal, the coefficients are obtained
from the dual vectors:
\begin{align}
a_{jr}(d)&=(v_j^{(0)})^Tu_r(d),
&
b_{jr}(d)&=(u_j^{(0)})^Tv_r(d).
\label{eq:adjacent_expansion_coefficients}
\end{align}
In what follows, we focus on the dominant
negativity-carrying mode and suppress its label $r$.
We express separation as
$\rho=d/d_{{\cancel{\mathcal{E}}}}$, where
$d_{{\cancel{\mathcal{E}}}}$ is the negativity-sphere
radius for the parameters under consideration.

To characterize participation in the adjacent basis,
we define the signed weights
\begin{align}
W_j(d)=a_j(d)b_j(d).
\label{eq:adjacent_mode_weights}
\end{align}
Canonical normalization implies
\begin{align}
\sum_{j=1}^{\ell}W_j(d)=u(d)^Tv(d)=1.
\label{eq:adjacent_weight_sum}
\end{align}
These weights are invariant under reciprocal rescaling
of either the target pair or any reference pair.
In particular, $u_j^{(0)}\to c_j u_j^{(0)}$ and
$v_j^{(0)}\to c_j^{-1}v_j^{(0)}$ induce
$a_j\to c_j^{-1}a_j$ and $b_j\to c_jb_j$,
leaving $W_j$ unchanged.

The weights have a direct canonical interpretation.
Define the parts of the target quadratures assembled
from the adjacent core by
\begin{align}
Q_C(d)&=\sum_{j\in C_0}a_j(d)Q_j^{(0)},
&
\Pi_C(d)&=\sum_{j\in C_0}b_j(d)\Pi_j^{(0)},
&
W_C(d)&=\sum_{j\in C_0}W_j(d).
\end{align}
The analogous halo weight satisfies $W_C+W_H=1$.
Thus $W_H$ measures the adjacent halo's contribution
to the canonical commutator of the target pair.
Neither $W_j$ nor $W_H$ is a probability or a contribution
to logarithmic negativity; both can be signed.
Individual halo weights also depend on the chosen
canonical completion. The sector sum is invariant
under changes of basis within the fixed halo sector
that preserve the separate position and momentum
structure.

The mode-space weights should be distinguished from
the spatial participation function $P_i(d)=u_i(d)v_i(d)$.
Substituting Eq.~\eqref{eq:adjacent_mode_expansion} gives
\begin{align}
P_i(d)
&=
\sum_j W_j(d)P_i^{(0,j)}
+
\sum_{j\neq k}a_j(d)b_k(d)
u_{i,j}^{(0)}v_{i,k}^{(0)},
\label{eq:spatial_participation_expansion}
\end{align}
where $u_{i,j}^{(0)}$ and $v_{i,j}^{(0)}$ are the
coefficients at site $i$ of reference mode $j$, and
$P_i^{(0,j)}=u_{i,j}^{(0)}v_{i,j}^{(0)}$.
The second sum contains cross terms between the position
component of one reference mode and the momentum
component of another. By biorthogonality, these terms
vanish upon summation over sites, but they can
redistribute participation within the interval and
substantially modify its spatial profile.
The weights $W_j(d)$ therefore describe the mode's
canonical composition without fully determining its
spatial shape.

\begin{figure}[t]
\begin{center}
\includegraphics[scale=0.38]{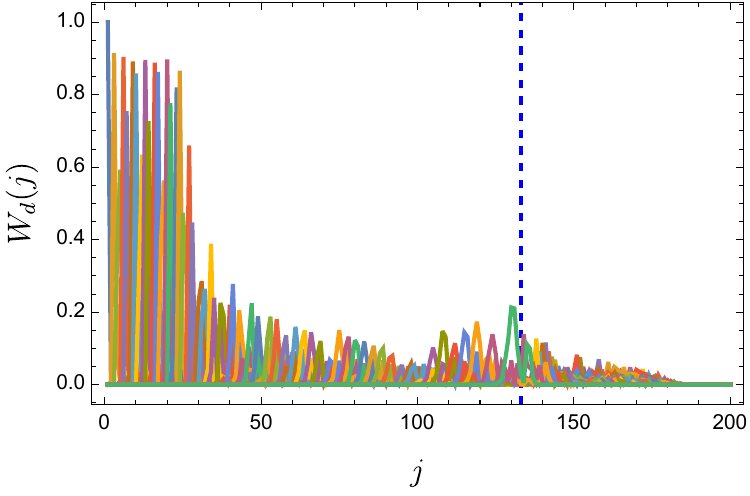}
\hspace{4mm}
\includegraphics[scale=0.38]{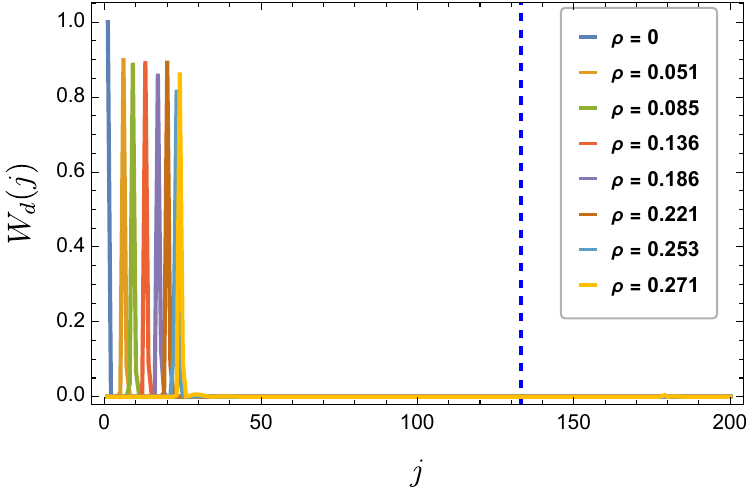}
\hspace{4mm}
\includegraphics[scale=0.38]{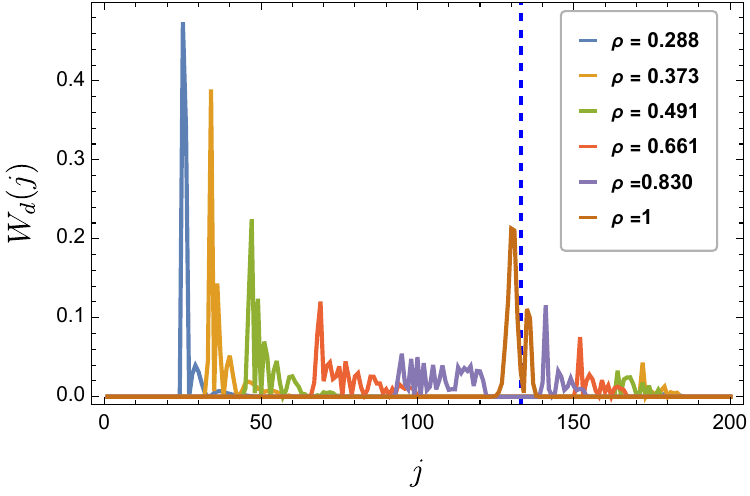}
\includegraphics[scale=0.38]{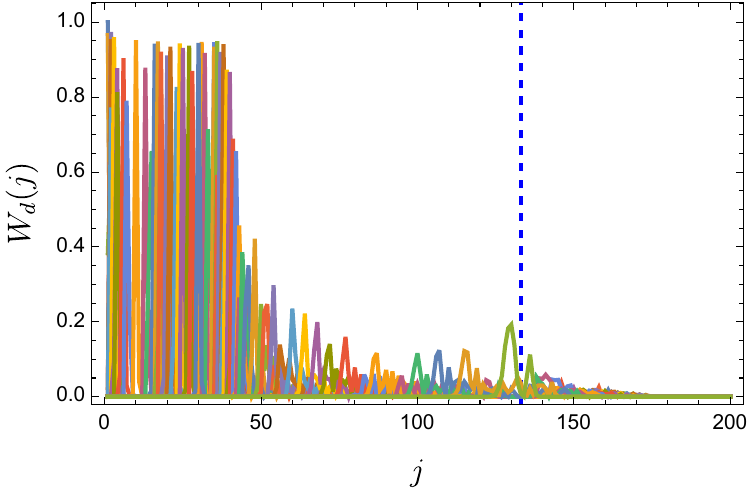}
\hspace{4mm}
\includegraphics[scale=0.38]{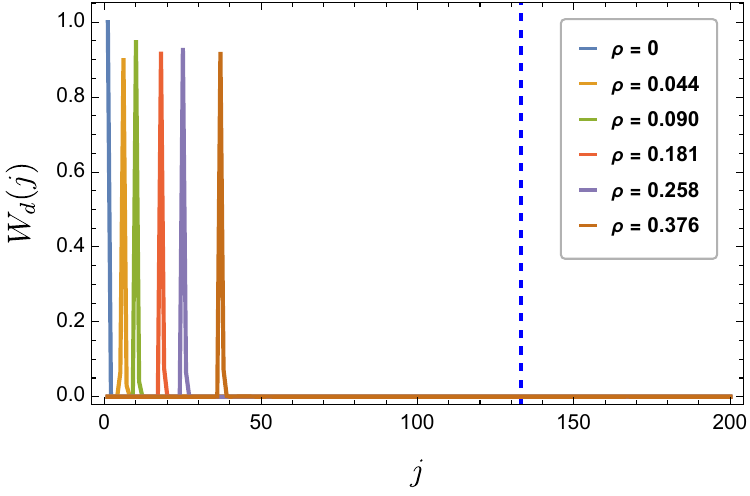}
\hspace{4mm}
\includegraphics[scale=0.38]{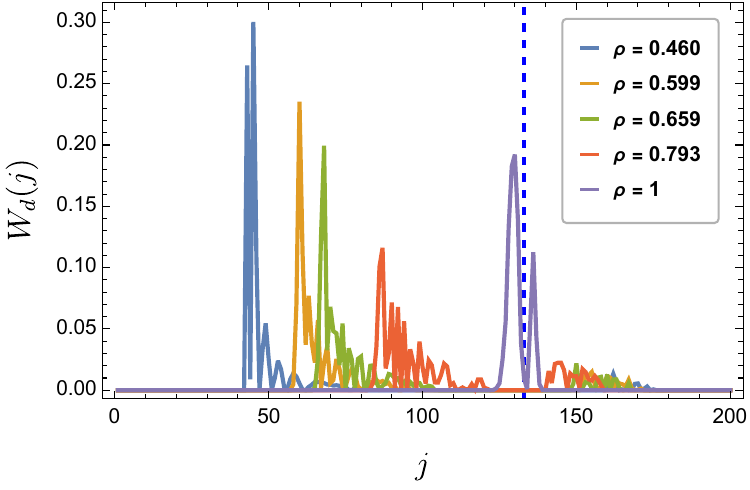}
\\
\includegraphics[scale=0.38]{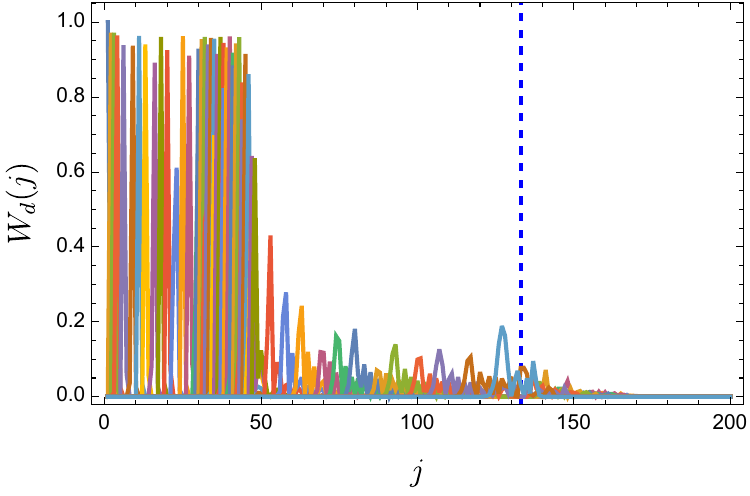}
\hspace{4mm}
\includegraphics[scale=0.38]{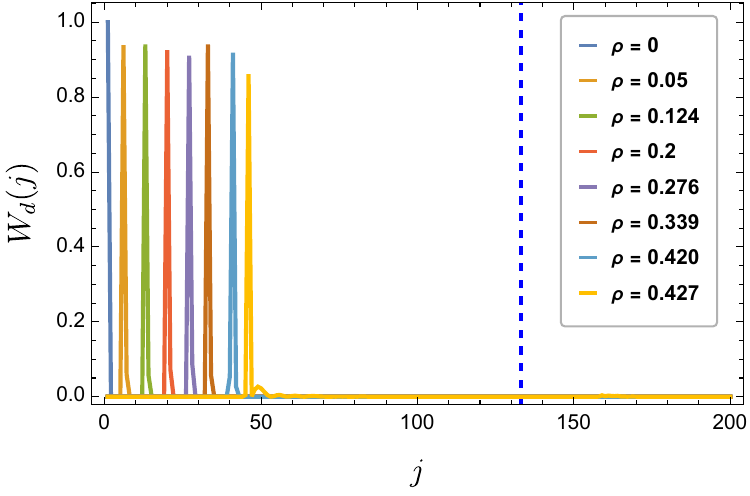}
\hspace{4mm}
\includegraphics[scale=0.38]{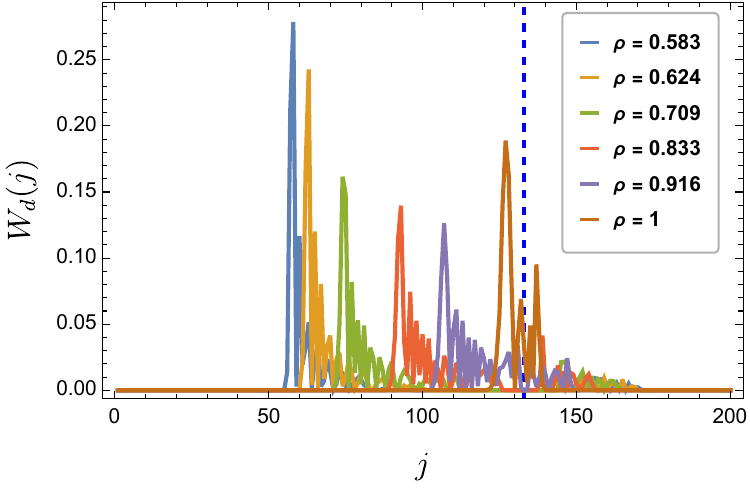}
\end{center}
\caption{
Adjacent-basis composition of the dominant
negativity-carrying mode for $\ell=200$ and $m=0.2, 0.4, 0.5$ (upper, middle and lower rows) with normalized separation
$\rho=d/d_{\cancel{\mathcal E}}$. $d_{\cancel{\mathcal E}}=1083, 744, 660$ for $m=0.2, 0.4, 0.5$. 
The left panels show the canonical weights
$W_j(d)=a_j(d)b_j(d)$ for the full set of separations
displayed. The central panels focus on the
boundary-localized regime, where a narrow peak moves
toward larger $j$, corresponding to progressively
weaker contributions to the adjacent logarithmic
negativity. The right panels focus on the collective
regime, where the distribution develops an asymmetric
tail toward higher-index reference modes, followed
by partial reconcentration near the negativity sphere.
Vertical dashed blue lines mark the adjacent
core-halo boundary.
}
\label{fig:adjacent_basis_flow}
\end{figure}

\begin{figure}[t]
	\begin{center}
        \includegraphics[scale=0.38]{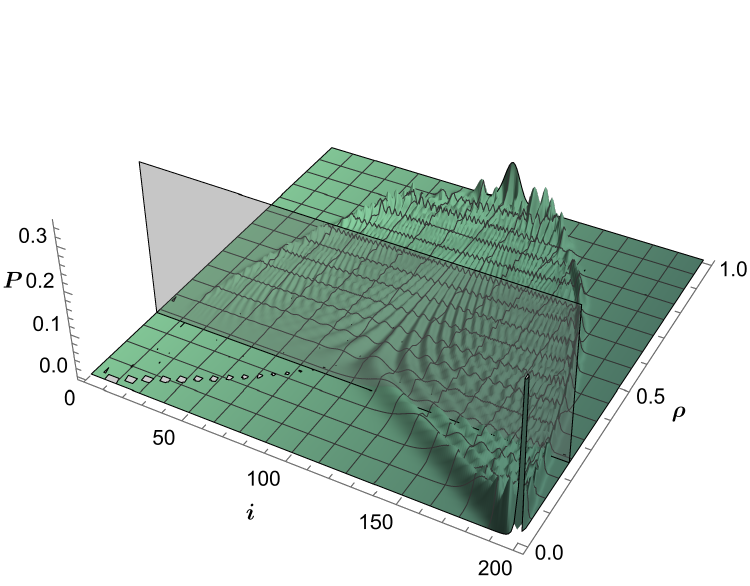}
        \hspace{2mm}
        \includegraphics[scale=.38]{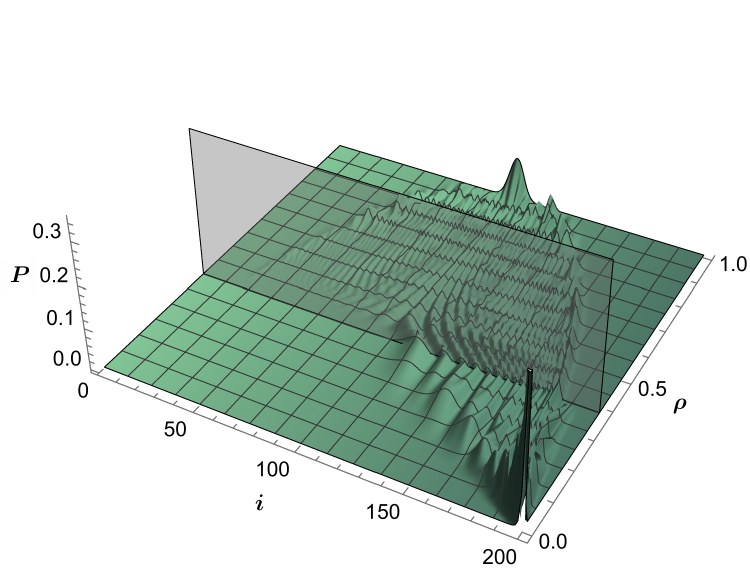}
        \hspace{2mm}
        \includegraphics[scale=.38]{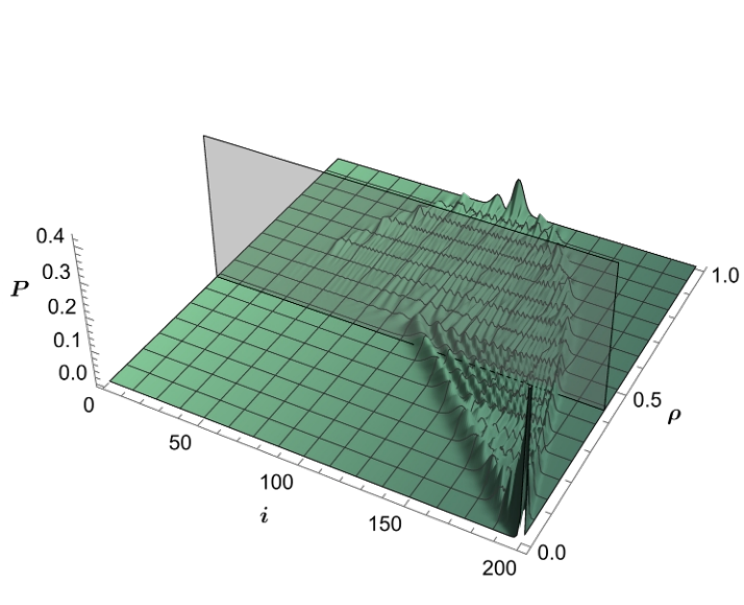}
        \\
        \includegraphics[scale=0.38]{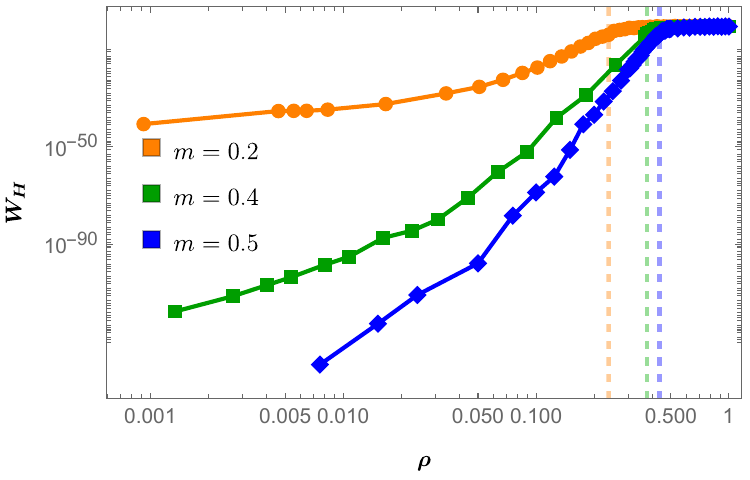}
	\end{center}
    \caption{
    The first row panels show the spatial separation flow as a function of $\rho$
from left for $m=0.2, 0.4, 0.5$, respectively. The bottom-left panel shows the net adjacent-halo weight $W_H(d)=\sum_{j\in H_0}W_j(d)$. Its growth indicates a crossover from negligible to appreciable halo participation; $W_H=10^{-4}$ is used as a numerical reference criterion rather than a sharp physical onset. The dotted vertical lines (corresponding to the transparent planes in the first-row panels) that mark this crossover are located at $\rho_*=0.237, 0.376, 0.436$ respectively. Complementary data for $W_C$ and $W_H$ are provided in Appendix~\ref{Appendix:SeparationFlow}.
}
	\label{fig:dominant_core_vs_d_W_H}
\end{figure}

Figure~\ref{fig:adjacent_basis_flow} shows how the
dominant negativity-carrying mode changes its
composition in this basis. At $\rho=0$, the target
mode coincides with the first reference mode,
so that $W_j(0)=\delta_{j1}$.
In the boundary-localized regime, the weights form
a narrow peak that moves toward larger $j$ as
separation increases. Its height remains close
to unity, indicating that a single adjacent
reference pair accounts for most of the target
mode's canonical normalization, while net
adjacent-halo participation remains negligible.
The dominant finite-separation mode thus
successively samples directions that contribute
less strongly to the adjacent logarithmic negativity.

In the collective regime, the narrow peak gives
way to an asymmetric distribution over several
reference modes. The peak continues to move toward
larger $j$, while a pronounced tail develops toward
still higher indices and contributions on the
lower-index side remain strongly suppressed.
This mixing accompanies the development of
approximately reflection-symmetric spatial profiles
within each interval and involves directions
belonging to the adjacent halo.
Near the negativity sphere, the distribution
becomes more concentrated again, showing that
spatial collectivity does not require monotonically
increasing spread in the adjacent basis.
The corresponding progression of adjacent-core
profiles with mode index connects the two flows,
but does not identify them mode by mode:
the finite-separation profile combines several
reference directions and includes the cross terms
in Eq.~\eqref{eq:spatial_participation_expansion}.

The comparison between different masses in
Figure~\ref{fig:adjacent_basis_flow} shows that increasing
the mass shifts the boundary-localized-to-collective
crossover toward larger adjacent-mode indices $j$.
Increasing the subsystem size makes the spike structure
in the boundary-localized regime and the crossover
between the two regimes more distinct.
The onset of collective participation is indicated by
the blue curve in the right panels of
Figure~\ref{fig:adjacent_basis_flow}. 
The numerical cost of consolidation and the adjacent-basis
expansion increases sharply as the mass decreases or the
subsystem size grows, preventing us from extending the
$\ell=200$ calculations in
Figure~\ref{fig:adjacent_basis_flow} to smaller masses.

Comparisons across masses and subsystem sizes indicate
that the extent of the boundary-localized regime is
controlled primarily by the dimensionless lattice mass
$m\epsilon$, rather than by $m\ell$.
As $m\epsilon$ decreases, the range of normalized
separations over which the dominant mode remains
boundary localized shrinks, and collective profiles
appear earlier in the separation flow.
In the massless regime examined, only the leading
adjacent mode retains a localized character, while
the dominant mode at nonzero resolved separation
develops collective participation.
The localized-to-collective crossover therefore depends
on the mass scale relative to the lattice cutoff.

The growth of adjacent-halo participation is quantified by $W_H$, shown in the lower panel of Figure~\ref{fig:dominant_core_vs_d_W_H} (see also Appendix~\ref{Appendix:SeparationFlow}). Its increase marks a crossover from negligible to appreciable net halo participation. This crossover is indicated by the transparent plane in the spatial separation flows in the upper panels. Within our numerical resolution, the crossover coincides with the onset of the collective regime.

The expansion makes the geometry dependence of the entanglement-carrying modes explicit.
Directions belonging to the halo at adjacency
can participate in the dominant NPT mode at
finite separation; their reference classification
does not preclude this role.
Together, these observations establish a systematic
reorganization through the adjacent-mode hierarchy,
while leaving the limiting halo weight under
continuum refinement as a separate question.

\subsection{Long-distance participation profiles and their UV sensitivity}
\label{subsec:last_mode_uvir}

We now follow the dominant negativity-carrying mode
toward the negativity sphere.
In the massless regime, its canonical components
and spatial participation develop markedly different
momentum-space structures.

\begin{figure}[h!]
\begin{center}
\includegraphics[scale=0.23]{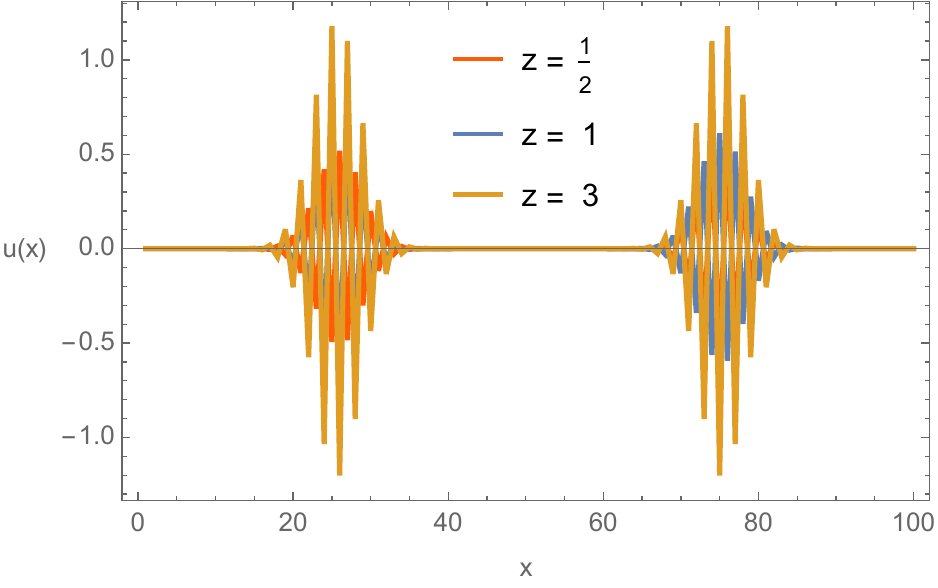}
\hspace{2mm}
\includegraphics[scale=0.23]{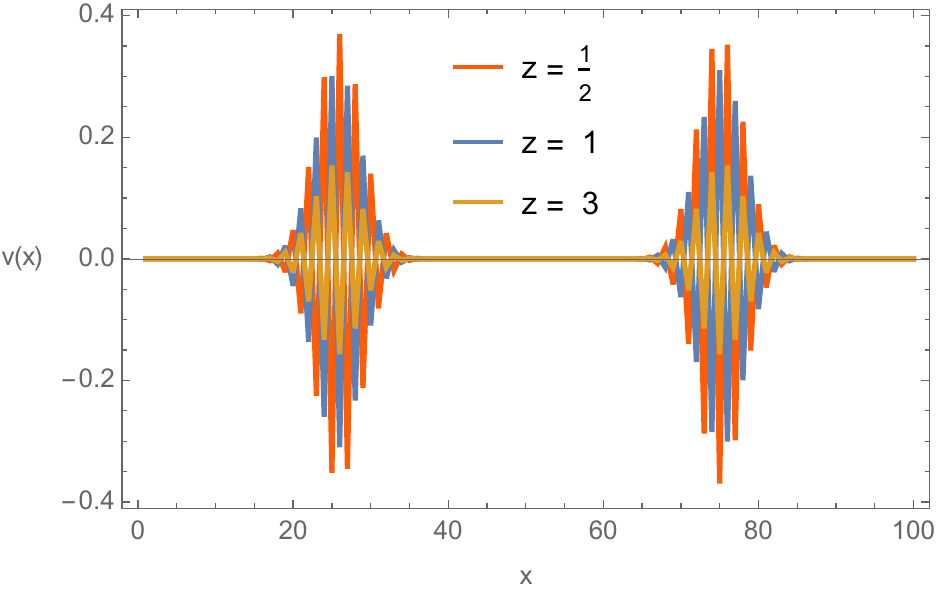}
\hspace{2mm}
\includegraphics[scale=0.23]{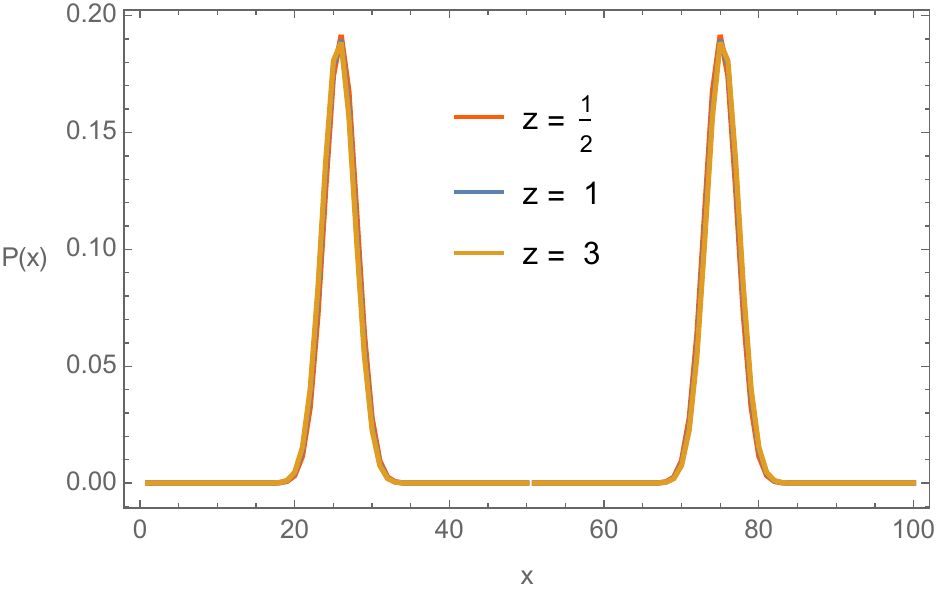}
\hspace{2mm}
\includegraphics[scale=0.23]{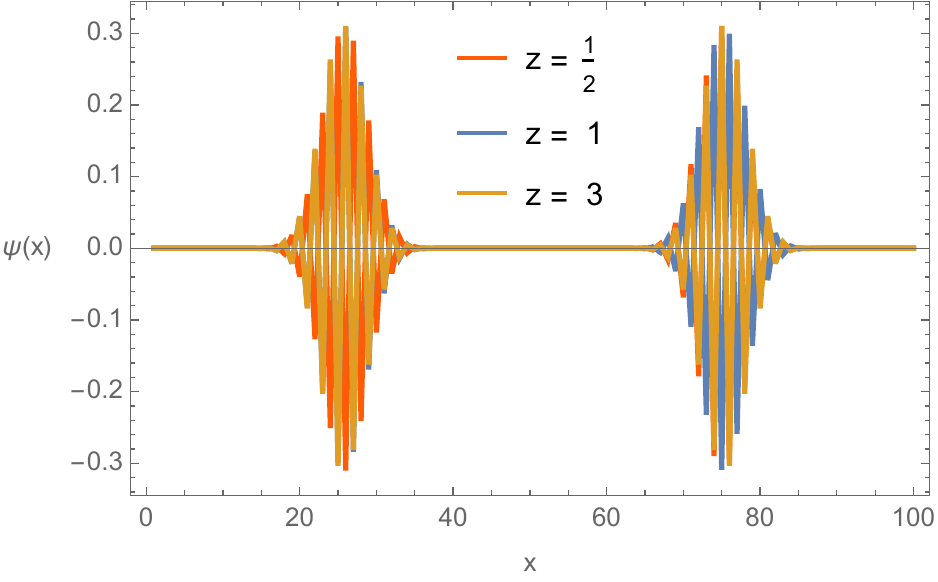}
\includegraphics[scale=0.23]{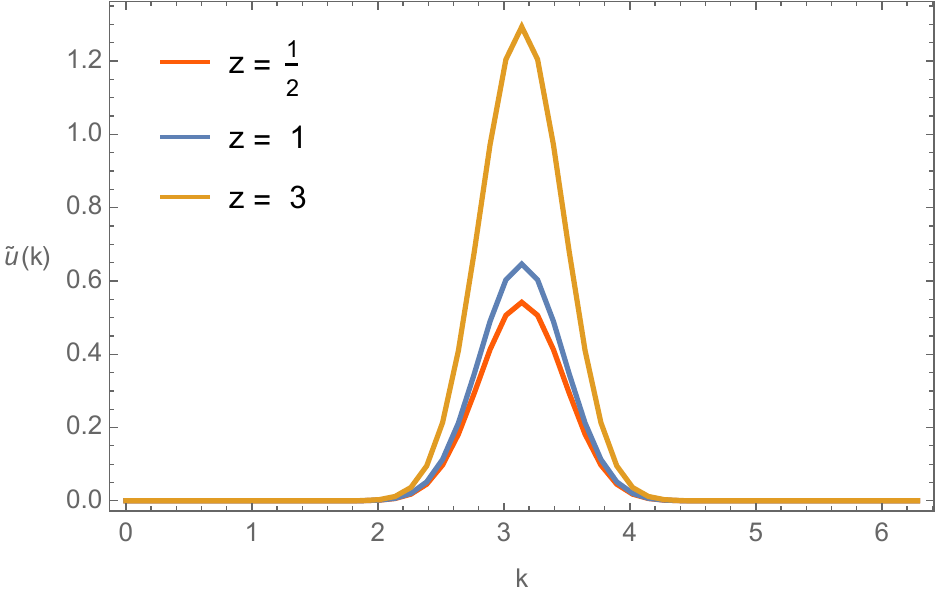}
\hspace{2mm}
\includegraphics[scale=0.23]{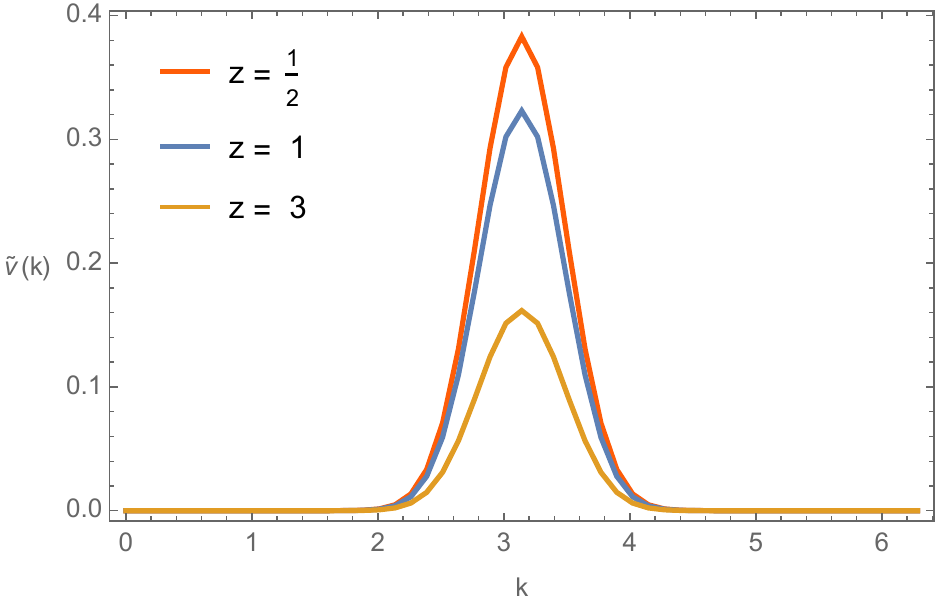}
\hspace{2mm}
\includegraphics[scale=0.23]{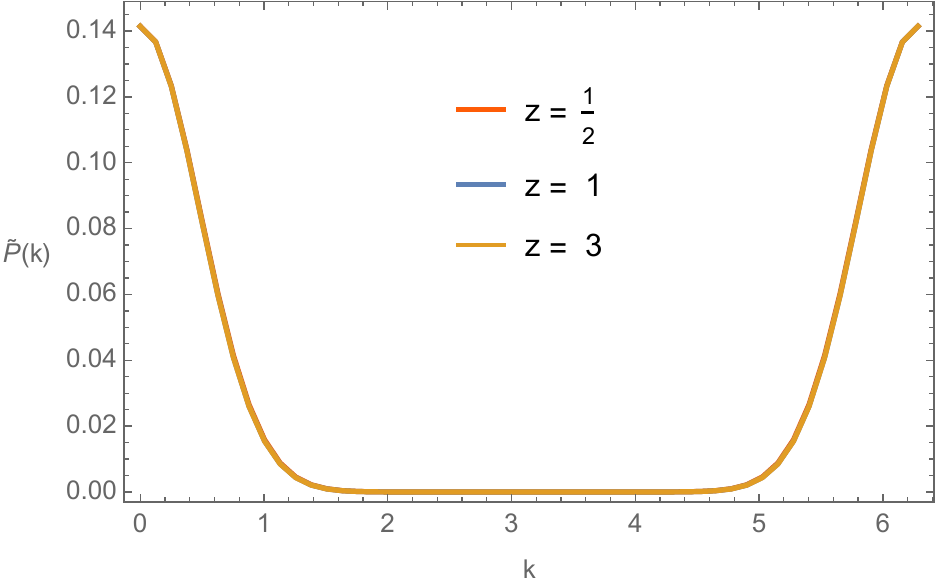}
\hspace{2mm}
\includegraphics[scale=0.23]{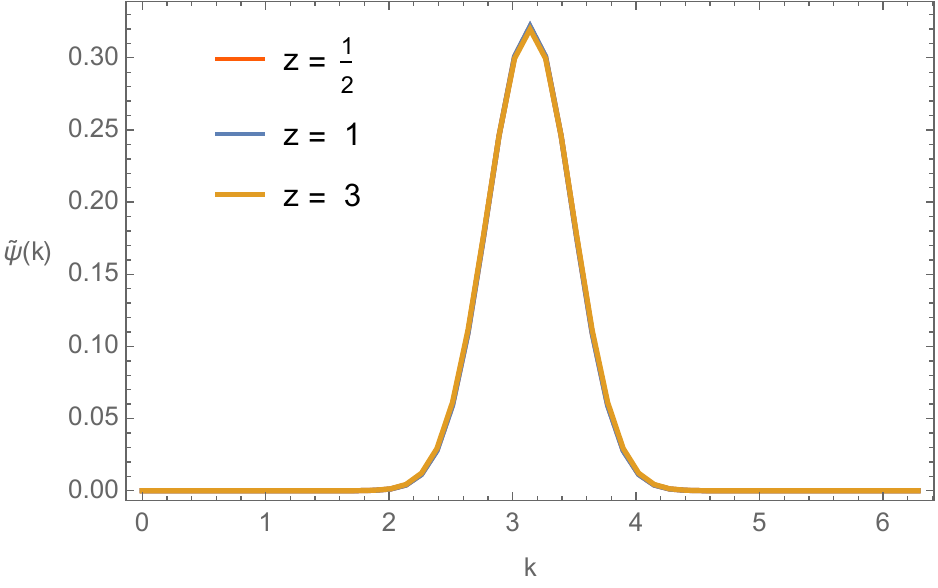}
\end{center}
\caption{
Mode structure near the negativity sphere in the
massless regime with $m=10^{-6}$ and $\ell=50$.
The columns display $u$, $v$, the participation
function $P=u\,v$, and the PT-mode wavefunction
$\Psi_{\rm SK}$, respectively, in real space
(top) and momentum space (bottom). In the real space plots, since the negativity sphere radius varies with $z$, we have truncated the separation between the subregions on the position axis.
}
\label{fig:modewise_dneg_massless}
\end{figure}

In the $d\to d_{{\cancel{\mathcal E}}}$ limit, as shown in figure \ref{fig:modewise_dneg_massless}, the canonical
components $u$ and $v$ form oscillatory wave packets
centered near the middle of each subregion.
The amplitude of $u$ increases with $z$, while that of $v$ decreases.
Both components are concentrated near the lattice
UV point $k=\pi$ in momentum space, with their peak
heights changing in the same respective directions.
The participation function behaves differently:
$P(x)=u(x)v(x)$ is positive, smooth, and center-peaked,
without the rapid sign oscillations of the separate
components. Its Fourier transform is strongly
IR dominated, with weight concentrated near $k=0$
and essentially no UV peak.
The point $k=2\pi$ represents the same IR point
at the repeated Brillouin-zone endpoint.

On the other hand, the Savage-Klco wavefunction $\Psi_{\rm SK}$ retains
the oscillatory wave-packet structure and UV dominance
of the PT mode.
For comparisons across $z$, we fix the arbitrary
overall sign of each canonical pair and of
$\Psi_{\rm SK}$ by requiring positive overlap with
a fixed reference profile.
The signs of $u$ and $v$ are changed together,
leaving $P=u\,v$ unchanged.
Figure~\ref{fig:modewise_dneg_massless} 
summarize the long-distance profiles.

\paragraph{Canonical structure and spatial participation.}

The contrast between $P$ and $\Psi_{\rm SK}$ reflects
the different properties they resolve.
For a covariance matrix with vanishing $q$--$p$
correlations, the PT symplectic eigenproblem implies
Eq.~\eqref{eq:uv}, relating $\Psi_{\rm SK}$ to a
canonical component of the same PT mode.
It therefore characterizes the mode's canonical
realization, rather than the spatial weight with
which the original oscillators participate in the
consolidated pair.
The latter is described by $P=u\,v$, which is
invariant under reciprocal rescaling of the conjugate
components and under their simultaneous sign reversal.
As discussed around Eq.~\eqref{modewise1}, $P$
resolves canonical participation, not a local density
of logarithmic negativity.

The coexistence of UV-dominated components and
IR-dominated participation is transparent in
Fourier space. Since $P$ is a pointwise product,
its Fourier transform is a convolution,
\begin{align}
\widetilde P(k)
=
\sum_q\widetilde u(q)\widetilde v(k-q),
\label{eq:P_convolution}
\end{align}
up to Fourier-normalization conventions.
Components concentrated near $q\simeq\pi$ can
therefore generate a contribution near
$k\simeq2\pi\equiv0$.
The smooth participation profile and the
oscillatory canonical components are thus
compatible properties of the same mode.
In particular, IR dominance of $\widetilde P$
does not imply that the mode can be constructed
from low-momentum field degrees of freedom alone.

This use of ``infrared'' should also be distinguished
from the zero-mode divergence discussed in the
previous section.
That divergence concerns the growth of
field-correlation amplitudes as the mass regulator
is removed, whereas the Fourier profile of $P$
describes the spatial variation of normalized
canonical participation.
Smooth, IR-dominated participation persists even
when the leading zero-mode divergence cancels
from the negativity-carrying sector.
It therefore neither contradicts that cancellation
nor identifies the divergent zero mode as the
source of the surviving negativity.

\paragraph{Scaling toward a common long-distance profile.}

\begin{figure}[h!]
\begin{center}
\includegraphics[scale=0.3]{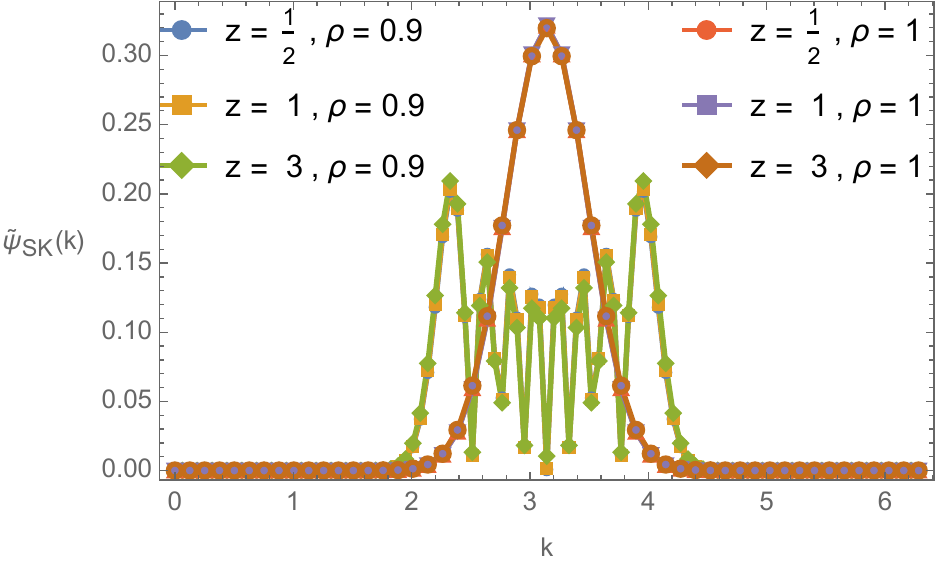}
\hspace{2mm}
\includegraphics[scale=0.3]{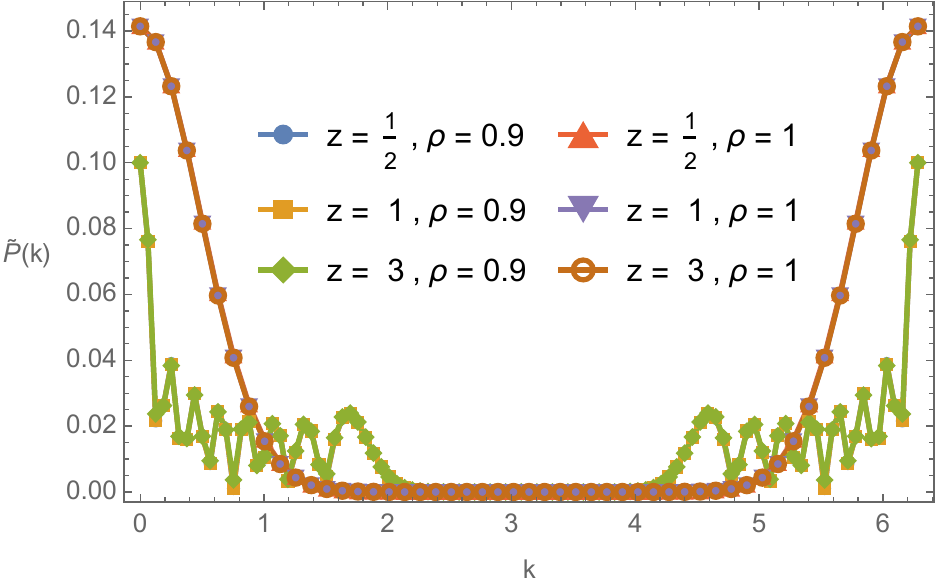}
\hspace{2mm}
\includegraphics[scale=0.3]{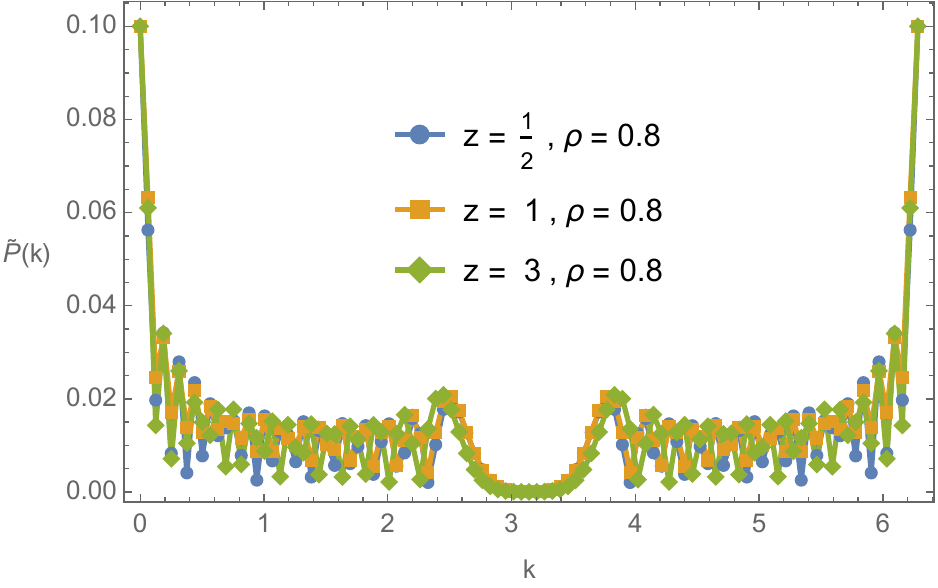}
\end{center}
\caption{
Comparison of long-distance momentum-space profiles
for $z=1/2,1,3$ at scaled separation
$\rho=d/d_{{\cancel{\mathcal E}}}(z)$.
The panels show the PT-mode wavefunction at
$\rho=0.9$ (left) and the normalized participation
profiles at $\rho=0.9$ (middle) and $\rho=0.8$
(right).
The participation profiles are already close
at $\rho=0.8$ and collapse within numerical
resolution at $\rho=0.9$, before the negativity
sphere is reached.
We use $m=10^{-6}$ and $\ell=100$,
with negativity radii $16720$, $11041$,
and $5079$ for $z=1/2,1,3$, respectively.
}
\label{fig:P_scaling}
\end{figure}

To compare theories with different negativity
ranges, we use the scaled separation
$\rho=d/d_{{\cancel{\mathcal E}}}(z)$. As Figure~\ref{fig:P_scaling} illustrates, for $z=1/2,1,3$, representing stiff and soft IR sectors, the normalized momentum-space
participation profiles $\widetilde P_z(k;\rho)$
are already close at $\rho=0.8$ and become
indistinguishable within the graphical
resolution of the comparison
at $\rho=0.9$ and at the resolved endpoint
denoted by $\rho=1$.
The data therefore support the asymptotic form
\begin{align}
\widetilde P_z(k;\rho)\longrightarrow P_*(k),
\qquad \rho\to1^-,
\label{eq:P_scaling}
\end{align}
for the exponents and massless regime studied.
The collapse before the final NPT/PPT crossing
indicates that the negativity radius sets,
empirically, a distance scale for the approach
to the common profile as well as the range
of NPT entanglement.
Changing $z$ substantially modifies that range,
while leaving the normalized asymptotic
participation profile approximately unchanged.

\paragraph {UV-controlled deformation.}
As we have shown in Section~\ref{sec:UVIRdeformations}, controlled deformation on UV, holding the IR fixed, helps to understand the effect of the UV sector on the negativity sphere radius. The collapse of the normalized participation profiles
across the dispersions studied does not establish
insensitivity to microscopic ultraviolet structure.
We therefore examine ultraviolet deformations that
preserve the leading infrared dispersion.

\begin{figure}[h!]
\begin{center}
\includegraphics[scale=0.35]{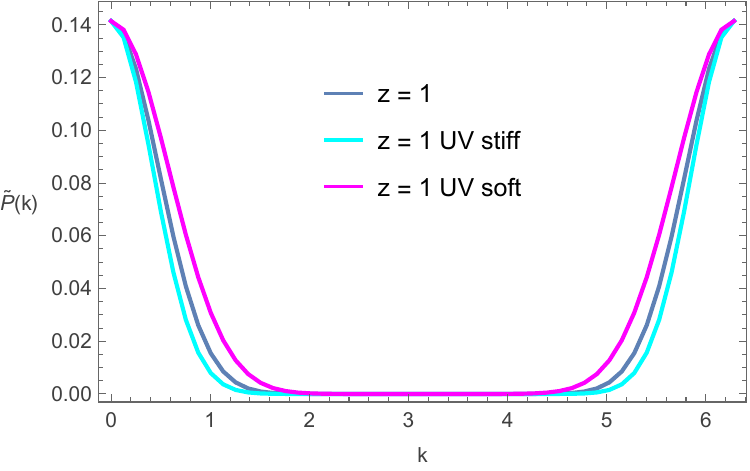}
\hspace{3mm}
\includegraphics[scale=0.35]{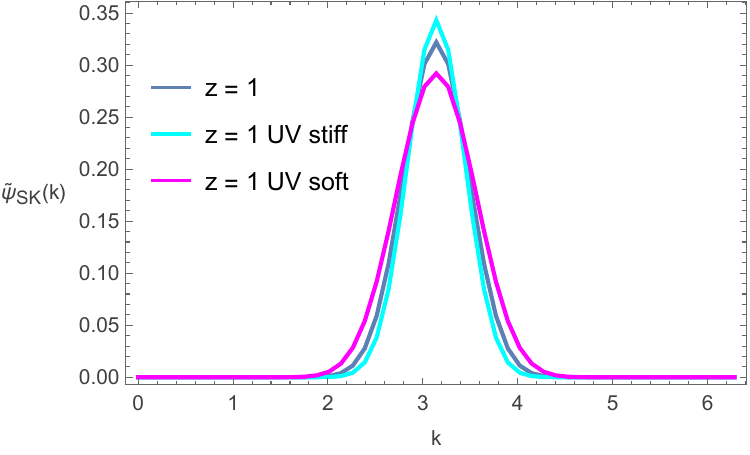}
\end{center}
\caption{
Effect of UV controlled deformations on $\widetilde P(k)$ and $\widetilde{\Psi}_{\rm SK}(k)$ for $z=1$. The UV stiff/soft correspond to $a=1,-1/8$ with 
$d_{{\cancel{\mathcal E}}}=4319, 2347$. We use $m=10^{-6}$ and $\ell=50$ in all plots and the undeformed theory corresponds to $d_{{\cancel{\mathcal E}}}=2723$.}
\label{fig:P_UV_deformation}
\end{figure}

Using the ultraviolet deformation introduced in Eq.~\ref{eq:UVdeformed_dispersion}, the canonical representation responds visibly: for $a>0$, the UV peak of $u$ becomes sharper, the high-momentum structure of $v$ changes in the opposite direction, and the UV peak of $\widetilde\Psi_{\rm SK}(k)$ is enhanced. The participation function remains IR-dominated. See Figure~\ref{fig:P_UV_deformation}. The values of $\widetilde P(k)$ at $k=0$ and $2\pi$ are unchanged, as required by the common normalization $\sum_iP_i=1$. Away from these equivalent endpoints, however, the deformation shifts the crossover between the IR lobe and intermediate momenta. In real space, it slightly broadens the central profile and reduces its peak. The persistence of IR dominance therefore coexists with UV sensitivity of the detailed participation profile. The opposite behavior happens for $a<0$.

\subsection*{Canonical components and spatial participation at adjacency}

Having related the spatial flows through the adjacent
basis, for the sake of completeness, we now compare the canonical components $u$ and $v$, their participation function $P=u\,v$, and the
Savage-Klco PT-mode wavefunction $\Psi_{\rm SK}$ introduced in Ref.~\cite{Klco:2021biu}
(see also Appendix~\ref{appendix:SKWavefunction}).
We begin with the dominant negativity-carrying mode
in the adjacent geometry.

\begin{figure}[h]
\begin{center}
\includegraphics[scale=0.3]{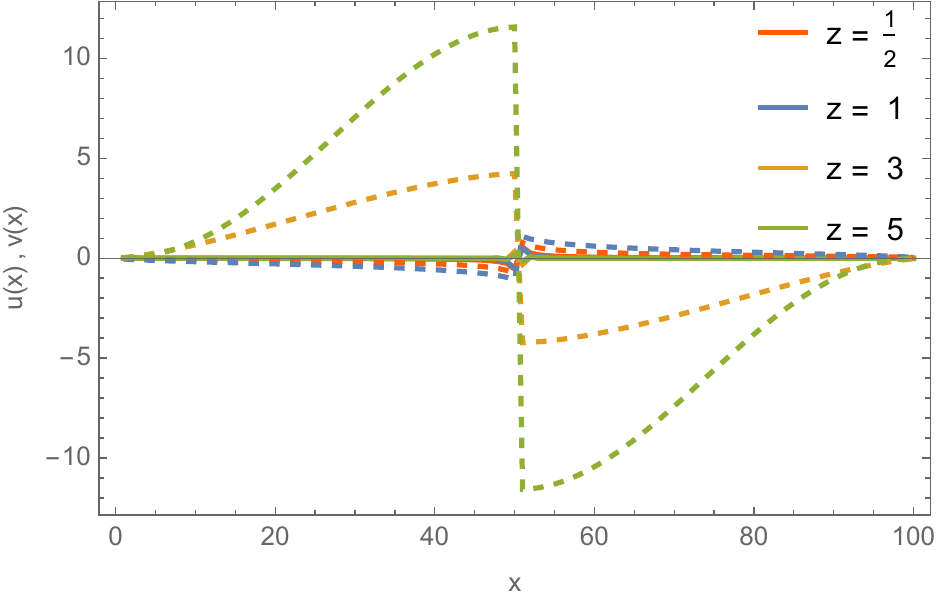}
\hspace{2mm}
\includegraphics[scale=0.3]{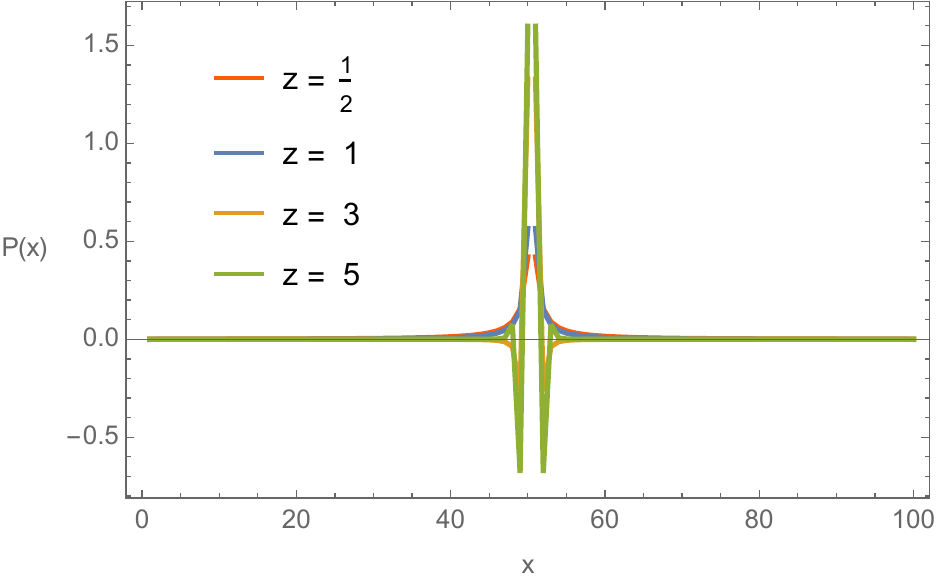}
\hspace{2mm}
\includegraphics[scale=0.3]{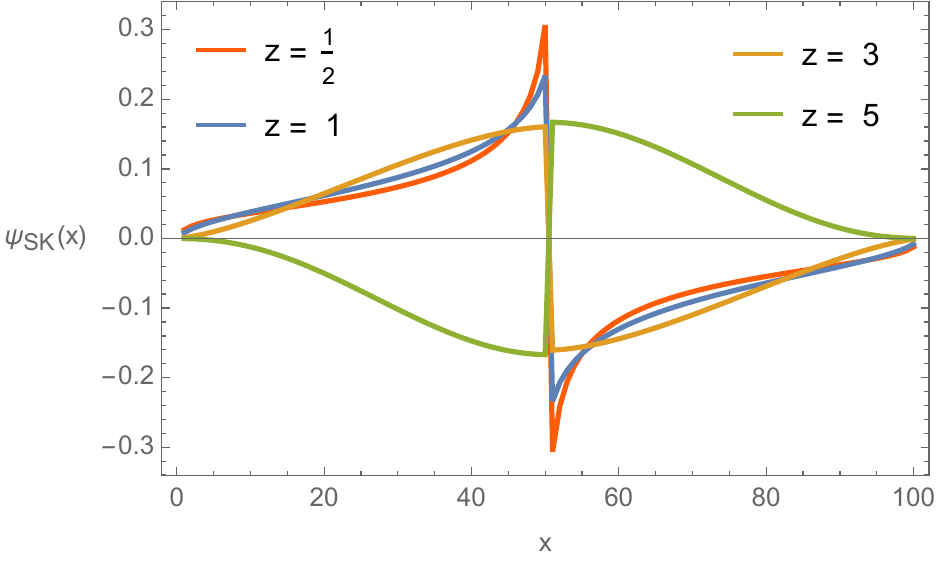}
\end{center}
\caption{
Mode structure at adjacency, $d=0$, in the massless
regime. The panels display the canonical components
$u$ (solid) and $v$ (dashed), the participation function
$P=u\,v$, and the PT-mode wavefunction $\Psi_{\rm SK}$
in real space.
We set $m=10^{-6}$ and $\ell=50$ throughout.
}
\label{fig:modewise_adjacent}
\end{figure}

Figure~\ref{fig:modewise_adjacent} displays these
quantities in real space.
Both canonical components are peaked at the shared
boundary, but their momentum-space profiles respond
differently to the dispersion.
The component $v$ remains IR dominated across the
exponents studied, with a nonzero UV contribution.
The component $u$ is IR dominated for $z\leq1$,
whereas a broad UV peak develops for $z>1$ and
sharpens as $z$ increases.

The participation function also changes character
across these regimes. For $z\leq1$, $P(x)$ is broadly
distributed around the shared boundary; for $z>1$,
it becomes sharply localized there, with a peak
height that increases with $z$.
Correspondingly, $\widetilde P(k)$ has a broad
IR--UV distribution for $z\leq1$.
Its IR maximum remains approximately fixed, while
the spectral weight near the lattice UV point
decreases as $z$ is reduced.
For $z>1$, a broad UV-centered maximum develops
and sharpens mildly with increasing $z$.

By contrast, $\Psi_{\rm SK}$ is less sensitive to
this change. Its real-space profile remains broadly
centered at the shared boundary and extends farther
into the subregion than $P(x)$.
In momentum space, it is IR dominated for all
exponents studied, with a small contribution near
$k=\pi$ that decreases as $z$ increases.

\subsection{Ultraviolet mode structure and collective participation}
\label{sec:UV_collective}

The preceding results refine the physical interpretation of the mode carrying the surviving long-distance entanglement. In the configurations where a single consolidated pair reproduces the entire logarithmic negativity, its canonical components retain pronounced ultraviolet structure, while its participation is smooth and dominated by long spatial scales. These observations distinguish the microscopic oscillations entering the canonical observables from the spatial distribution of the oscillators participating in the entangled pair. Identifying the mode solely through its ultraviolet components therefore leaves its collective spatial organization unresolved.

\begin{figure}[h!]
\begin{center}
\includegraphics[scale=0.32]{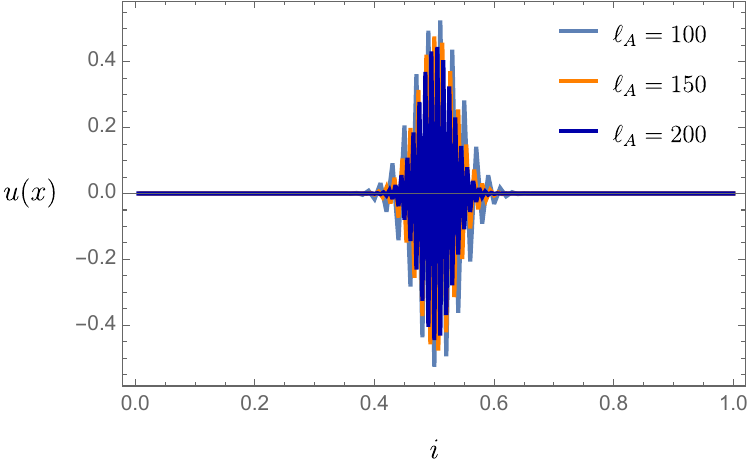}
\hspace{2mm}
\includegraphics[scale=0.32]{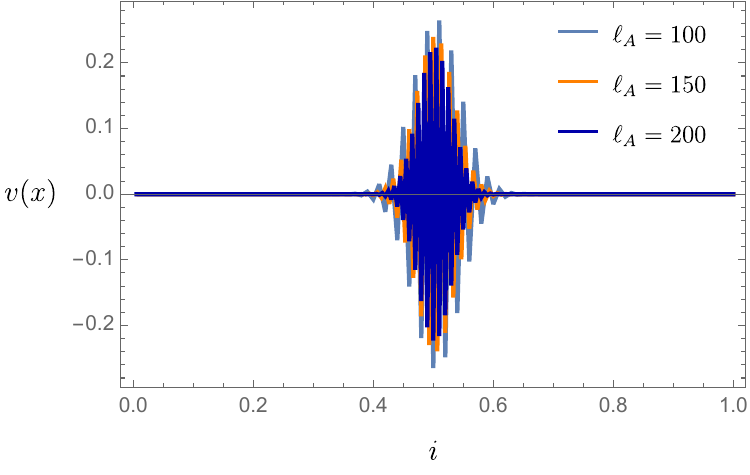}
\hspace{2mm}
\includegraphics[scale=0.32]{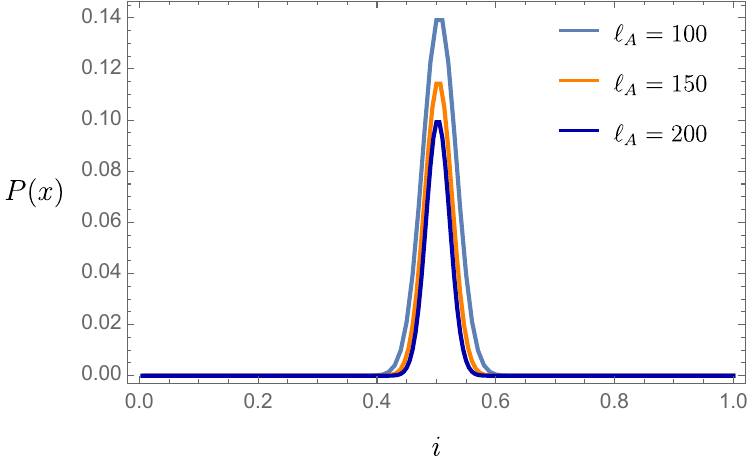}

\includegraphics[scale=0.32]{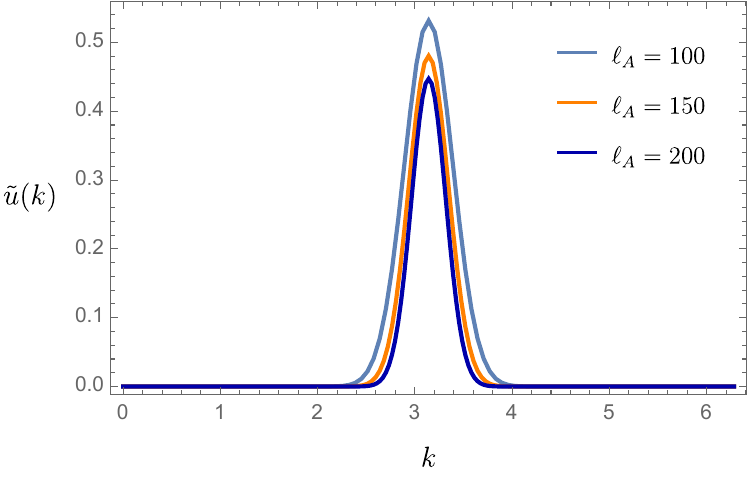}
\hspace{2mm}
\includegraphics[scale=0.32]{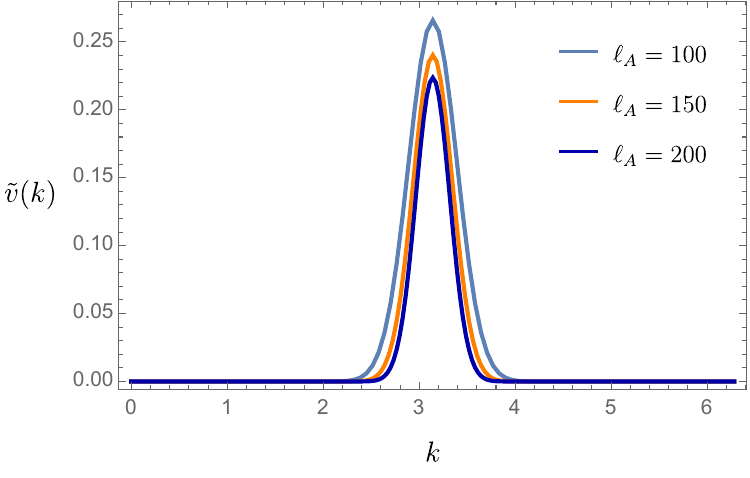}
\hspace{2mm}
\includegraphics[scale=0.32]{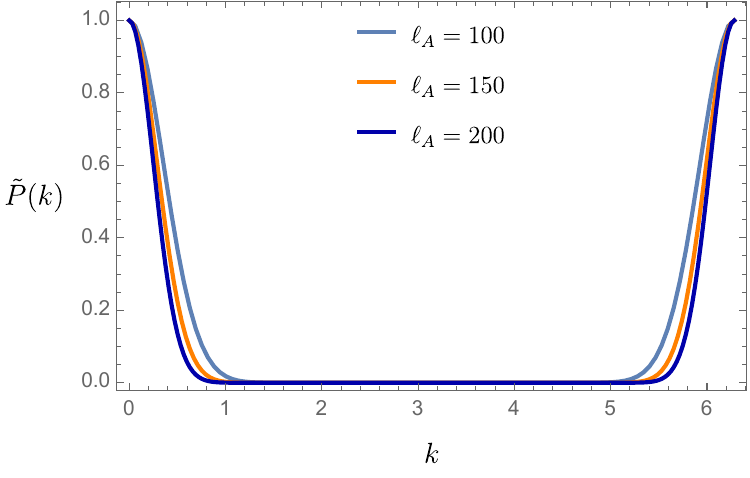}

\end{center}
\caption{
Mode structure of the entangled canonical modes with the smallest contribution to entanglement in a periodic chain for $m=10^{-6}$.
The columns display $u$, $v$, and the participation function $P=u\,v$, in real space (top) and momentum space (bottom).
}
\label{fig:modewise_Pure_massless}
\end{figure}

This distinction is already present in the pure-state setting studied by Botero and Reznik \cite{Botero:2004vpl}. As Figure~\ref{fig:modewise_Pure_massless} illustrates, the modewise decomposition identifies weakly entangled inner modes whose participation is concentrated near the midpoint of each region and whose canonical components exhibit short-wavelength oscillations. To connect this picture with our mixed-state results, we examine the real- and momentum-space participation of the least-entangled pair in the pure-state hierarchy. Although its canonical components contain substantial ultraviolet contributions (see the lower-left and lower-middle panels), its participation spectrum is concentrated at low momenta in the massless regime (see the lower-right panel). Thus, the distinction between ultraviolet canonical structure and infrared-dominated participation also occurs for a pure bipartition.

The participation scale depends systematically on the mass and system size. At fixed total size and bipartition, increasing the mass narrows the central participation profile and increases its high-momentum content. In the system-size comparisons considered here, with the relative sizes of the two regions held fixed, larger systems exhibit more pronounced infrared dominance of the participation spectrum. These observations show that the extent of collective participation is not determined by the short-wavelength character of the canonical components alone. Concentration around the midpoint is compatible with participation extending over many lattice sites; its spatial width must be distinguished from the oscillation wavelength of the components.

For separated mixed-state regions, our analysis establishes how this collective organization emerges as the entanglement decreases. The dominant pair evolves from boundary-localized participation toward a profile centered within each region. Its expansion in the adjacent-state canonical basis connects this evolution to a redistribution toward weaker, more extended reference modes, including adjacent-halo directions. Near the negativity radius, the participation profiles approach a common normalized form across the massless dispersions examined, even though the entanglement ranges remain theory-dependent.

These findings refine the ultraviolet mode picture developed in Refs.~\cite{Botero:2004vpl, Klco:2021biu}. The surviving entanglement combines microscopic oscillatory components with collective, long-wavelength participation. The ultraviolet sensitivity of its range and of its detailed profile remains important: as the deformation test above shows, infrared-dominated participation does not imply independence from the ultraviolet regulator. The additional structure established here is the spatial organization of that participation, its evolution with separation, and its common long-distance form within the family studied.

\section{Conclusions and discussion}
\label{sec:conclusions}
We have investigated how long-wavelength vacuum fluctuations and spatial separation affect the magnitude, range, and
spatial organization of entanglement shared by two finite
regions in Gaussian scalar vacua.
By combining correlation measures with a canonical-mode
analysis, we distinguish the infrared enhancement of total
correlations from the behavior of distillable entanglement
and identify the collective degrees of freedom carrying
the surviving entanglement.

The main results can be summarized as follows:
\begin{itemize}
	
	\item \textbf{Infrared sensitivity of logarithmic negativity.}
	We establish that logarithmic negativity remains uniformly
	bounded as the mass approaches zero for fixed finite
	regions at fixed lattice spacing, despite the infrared
	divergences of the field correlations.
	A two-site analysis illustrates the cancellation of
	the leading spatially uniform contribution, while
	the general bound does not require resolving the
	subleading divergences individually.
	Our numerical results for $z=1,3,5,7,9$ further support
	convergence to a finite massless limit with positive-power
	corrections. The detailed spectral mechanism governing this approach
	remains open, and finite infrared effects on the limiting
	negativity and its mass corrections are not excluded.
	
	\item \textbf{Divergent total correlations and bounded
		distillable entanglement.}
	The uniform bound on logarithmic negativity excludes a
	divergent distillation rate between fixed finite regions,
	even when mutual information diverges.
	Consequently, $E_D(A:B)/I(A:B)\to0$ as the mass approaches
	zero in these configurations.
	For bipartite Gaussian states, positive negativity
	guarantees distillability under unrestricted local
	operations and classical communication.
	This establishes a parametric separation between total
	correlations and distillable entanglement, without
	identifying the divergent contribution exclusively with
	classical correlations or bound entanglement, and without
	excluding finite infrared effects on distillable
	entanglement.
	
	\item \textbf{Spatial reorganization and the adjacent-mode
		hierarchy.}
	The negativity-carrying modes exhibit a crossover from
	boundary-localized to collective, approximately
	reflection-symmetric participation profiles.
	This evolution occurs both along the mode-index hierarchy
	at adjacency and as separation increases while following
	the dominant mode.
	Expansion in the complete adjacent-state canonical basis
	connects these observations: participation migrates toward
	weaker adjacent-core modes, while adjacent-halo directions
	become appreciably involved in the collective regime.
	The dimensionless lattice mass $m\epsilon$ controls the
	extent of the localized regime in the configurations
	examined. This regime shrinks as $m\epsilon$ decreases,
	with only the leading adjacent mode retaining a localized
	profile in the massless regime studied.
	
	\item \textbf{Canonical structure and spatial participation.}
	The canonical pair carrying the surviving long-distance
	logarithmic negativity exhibits smooth, \textit{infrared-dominated}
	participation despite its ultraviolet canonical structure.
	This distinction also appears in the pure-state modes
	examined here.
	It refines earlier high-frequency mode descriptions by
	revealing the collective spatial participation underlying
	the microscopic oscillations.
	
	\item \textbf{Universality across dispersions.}
	In the massless regime, the normalized long-distance
	participation profiles approach a universal form across
	the dynamical exponents studied when separation is measured
	in units of the negativity radius.
	The collapse is already visible before the NPT/PPT
	threshold. In the massive regime, profiles at fixed
	nonzero $m\ell$ retain a dependence on the dynamical
	exponent.
	
	\item \textbf{Entanglement range.}
	Across the exponents examined, the massless
	negativity radius exhibits the common scaling
	$d_{\cancel{\mathcal{E}}}\simeq\gamma(z)\ell^2/\epsilon$,
	with a dispersion-dependent coefficient.
	This extends the quadratic scaling previously
	established for the relativistic harmonic chain
	to the other dispersion exponents studied.
	
\end{itemize}

The infrared analysis and the mode-wise description address
different aspects of the same reduced state.
The former examines how long-wavelength fluctuations enter
the correlation measures, while the latter identifies the
canonical degrees of freedom participating in the negativity.
Together, they show that \textit{stronger total correlations need
not imply greater logarithmic negativity}, and that
increasing separation changes the composition of the
dominant entanglement mode as well as its contribution
to the negativity.

The adjacent-basis expansion shows that spatial collectivity
need not correspond to a monotonically increasing spread
over reference modes: near the negativity sphere, the
canonical weights can become more concentrated even while
participation remains collective.
These weights resolve canonical normalization rather than
fractions of logarithmic negativity.
Cross terms between reference modes also contribute to
the spatial profile, so the expansion connects the two
flows without identifying them mode by mode.

The distinction between canonical structure and
participation refines the earlier high-frequency mode
descriptions of Botero--Reznik and Klco--Savage
\cite{Botero:2004vpl,Klco:2021biu}.
Ultraviolet-dominated canonical components do not fully
describe how the original oscillators participate
collectively in the entangled pair.
Conversely, infrared-dominated participation does not
imply that the pair can be constructed from low-momentum
field observables alone.
The tested ultraviolet deformation preserves this
distinction while modifying the participation profile
and the entanglement range.

These results complement contour approaches to the spatial
structure of Gaussian entanglement
\cite{Vidal:2014aal,Coser:2017dtb,Zambotti:2026vpm}.
Here the resolved quantity is the canonical participation
of an individual consolidated mode, rather than a spatial
decomposition of the total logarithmic negativity.
A direct comparison of these profiles with negativity
contours would help determine which features of their
separation dependence are shared by the two diagnostics.

The observed profile collapse raises a question beyond
the existence of scale invariance: why should theories
with different dynamical exponents share the same
normalized long-distance participation profile?
Scale invariance alone does not require their scaling
functions to coincide.
The breakdown of the collapse under mass deformations
provides a constraint on an analytical explanation.

An analytical explanation of the common quadratic scaling
of the negativity radius must account for both the
dispersion and the spatial regulator.
Extrapolation beyond the numerical window also requires
control of the finite-mass crossover: both $m\ell$ and
$m d_{\cancel{\mathcal{E}}}$ must remain small to follow
the massless scaling examined here.

In the massless configurations examined, controlled
dispersion deformations show that stiffening either
momentum sector extends the negativity-sphere radius,
whereas softening it shortens the radius, with the
complementary sector preserved as specified in the
construction. These responses are consistent with
infrared softening outweighing ultraviolet stiffening
along the near-massless Lifshitz family, but do not
quantitatively separate their contributions or establish
a general monotonicity principle.

The complementary singular-value analysis in
Appendix~\ref{app:correlation_channels} reveals
an increasingly pronounced hierarchy of cross-region
correlations at large separation.
This hierarchy does not determine the number of
NPT modes, which depends on the full covariance
matrix, including its local blocks.
Cross-region correlations persist beyond the
negativity sphere, which marks the disappearance
of NPT entanglement rather than of correlations.

Our infrared results also invite comparison with the
resources required to prepare a Gaussian vacuum.
Optical preparation of the harmonic-chain vacuum from
squeezed inputs and a passive network requires divergent
squeezing as the massless limit is approached
(see Appendix A of \cite{Mollabashi:2024gik}).
By contrast, logarithmic negativity between fixed finite
regions at fixed lattice spacing remains bounded and
therefore bounds their distillable entanglement even
as mutual information diverges.
These observations concern different operational tasks:
preparing the global vacuum and distilling entanglement
after its complement has been discarded.
They do not identify the divergent squeezing resource
with the hidden entanglement of a particular core-halo
preparation.
It remains an open question how the infrared preparation
requirements are distributed among local squeezing,
entangled mode pairs, and correlated noise.

Our results motivate investigating the preparation resources
of Lifshitz vacua in continuous-variable quantum simulators.
The pronounced enhancement of the preparation-halo
entanglement at larger $z$ suggests an increased resource
requirement for preparation procedures based on the
decomposition employed here.
Alternative Gaussian decompositions can, however,
reduce preparation-resource bounds, as demonstrated
by minimum-noise filtering \cite{Gao:2024esb}.
Applying such methods to Lifshitz vacua would determine
whether the observed enhancement persists under improved
preparation procedures and how it translates into
requirements such as squeezing and entangling operations.

A natural extension is to investigate the spatial
organization of entanglement in fermionic Gaussian
states. Pure bipartite states admit a decomposition
into entangled mode pairs under local fermionic
canonical transformations \cite{Botero:2004twx},
providing a starting point for examining the mixed
states associated with two retained intervals.
For such states, the conventional partial transpose
generally produces a sum of two Gaussian operators
rather than a single Gaussian operator
\cite{Eisler:2015tgq}.
The partial-time-reversal construction instead
preserves Gaussianity at the operator level and
defines a fermionic negativity accessible through
covariance-matrix methods
\cite{Shapourian:2016cqu,Shapourian:2018ozl}.
Building on studies of negativity in Dirac
\cite{Herzog:2016ohd,Coser:2015eba}
and Lifshitz fermion theories \cite{Vasli:2024mrf},
one could investigate whether mixed-state
entanglement admits an analogous modewise
organization and how the relevant canonical
profiles change with separation.
This would test which aspects of the
boundary-to-collective crossover and the distinction
between canonical structure and participation
extend to fermions.

Several concrete questions remain open.
An analytical description of the long-distance
participation profile would help explain the observed
collapse across Lifshitz exponents and its sensitivity
to UV deformations.
A systematic study of the adjacent-mode expansion for
$z\neq1$, beyond the analysis presented here, is also
needed to determine how the dispersion affects peak
migration, mixing among reference modes, and the
contribution of the adjacent halo.
The behavior of this halo contribution under increasing
spatial resolution should be examined separately at
fixed physical interval sizes and separation, and at
fixed $\rho=d/d_{\cancel{\mathcal{E}}}$, since the latter
follows a cutoff-dependent separation.
Finally, determining how subleading cross-region
correlation channels affect the smallest PT symplectic
eigenvalue would clarify their role in setting the
negativity threshold.
These investigations would connect the observed
reorganization of canonical modes more directly
to the persistence and eventual loss of NPT
entanglement.

\paragraph{Acknowledgments}
We thank Tadashi Takayanagi for helpful discussions, Natalie Klco for helpful correspondence concerning our early results, and Saleh Rahimi-Keshari for comments on the draft. We also acknowledge discussions with ChatGPT.

\bibliographystyle{unsrturl}
\bibliography{bibliography}

\begin{thebibliography}{100}

\bibitem{Feynman:1981tf}
Richard~P. Feynman.
\newblock {Simulating physics with computers}.
\newblock {\em Int. J. Theor. Phys.}, 21:467--488, 1982.
\newblock \href {https://doi.org/10.1007/BF02650179}
  {\path{doi:10.1007/BF02650179}}.

\bibitem{Bennett:1992zzb}
Charles~H. Bennett and Stephen~J. Wiesner.
\newblock {Communication via one- and two-particle operators on
  Einstein-Podolsky-Rosen states}.
\newblock {\em Phys. Rev. Lett.}, 69:2881--2884, 1992.
\newblock \href {https://doi.org/10.1103/PhysRevLett.69.2881}
  {\path{doi:10.1103/PhysRevLett.69.2881}}.

\bibitem{Bennett:1992tv}
Charles~H. Bennett, Gilles Brassard, Claude Crepeau, Richard Jozsa, Asher
  Peres, and William~K. Wootters.
\newblock {Teleporting an unknown quantum state via dual classical and
  Einstein-Podolsky-Rosen channels}.
\newblock {\em Phys. Rev. Lett.}, 70:1895--1899, 1993.
\newblock \href {https://doi.org/10.1103/PhysRevLett.70.1895}
  {\path{doi:10.1103/PhysRevLett.70.1895}}.

\bibitem{Bennett:2014rmv}
Charles~H. Bennett and Gilles Brassard.
\newblock {Quantum cryptography: Public key distribution and coin tossing}.
\newblock {\em Theor. Comput. Sci.}, 560:7--11, 2014.
\newblock \href {https://arxiv.org/abs/2003.06557} {\path{arXiv:2003.06557}},
  \href {https://doi.org/10.1016/j.tcs.2014.05.025}
  {\path{doi:10.1016/j.tcs.2014.05.025}}.

\bibitem{Amico:2007ag}
Luigi Amico, Rosario Fazio, Andreas Osterloh, and Vlatko Vedral.
\newblock {Entanglement in many-body systems}.
\newblock {\em Rev. Mod. Phys.}, 80:517--576, 2008.
\newblock \href {https://arxiv.org/abs/quant-ph/0703044}
  {\path{arXiv:quant-ph/0703044}}, \href
  {https://doi.org/10.1103/RevModPhys.80.517}
  {\path{doi:10.1103/RevModPhys.80.517}}.

\bibitem{Laflorencie:2015eck}
Nicolas Laflorencie.
\newblock {Quantum entanglement in condensed matter systems}.
\newblock {\em Phys. Rept.}, 643:1--59, 2016.
\newblock \href {https://arxiv.org/abs/1512.03388} {\path{arXiv:1512.03388}},
  \href {https://doi.org/10.1016/j.physrep.2016.06.008}
  {\path{doi:10.1016/j.physrep.2016.06.008}}.

\bibitem{Casini:2009sr}
H.~Casini and M.~Huerta.
\newblock {Entanglement entropy in free quantum field theory}.
\newblock {\em J. Phys. A}, 42:504007, 2009.
\newblock \href {https://arxiv.org/abs/0905.2562} {\path{arXiv:0905.2562}},
  \href {https://doi.org/10.1088/1751-8113/42/50/504007}
  {\path{doi:10.1088/1751-8113/42/50/504007}}.

\bibitem{Casini:2022rlv}
Horacio Casini and Marina Huerta.
\newblock {Lectures on entanglement in quantum field theory}.
\newblock {\em PoS}, TASI2021:002, 2023.
\newblock \href {https://arxiv.org/abs/2201.13310} {\path{arXiv:2201.13310}},
  \href {https://doi.org/10.22323/1.403.0002} {\path{doi:10.22323/1.403.0002}}.

\bibitem{Ryu:2006bv}
Shinsei Ryu and Tadashi Takayanagi.
\newblock {Holographic derivation of entanglement entropy from AdS/CFT}.
\newblock {\em Phys. Rev. Lett.}, 96:181602, 2006.
\newblock \href {https://arxiv.org/abs/hep-th/0603001}
  {\path{arXiv:hep-th/0603001}}, \href
  {https://doi.org/10.1103/PhysRevLett.96.181602}
  {\path{doi:10.1103/PhysRevLett.96.181602}}.

\bibitem{Solodukhin:2011gn}
Sergey~N. Solodukhin.
\newblock {Entanglement entropy of black holes}.
\newblock {\em Living Rev. Rel.}, 14:8, 2011.
\newblock \href {https://arxiv.org/abs/1104.3712} {\path{arXiv:1104.3712}},
  \href {https://doi.org/10.12942/lrr-2011-8} {\path{doi:10.12942/lrr-2011-8}}.

\bibitem{Almheiri:2020cfm}
Ahmed Almheiri, Thomas Hartman, Juan Maldacena, Edgar Shaghoulian, and
  Amirhossein Tajdini.
\newblock {The entropy of Hawking radiation}.
\newblock {\em Rev. Mod. Phys.}, 93(3):035002, 2021.
\newblock \href {https://arxiv.org/abs/2006.06872} {\path{arXiv:2006.06872}},
  \href {https://doi.org/10.1103/RevModPhys.93.035002}
  {\path{doi:10.1103/RevModPhys.93.035002}}.

\bibitem{Bombelli:1986rw}
Luca Bombelli, Rabinder~K. Koul, Joohan Lee, and Rafael~D. Sorkin.
\newblock {A Quantum Source of Entropy for Black Holes}.
\newblock {\em Phys. Rev. D}, 34:373--383, 1986.
\newblock \href {https://doi.org/10.1103/PhysRevD.34.373}
  {\path{doi:10.1103/PhysRevD.34.373}}.

\bibitem{Srednicki:1993im}
Mark Srednicki.
\newblock {Entropy and area}.
\newblock {\em Phys. Rev. Lett.}, 71:666--669, 1993.
\newblock \href {https://arxiv.org/abs/hep-th/9303048}
  {\path{arXiv:hep-th/9303048}}, \href
  {https://doi.org/10.1103/PhysRevLett.71.666}
  {\path{doi:10.1103/PhysRevLett.71.666}}.

\bibitem{Plenio:2004he}
M.~B. Plenio, J.~Eisert, J.~Dreissig, and M.~Cramer.
\newblock {Entropy, entanglement, and area: analytical results for harmonic
  lattice systems}.
\newblock {\em Phys. Rev. Lett.}, 94:060503, 2005.
\newblock \href {https://arxiv.org/abs/quant-ph/0405142}
  {\path{arXiv:quant-ph/0405142}}, \href
  {https://doi.org/10.1103/PhysRevLett.94.060503}
  {\path{doi:10.1103/PhysRevLett.94.060503}}.

\bibitem{Hastings:2007iok}
M.~B. Hastings.
\newblock {An area law for one-dimensional quantum systems}.
\newblock {\em J. Stat. Mech.}, 0708:P08024, 2007.
\newblock \href {https://arxiv.org/abs/0705.2024} {\path{arXiv:0705.2024}},
  \href {https://doi.org/10.1088/1742-5468/2007/08/P08024}
  {\path{doi:10.1088/1742-5468/2007/08/P08024}}.

\bibitem{Eisert:2008ur}
J.~Eisert, M.~Cramer, and M.~B. Plenio.
\newblock {Colloquium: Area laws for the entanglement entropy}.
\newblock {\em Rev. Mod. Phys.}, 82:277--306, 2010.
\newblock \href {https://arxiv.org/abs/0808.3773} {\path{arXiv:0808.3773}},
  \href {https://doi.org/10.1103/RevModPhys.82.277}
  {\path{doi:10.1103/RevModPhys.82.277}}.

\bibitem{Groisman:2005dbo}
Berry Groisman, Sandu Popescu, and Andreas Winter.
\newblock {Quantum, classical, and total amount of correlations in a quantum
  state}.
\newblock {\em Phys. Rev. A}, 72(3):032317, 2005.
\newblock \href {https://arxiv.org/abs/quant-ph/0410091}
  {\path{arXiv:quant-ph/0410091}}, \href
  {https://doi.org/10.1103/PhysRevA.72.032317}
  {\path{doi:10.1103/PhysRevA.72.032317}}.

\bibitem{Wolf:2007tdq}
Michael~M. Wolf, Frank Verstraete, Matthew~B. Hastings, and J.~Ignacio Cirac.
\newblock {Area Laws in Quantum Systems: Mutual Information and Correlations}.
\newblock {\em Phys. Rev. Lett.}, 100(7):070502, 2008.
\newblock \href {https://arxiv.org/abs/0704.3906} {\path{arXiv:0704.3906}},
  \href {https://doi.org/10.1103/PhysRevLett.100.070502}
  {\path{doi:10.1103/PhysRevLett.100.070502}}.

\bibitem{Vidal:2002zz}
G.~Vidal and R.~F. Werner.
\newblock {Computable measure of entanglement}.
\newblock {\em Phys. Rev. A}, 65:032314, 2002.
\newblock \href {https://arxiv.org/abs/quant-ph/0102117}
  {\path{arXiv:quant-ph/0102117}}, \href
  {https://doi.org/10.1103/PhysRevA.65.032314}
  {\path{doi:10.1103/PhysRevA.65.032314}}.

\bibitem{Calabrese:2004eu}
Pasquale Calabrese and John~L. Cardy.
\newblock {Entanglement entropy and quantum field theory}.
\newblock {\em J. Stat. Mech.}, 0406:P06002, 2004.
\newblock \href {https://arxiv.org/abs/hep-th/0405152}
  {\path{arXiv:hep-th/0405152}}, \href
  {https://doi.org/10.1088/1742-5468/2004/06/P06002}
  {\path{doi:10.1088/1742-5468/2004/06/P06002}}.

\bibitem{Calabrese:2009ez}
Pasquale Calabrese, John Cardy, and Erik Tonni.
\newblock {Entanglement entropy of two disjoint intervals in conformal field
  theory}.
\newblock {\em J. Stat. Mech.}, 0911:P11001, 2009.
\newblock \href {https://arxiv.org/abs/0905.2069} {\path{arXiv:0905.2069}},
  \href {https://doi.org/10.1088/1742-5468/2009/11/P11001}
  {\path{doi:10.1088/1742-5468/2009/11/P11001}}.

\bibitem{Calabrese:2012ew}
Pasquale Calabrese, John Cardy, and Erik Tonni.
\newblock {Entanglement negativity in quantum field theory}.
\newblock {\em Phys. Rev. Lett.}, 109:130502, 2012.
\newblock \href {https://arxiv.org/abs/1206.3092} {\path{arXiv:1206.3092}},
  \href {https://doi.org/10.1103/PhysRevLett.109.130502}
  {\path{doi:10.1103/PhysRevLett.109.130502}}.

\bibitem{Furukawa:2008uk}
Shunsuke Furukawa, Vincent Pasquier, and Jun'ichi Shiraishi.
\newblock {Mutual Information and Compactification Radius in a c=1 Critical
  Phase in One Dimension}.
\newblock {\em Phys. Rev. Lett.}, 102:170602, 2009.
\newblock \href {https://arxiv.org/abs/0809.5113} {\path{arXiv:0809.5113}},
  \href {https://doi.org/10.1103/PhysRevLett.102.170602}
  {\path{doi:10.1103/PhysRevLett.102.170602}}.

\bibitem{Alba:2009ek}
Vincenzo Alba, Luca Tagliacozzo, and Pasquale Calabrese.
\newblock {Entanglement entropy of two disjoint blocks in critical Ising
  models}.
\newblock {\em Phys. Rev. B}, 81:060411, 2010.
\newblock \href {https://arxiv.org/abs/0910.0706} {\path{arXiv:0910.0706}},
  \href {https://doi.org/10.1103/PhysRevB.81.060411}
  {\path{doi:10.1103/PhysRevB.81.060411}}.

\bibitem{Calabrese:2013mi}
Pasquale Calabrese, Luca Tagliacozzo, and Erik Tonni.
\newblock {Entanglement negativity in the critical Ising chain}.
\newblock {\em J. Stat. Mech.}, 1305:P05002, 2013.
\newblock \href {https://arxiv.org/abs/1302.1113} {\path{arXiv:1302.1113}},
  \href {https://doi.org/10.1088/1742-5468/2013/05/P05002}
  {\path{doi:10.1088/1742-5468/2013/05/P05002}}.

\bibitem{Audenaert:2002xfl}
K.~Audenaert, J.~Eisert, M.~B. Plenio, and R.~F. Werner.
\newblock {Entanglement Properties of the Harmonic Chain}.
\newblock {\em Phys. Rev. A}, 66:042327, 2002.
\newblock \href {https://arxiv.org/abs/quant-ph/0205025}
  {\path{arXiv:quant-ph/0205025}}, \href
  {https://doi.org/10.1103/PhysRevA.66.042327}
  {\path{doi:10.1103/PhysRevA.66.042327}}.

\bibitem{Marcovitch:2008sxc}
S.~Marcovitch, A.~Retzker, M.~B. Plenio, and B.~Reznik.
\newblock {Critical and noncritical long-range entanglement in Klein-Gordon
  fields}.
\newblock {\em Phys. Rev. A}, 80:012325, 2009.
\newblock \href {https://arxiv.org/abs/0811.1288} {\path{arXiv:0811.1288}},
  \href {https://doi.org/10.1103/PhysRevA.80.012325}
  {\path{doi:10.1103/PhysRevA.80.012325}}.

\bibitem{Anders:2008lus}
Janet Anders.
\newblock {Thermal state entanglement in harmonic lattices}.
\newblock {\em Phys. Rev. A}, 77(6):062102, 2008.
\newblock \href {https://arxiv.org/abs/0803.1102} {\path{arXiv:0803.1102}},
  \href {https://doi.org/10.1103/PhysRevA.77.062102}
  {\path{doi:10.1103/PhysRevA.77.062102}}.

\bibitem{Calabrese:2014yza}
Pasquale Calabrese, John Cardy, and Erik Tonni.
\newblock {Finite temperature entanglement negativity in conformal field
  theory}.
\newblock {\em J. Phys. A}, 48(1):015006, 2015.
\newblock \href {https://arxiv.org/abs/1408.3043} {\path{arXiv:1408.3043}},
  \href {https://doi.org/10.1088/1751-8113/48/1/015006}
  {\path{doi:10.1088/1751-8113/48/1/015006}}.

\bibitem{DeNobili:2015dla}
Cristiano De~Nobili, Andrea Coser, and Erik Tonni.
\newblock {Entanglement entropy and negativity of disjoint intervals in CFT:
  Some numerical extrapolations}.
\newblock {\em J. Stat. Mech.}, 1506(6):P06021, 2015.
\newblock \href {https://arxiv.org/abs/1501.04311} {\path{arXiv:1501.04311}},
  \href {https://doi.org/10.1088/1742-5468/2015/06/P06021}
  {\path{doi:10.1088/1742-5468/2015/06/P06021}}.

\bibitem{Eisler:2016bfd}
Viktor Eisler and Zolt{\'a}n Zimbor{\'a}s.
\newblock {Entanglement negativity in two-dimensional free lattice models}.
\newblock {\em Phys. Rev. B}, 93(11):115148, 2016.
\newblock \href {https://arxiv.org/abs/1511.08819} {\path{arXiv:1511.08819}},
  \href {https://doi.org/10.1103/PhysRevB.93.115148}
  {\path{doi:10.1103/PhysRevB.93.115148}}.

\bibitem{Zhou:2016ykv}
Tianci Zhou, Xiao Chen, Thomas Faulkner, and Eduardo Fradkin.
\newblock {Entanglement entropy and mutual information of circular entangling
  surfaces in the 2 + 1-dimensional quantum Lifshitz model}.
\newblock {\em J. Stat. Mech.}, 1609(9):093101, 2016.
\newblock \href {https://arxiv.org/abs/1607.01771} {\path{arXiv:1607.01771}},
  \href {https://doi.org/10.1088/1742-5468/2016/09/093101}
  {\path{doi:10.1088/1742-5468/2016/09/093101}}.

\bibitem{MohammadiMozaffar:2017chk}
M.~Reza Mohammadi~Mozaffar and Ali Mollabashi.
\newblock {Logarithmic Negativity in Lifshitz Harmonic Models}.
\newblock {\em J. Stat. Mech.}, 1805(5):053113, 2018.
\newblock \href {https://arxiv.org/abs/1712.03731} {\path{arXiv:1712.03731}},
  \href {https://doi.org/10.1088/1742-5468/aac135}
  {\path{doi:10.1088/1742-5468/aac135}}.

\bibitem{Angel-Ramelli:2020wfo}
J.~Angel-Ramelli, C.~Berthiere, V.~Giangreco~M. Puletti, and L.~Thorlacius.
\newblock {Logarithmic Negativity in Quantum Lifshitz Theories}.
\newblock {\em JHEP}, 09:011, 2020.
\newblock \href {https://arxiv.org/abs/2002.05713} {\path{arXiv:2002.05713}},
  \href {https://doi.org/10.1007/JHEP09(2020)011}
  {\path{doi:10.1007/JHEP09(2020)011}}.

\bibitem{Klco:2021biu}
Natalie Klco and Martin~J. Savage.
\newblock {Entanglement Spheres and a UV-IR Connection in Effective Field
  Theories}.
\newblock {\em Phys. Rev. Lett.}, 127(21):211602, 2021.
\newblock \href {https://arxiv.org/abs/2103.14999} {\path{arXiv:2103.14999}},
  \href {https://doi.org/10.1103/PhysRevLett.127.211602}
  {\path{doi:10.1103/PhysRevLett.127.211602}}.

\bibitem{Gao:2024nbl}
Boyu Gao and Natalie Klco.
\newblock {Detecting spacelike vacuum entanglement at all distances and
  promoting negativity to a necessary and sufficient entanglement measure in
  many-body regimes}.
\newblock {\em Phys. Rev. A}, 112(1):012430, 2025.
\newblock \href {https://arxiv.org/abs/2410.04162} {\path{arXiv:2410.04162}},
  \href {https://doi.org/10.1103/m9w1-ppqz} {\path{doi:10.1103/m9w1-ppqz}}.

\bibitem{access1}
Geza Giedke, Lu-Ming Duan, J.~Ignacio Cirac, and Peter Zoller.
\newblock {Entanglement Criteria for All Bipartite Gaussian States }.
\newblock {\em Phys. Rev. Lett.}, 87:167904, 2001.
\newblock \href {https://arxiv.org/abs/quant-ph/0104072}
  {\path{arXiv:quant-ph/0104072}}, \href
  {https://doi.org/10.1103/PhysRevLett.87.167904}
  {\path{doi:10.1103/PhysRevLett.87.167904}}.

\bibitem{Calabrese:2010he}
Pasquale Calabrese, John Cardy, and Erik Tonni.
\newblock {Entanglement entropy of two disjoint intervals in conformal field
  theory II}.
\newblock {\em J. Stat. Mech.}, 1101:P01021, 2011.
\newblock \href {https://arxiv.org/abs/1011.5482} {\path{arXiv:1011.5482}},
  \href {https://doi.org/10.1088/1742-5468/2011/01/P01021}
  {\path{doi:10.1088/1742-5468/2011/01/P01021}}.

\bibitem{Cardy:2013nua}
John Cardy.
\newblock {Some results on the mutual information of disjoint regions in higher
  dimensions}.
\newblock {\em J. Phys. A}, 46:285402, 2013.
\newblock \href {https://arxiv.org/abs/1304.7985} {\path{arXiv:1304.7985}},
  \href {https://doi.org/10.1088/1751-8113/46/28/285402}
  {\path{doi:10.1088/1751-8113/46/28/285402}}.

\bibitem{Agon:2015ftl}
Cesar Ag{\'o}n and Thomas Faulkner.
\newblock {Quantum Corrections to Holographic Mutual Information}.
\newblock {\em JHEP}, 08:118, 2016.
\newblock \href {https://arxiv.org/abs/1511.07462} {\path{arXiv:1511.07462}},
  \href {https://doi.org/10.1007/JHEP08(2016)118}
  {\path{doi:10.1007/JHEP08(2016)118}}.

\bibitem{Shiba:2012np}
Noburo Shiba.
\newblock {Entanglement Entropy of Two Spheres}.
\newblock {\em JHEP}, 07:100, 2012.
\newblock \href {https://arxiv.org/abs/1201.4865} {\path{arXiv:1201.4865}},
  \href {https://doi.org/10.1007/JHEP07(2012)100}
  {\path{doi:10.1007/JHEP07(2012)100}}.

\bibitem{Calabrese:2012nk}
Pasquale Calabrese, John Cardy, and Erik Tonni.
\newblock {Entanglement negativity in extended systems: A field theoretical
  approach}.
\newblock {\em J. Stat. Mech.}, 1302:P02008, 2013.
\newblock \href {https://arxiv.org/abs/1210.5359} {\path{arXiv:1210.5359}},
  \href {https://doi.org/10.1088/1742-5468/2013/02/P02008}
  {\path{doi:10.1088/1742-5468/2013/02/P02008}}.

\bibitem{Parez:2023xpj}
Gilles Parez and William Witczak-Krempa.
\newblock {Entanglement negativity between separated regions in quantum
  critical systems}.
\newblock {\em Phys. Rev. Res.}, 6(2):023125, 2024.
\newblock \href {https://arxiv.org/abs/2310.15273} {\path{arXiv:2310.15273}},
  \href {https://doi.org/10.1103/PhysRevResearch.6.023125}
  {\path{doi:10.1103/PhysRevResearch.6.023125}}.

\bibitem{Arias:2026bqh}
Malen Arias, Marina Huerta, Andrei Rotaru, and Erik Tonni.
\newblock {Entanglement negativity for a free scalar chiral current}.
\newblock 1 2026.
\newblock \href {https://arxiv.org/abs/2601.04995} {\path{arXiv:2601.04995}}.

\bibitem{Botero:2003wov}
Alonso Botero and Benni Reznik.
\newblock {Mode-Wise Entanglement of Gaussian States}.
\newblock {\em Phys. Rev. A}, 67:052311, 2003.
\newblock \href {https://arxiv.org/abs/quant-ph/0209026}
  {\path{arXiv:quant-ph/0209026}}, \href
  {https://doi.org/10.1103/PhysRevA.67.052311}
  {\path{doi:10.1103/PhysRevA.67.052311}}.

\bibitem{Botero:2004vpl}
Alonso Botero and Benni Reznik.
\newblock {Spatial structures and localization of vacuum entanglement in the
  linear harmonic chain}.
\newblock {\em Phys. Rev. A}, 70(5):052329, 2004.
\newblock \href {https://arxiv.org/abs/quant-ph/0403233}
  {\path{arXiv:quant-ph/0403233}}, \href
  {https://doi.org/10.1103/PhysRevA.70.052329}
  {\path{doi:10.1103/PhysRevA.70.052329}}.

\bibitem{Klco:2021cxq}
Natalie Klco, D.~H. Beck, and Martin~J. Savage.
\newblock {Entanglement structures in quantum field theories: Negativity cores
  and bound entanglement in the vacuum}.
\newblock {\em Phys. Rev. A}, 107(1):012415, 2023.
\newblock \href {https://arxiv.org/abs/2110.10736} {\path{arXiv:2110.10736}},
  \href {https://doi.org/10.1103/PhysRevA.107.012415}
  {\path{doi:10.1103/PhysRevA.107.012415}}.

\bibitem{Gao:2024esb}
Boyu Gao and Natalie Klco.
\newblock {Partial-transpose-guided entanglement classes and minimum noise
  filtering in many-body Gaussian quantum systems}.
\newblock {\em Phys. Rev. A}, 109(6):062413, 2024.
\newblock \href {https://arxiv.org/abs/2402.13881} {\path{arXiv:2402.13881}},
  \href {https://doi.org/10.1103/PhysRevA.109.062413}
  {\path{doi:10.1103/PhysRevA.109.062413}}.

\bibitem{Perche:2026ams}
T.~Rick Perche and Patricia Ribes-Metidieri.
\newblock {Local Vacuum Entanglement through Most Entangled Modes}.
\newblock 9 2026.
\newblock \href {https://arxiv.org/abs/2609.28621} {\path{arXiv:2609.28621}}.

\bibitem{Vidal:2014aal}
Guifre Vidal and Yangang Chen.
\newblock {Entanglement contour}.
\newblock {\em J. Stat. Mech.}, 2014(10):P10011, 2014.
\newblock \href {https://arxiv.org/abs/1406.1471} {\path{arXiv:1406.1471}},
  \href {https://doi.org/10.1088/1742-5468/2014/10/P10011}
  {\path{doi:10.1088/1742-5468/2014/10/P10011}}.

\bibitem{Kudler-Flam:2019nhr}
Jonah Kudler-Flam, Hassan Shapourian, and Shinsei Ryu.
\newblock {The negativity contour: a quasi-local measure of entanglement for
  mixed states}.
\newblock {\em SciPost Phys.}, 8(4):063, 2020.
\newblock \href {https://arxiv.org/abs/1908.07540} {\path{arXiv:1908.07540}},
  \href {https://doi.org/10.21468/SciPostPhys.8.4.063}
  {\path{doi:10.21468/SciPostPhys.8.4.063}}.

\bibitem{Frerot:2015irl}
Ir{\'e}n{\'e}e Fr{\'e}rot and Tommaso Roscilde.
\newblock {Area law and its violation: A microscopic inspection into the
  structure of entanglement and fluctuations}.
\newblock {\em Phys. Rev. B}, 92:115129, 2015.
\newblock \href {https://arxiv.org/abs/1506.00545} {\path{arXiv:1506.00545}},
  \href {https://doi.org/10.1103/PhysRevB.92.115129}
  {\path{doi:10.1103/PhysRevB.92.115129}}.

\bibitem{Coser:2017dtb}
Andrea Coser, Cristiano De~Nobili, and Erik Tonni.
\newblock {A contour for the entanglement entropies in harmonic lattices}.
\newblock {\em J. Phys. A}, 50(31):314001, 2017.
\newblock \href {https://arxiv.org/abs/1701.08427} {\path{arXiv:1701.08427}},
  \href {https://doi.org/10.1088/1751-8121/aa7902}
  {\path{doi:10.1088/1751-8121/aa7902}}.

\bibitem{Zambotti:2026vpm}
Gioele Zambotti and Erik Tonni.
\newblock {A contour for the entanglement negativity of bosonic Gaussian
  states}.
\newblock {\em SciPost Phys.}, 21:027, 2026.
\newblock \href {https://arxiv.org/abs/2602.18099} {\path{arXiv:2602.18099}},
  \href {https://doi.org/10.21468/SciPostPhys.21.1.027}
  {\path{doi:10.21468/SciPostPhys.21.1.027}}.

\bibitem{Adesso:2007tx}
Gerardo Adesso and Fabrizio Illuminati.
\newblock {Entanglement in continuous variable systems: Recent advances and
  current perspectives}.
\newblock {\em J. Phys. A}, 40:7821--7880, 2007.
\newblock \href {https://arxiv.org/abs/quant-ph/0701221}
  {\path{arXiv:quant-ph/0701221}}, \href
  {https://doi.org/10.1088/1751-8113/40/28/S01}
  {\path{doi:10.1088/1751-8113/40/28/S01}}.

\bibitem{Zhong:2020iql}
Han-Sen Zhong et~al.
\newblock {Quantum computational advantage using photons}.
\newblock {\em Science}, 370(6523):1460--1463, 2020.
\newblock \href {https://arxiv.org/abs/2012.01625} {\path{arXiv:2012.01625}},
  \href {https://doi.org/10.1126/science.abe8770}
  {\path{doi:10.1126/science.abe8770}}.

\bibitem{Zhong:2021dml}
Han-Sen Zhong et~al.
\newblock {Phase-Programmable Gaussian Boson Sampling Using Stimulated Squeezed
  Light}.
\newblock {\em Phys. Rev. Lett.}, 127:180502, 2021.
\newblock \href {https://arxiv.org/abs/2106.15534} {\path{arXiv:2106.15534}},
  \href {https://doi.org/10.1103/PhysRevLett.127.180502}
  {\path{doi:10.1103/PhysRevLett.127.180502}}.

\bibitem{Madsen:2022uqz}
Lars~S. Madsen et~al.
\newblock {Quantum computational advantage with a programmable photonic
  processor}.
\newblock {\em Nature}, 606(7912):75--81, 2022.
\newblock \href {https://doi.org/10.1038/s41586-022-04725-x}
  {\path{doi:10.1038/s41586-022-04725-x}}.

\bibitem{Plenio:2005cwa}
M.~B. Plenio.
\newblock {Logarithmic Negativity: A Full Entanglement Monotone That is not
  Convex}.
\newblock {\em Phys. Rev. Lett.}, 95:090503, 2005.
\newblock \href {https://arxiv.org/abs/quant-ph/0505071}
  {\path{arXiv:quant-ph/0505071}}, \href
  {https://doi.org/10.1103/PhysRevLett.95.090503}
  {\path{doi:10.1103/PhysRevLett.95.090503}}.

\bibitem{Simon:1999lfr}
R.~Simon.
\newblock {Peres-Horodecki Separability Criterion for Continuous Variable
  Systems}.
\newblock {\em Phys. Rev. Lett.}, 84:2726--2729, 2000.
\newblock \href {https://arxiv.org/abs/quant-ph/9909044}
  {\path{arXiv:quant-ph/9909044}}, \href
  {https://doi.org/10.1103/PhysRevLett.84.2726}
  {\path{doi:10.1103/PhysRevLett.84.2726}}.

\bibitem{Horodecki:1996nc}
Michal Horodecki, Pawel Horodecki, and Ryszard Horodecki.
\newblock {On the necessary and sufficient conditions for separability of mixed
  quantum states}.
\newblock {\em Phys. Lett. A}, 223:1, 1996.
\newblock \href {https://arxiv.org/abs/quant-ph/9605038}
  {\path{arXiv:quant-ph/9605038}}, \href
  {https://doi.org/10.1016/S0375-9601(96)00706-2}
  {\path{doi:10.1016/S0375-9601(96)00706-2}}.

\bibitem{Peres:1996dw}
Asher Peres.
\newblock {Separability criterion for density matrices}.
\newblock {\em Phys. Rev. Lett.}, 77:1413--1415, 1996.
\newblock \href {https://arxiv.org/abs/quant-ph/9604005}
  {\path{arXiv:quant-ph/9604005}}, \href
  {https://doi.org/10.1103/PhysRevLett.77.1413}
  {\path{doi:10.1103/PhysRevLett.77.1413}}.

\bibitem{serafini}
Alessio Serafini.
\newblock {\em {Quantum Continuous Variables}}.
\newblock CRC Press, 10 2017.
\newblock \href {https://doi.org/10.1201/9781315118727}
  {\path{doi:10.1201/9781315118727}}.

\bibitem{MohammadiMozaffar:2017nri}
M.~Reza Mohammadi~Mozaffar and Ali Mollabashi.
\newblock {Entanglement in Lifshitz-type Quantum Field Theories}.
\newblock {\em JHEP}, 07:120, 2017.
\newblock \href {https://arxiv.org/abs/1705.00483} {\path{arXiv:1705.00483}},
  \href {https://doi.org/10.1007/JHEP07(2017)120}
  {\path{doi:10.1007/JHEP07(2017)120}}.

\bibitem{He:2017wla}
Temple He, Javier~M. Magan, and Stefan Vandoren.
\newblock {Entanglement Entropy in Lifshitz Theories}.
\newblock {\em SciPost Phys.}, 3(5):034, 2017.
\newblock \href {https://arxiv.org/abs/1705.01147} {\path{arXiv:1705.01147}},
  \href {https://doi.org/10.21468/SciPostPhys.3.5.034}
  {\path{doi:10.21468/SciPostPhys.3.5.034}}.

\bibitem{Lifshitz1}
Evgeny Lifshitz.
\newblock On the theory of second-order phase transitions i.
\newblock {\em Zh. Eksp. Teor. Fiz.}, 11:255, 1941.

\bibitem{Lifshitz2}
Evgeny Lifshitz.
\newblock On the theory of second-order phase transitions ii.
\newblock {\em Zh. Eksp. Teor. Fiz.}, 11:269, 1941.

\bibitem{Hartnoll:2009sz}
Sean~A. Hartnoll.
\newblock {Lectures on holographic methods for condensed matter physics}.
\newblock {\em Class. Quant. Grav.}, 26:224002, 2009.
\newblock \href {https://arxiv.org/abs/0903.3246} {\path{arXiv:0903.3246}},
  \href {https://doi.org/10.1088/0264-9381/26/22/224002}
  {\path{doi:10.1088/0264-9381/26/22/224002}}.

\bibitem{Sachdev}
Subir Sachdev.
\newblock {\em {Quantum Phase Transitions}}.
\newblock Cambridge University Press, 4 2011.
\newblock \href {https://doi.org/10.1017/cbo9780511973765}
  {\path{doi:10.1017/cbo9780511973765}}.

\bibitem{Ardonne:2003wa}
Eddy Ardonne, Paul Fendley, and Eduardo Fradkin.
\newblock {Topological order and conformal quantum critical points}.
\newblock {\em Annals Phys.}, 310:493--551, 2004.
\newblock \href {https://arxiv.org/abs/cond-mat/0311466}
  {\path{arXiv:cond-mat/0311466}}, \href
  {https://doi.org/10.1016/j.aop.2004.01.004}
  {\path{doi:10.1016/j.aop.2004.01.004}}.

\bibitem{Fradkin:2006mb}
Eduardo Fradkin and Joel~E. Moore.
\newblock {Entanglement entropy of 2D conformal quantum critical points:
  hearing the shape of a quantum drum}.
\newblock {\em Phys. Rev. Lett.}, 97:050404, 2006.
\newblock \href {https://arxiv.org/abs/cond-mat/0605683}
  {\path{arXiv:cond-mat/0605683}}, \href
  {https://doi.org/10.1103/PhysRevLett.97.050404}
  {\path{doi:10.1103/PhysRevLett.97.050404}}.

\bibitem{Hsu:2008af}
Benjamin Hsu, Michael Mulligan, Eduardo Fradkin, and Eun-Ah Kim.
\newblock {Universal entanglement entropy in 2D conformal quantum critical
  points}.
\newblock {\em Phys. Rev. B}, 79:115421, 2009.
\newblock \href {https://arxiv.org/abs/0812.0203} {\path{arXiv:0812.0203}},
  \href {https://doi.org/10.1103/PhysRevB.79.115421}
  {\path{doi:10.1103/PhysRevB.79.115421}}.

\bibitem{Fradkin:2009dus}
Eduardo Fradkin.
\newblock {Scaling of Entanglement Entropy at 2D quantum Lifshitz fixed points
  and topological fluids}.
\newblock {\em J. Phys. A}, 42:504011, 2009.
\newblock \href {https://arxiv.org/abs/0906.1569} {\path{arXiv:0906.1569}},
  \href {https://doi.org/10.1088/1751-8113/42/50/504011}
  {\path{doi:10.1088/1751-8113/42/50/504011}}.

\bibitem{Solodukhin:2009sk}
Sergey~N. Solodukhin.
\newblock {Entanglement Entropy in Non-Relativistic Field Theories}.
\newblock {\em JHEP}, 04:101, 2010.
\newblock \href {https://arxiv.org/abs/0909.0277} {\path{arXiv:0909.0277}},
  \href {https://doi.org/10.1007/JHEP04(2010)101}
  {\path{doi:10.1007/JHEP04(2010)101}}.

\bibitem{Oshikawa:2010kv}
Masaki Oshikawa.
\newblock {Boundary Conformal Field Theory and Entanglement Entropy in
  Two-Dimensional Quantum Lifshitz Critical Point}.
\newblock 7 2010.
\newblock \href {https://arxiv.org/abs/1007.3739} {\path{arXiv:1007.3739}}.

\bibitem{Zaletel:2011ir}
Michael~P. Zaletel, Jens~H. Bardarson, and Joel~E. Moore.
\newblock {Logarithmic terms in entanglement entropies of 2D quantum critical
  points and Shannon entropies of spin chains}.
\newblock {\em Phys. Rev. Lett.}, 107:020402, 2011.
\newblock \href {https://arxiv.org/abs/1103.5452} {\path{arXiv:1103.5452}},
  \href {https://doi.org/10.1103/PhysRevLett.107.020402}
  {\path{doi:10.1103/PhysRevLett.107.020402}}.

\bibitem{Rajabpour:2014osa}
M.~A. Rajabpour and S.~Sotiriadis.
\newblock {Quantum quench in long-range field theories}.
\newblock {\em Phys. Rev. B}, 91(4):045131, 2015.
\newblock \href {https://arxiv.org/abs/1409.6558} {\path{arXiv:1409.6558}},
  \href {https://doi.org/10.1103/PhysRevB.91.045131}
  {\path{doi:10.1103/PhysRevB.91.045131}}.

\bibitem{Chen:2016kjp}
Xiao Chen, William Witczak-Krempa, Thomas Faulkner, and Eduardo Fradkin.
\newblock {Two-cylinder entanglement entropy under a twist}.
\newblock {\em J. Stat. Mech.}, 1704(4):043104, 2017.
\newblock \href {https://arxiv.org/abs/1611.01847} {\path{arXiv:1611.01847}},
  \href {https://doi.org/10.1088/1742-5468/aa668a}
  {\path{doi:10.1088/1742-5468/aa668a}}.

\bibitem{Chen:2017txi}
Xiao Chen, Eduardo Fradkin, and William Witczak-Krempa.
\newblock {Quantum spin chains with multiple dynamics}.
\newblock {\em Phys. Rev. B}, 96(18):180402, 2017.
\newblock \href {https://arxiv.org/abs/1706.02304} {\path{arXiv:1706.02304}},
  \href {https://doi.org/10.1103/PhysRevB.96.180402}
  {\path{doi:10.1103/PhysRevB.96.180402}}.

\bibitem{MohammadiMozaffar:2018vmk}
M.~Reza Mohammadi~Mozaffar and Ali Mollabashi.
\newblock {Entanglement Evolution in Lifshitz-type Scalar Theories}.
\newblock {\em JHEP}, 01:137, 2019.
\newblock \href {https://arxiv.org/abs/1811.11470} {\path{arXiv:1811.11470}},
  \href {https://doi.org/10.1007/JHEP01(2019)137}
  {\path{doi:10.1007/JHEP01(2019)137}}.

\bibitem{Angel-Ramelli:2019nji}
J.~Angel-Ramelli, V.~Giangreco~M. Puletti, and L.~Thorlacius.
\newblock {Entanglement Entropy in Generalised Quantum Lifshitz Models}.
\newblock {\em JHEP}, 08:072, 2019.
\newblock \href {https://arxiv.org/abs/1906.08252} {\path{arXiv:1906.08252}},
  \href {https://doi.org/10.1007/JHEP08(2019)072}
  {\path{doi:10.1007/JHEP08(2019)072}}.

\bibitem{Angel-Ramelli:2020xvd}
J.~Angel-Ramelli.
\newblock {Entanglement Entropy of Excited States in the Quantum Lifshitz
  Model}.
\newblock {\em J. Stat. Mech.}, 2101:013102, 2021.
\newblock \href {https://arxiv.org/abs/2009.02283} {\path{arXiv:2009.02283}},
  \href {https://doi.org/10.1088/1742-5468/abcd35}
  {\path{doi:10.1088/1742-5468/abcd35}}.

\bibitem{Hartmann:2021vrt}
Dion Hartmann, Kevin Kavanagh, and Stefan Vandoren.
\newblock {Entanglement entropy with Lifshitz fermions}.
\newblock {\em SciPost Phys.}, 11(2):031, 2021.
\newblock \href {https://arxiv.org/abs/2104.10913} {\path{arXiv:2104.10913}},
  \href {https://doi.org/10.21468/SciPostPhys.11.2.031}
  {\path{doi:10.21468/SciPostPhys.11.2.031}}.

\bibitem{Mozaffar:2021nex}
M.~Reza~Mohammadi Mozaffar and Ali Mollabashi.
\newblock {Time scaling of entanglement in integrable scale-invariant
  theories}.
\newblock {\em Phys. Rev. Res.}, 4(2):L022010, 2022.
\newblock \href {https://arxiv.org/abs/2106.14700} {\path{arXiv:2106.14700}},
  \href {https://doi.org/10.1103/PhysRevResearch.4.L022010}
  {\path{doi:10.1103/PhysRevResearch.4.L022010}}.

\bibitem{Berthiere:2023bwn}
Cl{\'e}ment Berthiere, Bin Chen, and Hongjie Chen.
\newblock {Reflected entropy and Markov gap in Lifshitz theories}.
\newblock {\em JHEP}, 09:160, 2023.
\newblock \href {https://arxiv.org/abs/2307.12247} {\path{arXiv:2307.12247}},
  \href {https://doi.org/10.1007/JHEP09(2023)160}
  {\path{doi:10.1007/JHEP09(2023)160}}.

\bibitem{Vasli:2024mrf}
Mohammad~Javad Vasli, Komeil Babaei~Velni, M.~Reza Mohammadi~Mozaffar, and Ali
  Mollabashi.
\newblock {Entanglement in Lifshitz fermion theories}.
\newblock {\em JHEP}, 09:122, 2024.
\newblock \href {https://arxiv.org/abs/2405.18097} {\path{arXiv:2405.18097}},
  \href {https://doi.org/10.1007/JHEP09(2024)122}
  {\path{doi:10.1007/JHEP09(2024)122}}.

\bibitem{MohammadiMozaffar:2026osc}
M.~Reza Mohammadi~Mozaffar and Ali Mollabashi.
\newblock {Symmetry resolved entanglement in Lifshitz field theories}.
\newblock {\em JHEP}, 07:234, 2026.
\newblock \href {https://arxiv.org/abs/2604.19082} {\path{arXiv:2604.19082}},
  \href {https://doi.org/10.1007/JHEP07(2026)234}
  {\path{doi:10.1007/JHEP07(2026)234}}.

\bibitem{PhysRevLett.87.167904}
G.~Giedke, B.~Kraus, M.~Lewenstein, and J.~I. Cirac.
\newblock Entanglement criteria for all bipartite gaussian states.
\newblock {\em Phys. Rev. Lett.}, 87:167904, Oct 2001.
\newblock URL: \url{https://link.aps.org/doi/10.1103/PhysRevLett.87.167904},
  \href {https://doi.org/10.1103/PhysRevLett.87.167904}
  {\path{doi:10.1103/PhysRevLett.87.167904}}.

\bibitem{Klco:2019yrb}
Natalie Klco and Martin~J. Savage.
\newblock {Systematically Localizable Operators for Quantum Simulations of
  Quantum Field Theories}.
\newblock {\em Phys. Rev. A}, 102(1):012619, 2020.
\newblock \href {https://arxiv.org/abs/1912.03577} {\path{arXiv:1912.03577}},
  \href {https://doi.org/10.1103/PhysRevA.102.012619}
  {\path{doi:10.1103/PhysRevA.102.012619}}.

\bibitem{Klco:2020rga}
Natalie Klco and Martin~J. Savage.
\newblock {Geometric quantum information structure in quantum fields and their
  lattice simulation}.
\newblock {\em Phys. Rev. D}, 103(6):065007, 2021.
\newblock \href {https://arxiv.org/abs/2008.03647} {\path{arXiv:2008.03647}},
  \href {https://doi.org/10.1103/PhysRevD.103.065007}
  {\path{doi:10.1103/PhysRevD.103.065007}}.

\bibitem{Yazdi:2016cxn}
Yasaman~K. Yazdi.
\newblock {Zero Modes and Entanglement Entropy}.
\newblock {\em JHEP}, 04:140, 2017.
\newblock \href {https://arxiv.org/abs/1608.04744} {\path{arXiv:1608.04744}},
  \href {https://doi.org/10.1007/JHEP04(2017)140}
  {\path{doi:10.1007/JHEP04(2017)140}}.

\bibitem{Casini:2005zv}
H.~Casini and M.~Huerta.
\newblock {Entanglement and alpha entropies for a massive scalar field in two
  dimensions}.
\newblock {\em J. Stat. Mech.}, 0512:P12012, 2005.
\newblock \href {https://arxiv.org/abs/cond-mat/0511014}
  {\path{arXiv:cond-mat/0511014}}, \href
  {https://doi.org/10.1088/1742-5468/2005/12/P12012}
  {\path{doi:10.1088/1742-5468/2005/12/P12012}}.

\bibitem{Pye:2015tta}
Jason Pye, William Donnelly, and Achim Kempf.
\newblock {Locality and entanglement in bandlimited quantum field theory}.
\newblock {\em Phys. Rev. D}, 92(10):105022, 2015.
\newblock \href {https://arxiv.org/abs/1508.05953} {\path{arXiv:1508.05953}},
  \href {https://doi.org/10.1103/PhysRevD.92.105022}
  {\path{doi:10.1103/PhysRevD.92.105022}}.

\bibitem{Jain:2021ppx}
Parul Jain, S.~Mahesh Chandran, and S.~Shankaranarayanan.
\newblock {Log to log-log crossover of entanglement in $(1+1)-$ dimensional
  massive scalar field}.
\newblock {\em Phys. Rev. D}, 103:125008, 2021.
\newblock \href {https://arxiv.org/abs/2103.01772} {\path{arXiv:2103.01772}},
  \href {https://doi.org/10.1103/PhysRevD.103.125008}
  {\path{doi:10.1103/PhysRevD.103.125008}}.

\bibitem{Blondeau-Fournier:2015yoa}
Olivier Blondeau-Fournier, Olalla~A. Castro-Alvaredo, and Benjamin Doyon.
\newblock {Universal scaling of the logarithmic negativity in massive quantum
  field theory}.
\newblock {\em J. Phys. A}, 49(12):125401, 2016.
\newblock \href {https://arxiv.org/abs/1508.04026} {\path{arXiv:1508.04026}},
  \href {https://doi.org/10.1088/1751-8113/49/12/125401}
  {\path{doi:10.1088/1751-8113/49/12/125401}}.

\bibitem{Marshall:2015mna}
Kevin Marshall, Raphael Pooser, George Siopsis, and Christian Weedbrook.
\newblock {Quantum simulation of quantum field theory using continuous
  variables}.
\newblock {\em Phys. Rev. A}, 92(6):063825, 2015.
\newblock \href {https://arxiv.org/abs/1503.08121} {\path{arXiv:1503.08121}},
  \href {https://doi.org/10.1103/PhysRevA.92.063825}
  {\path{doi:10.1103/PhysRevA.92.063825}}.

\bibitem{Bauer:2021sgi}
Christian~W. Bauer, Plato Deliyannis, Marat Freytsis, and Benjamin Nachman.
\newblock {Practical considerations for the preparation of multivariate
  Gaussian states on quantum computers}.
\newblock 9 2021.
\newblock \href {https://arxiv.org/abs/2109.10918} {\path{arXiv:2109.10918}}.

\bibitem{Reznik:2002fz}
Benni Reznik.
\newblock {Entanglement from the vacuum}.
\newblock {\em Found. Phys.}, 33:167--176, 2003.
\newblock \href {https://arxiv.org/abs/quant-ph/0212044}
  {\path{arXiv:quant-ph/0212044}}, \href
  {https://doi.org/10.1023/A:1022875910744}
  {\path{doi:10.1023/A:1022875910744}}.

\bibitem{Reznik:2003mnx}
Benni Reznik, Alex Retzker, and Jonathan Silman.
\newblock {Violating Bell's inequalities in the vacuum}.
\newblock {\em Phys. Rev. A}, 71:042104, 2005.
\newblock \href {https://arxiv.org/abs/quant-ph/0310058}
  {\path{arXiv:quant-ph/0310058}}, \href
  {https://doi.org/10.1103/PhysRevA.71.042104}
  {\path{doi:10.1103/PhysRevA.71.042104}}.

\bibitem{Klco:2023ojt}
Natalie Klco and D.~H. Beck.
\newblock {Entanglement structures in quantum field theories. II. Distortions
  of vacuum correlations through the lens of local observers}.
\newblock {\em Phys. Rev. A}, 108(1):012429, 2023.
\newblock \href {https://arxiv.org/abs/2304.04143} {\path{arXiv:2304.04143}},
  \href {https://doi.org/10.1103/PhysRevA.108.012429}
  {\path{doi:10.1103/PhysRevA.108.012429}}.

\bibitem{Giedke:2000ruw}
Geza Giedke, Lu-Ming Duan, J.~Ignacio Cirac, and Peter Zoller.
\newblock {All inseparable two-mode Gaussian continuous variable states are
  distillable}.
\newblock 8 2000.
\newblock \href {https://arxiv.org/abs/quant-ph/0007061}
  {\path{arXiv:quant-ph/0007061}}.

\bibitem{Giedke:2003wns}
G.~Giedke, M.~M. Wolf, O.~Krueger, R.~F. Werner, and J.~I. Cirac.
\newblock {Entanglement of formation for symmetric Gaussian states}.
\newblock {\em Phys. Rev. Lett.}, 91:107901, 2003.
\newblock \href {https://arxiv.org/abs/quant-ph/0304042}
  {\path{arXiv:quant-ph/0304042}}, \href
  {https://doi.org/10.1103/PhysRevLett.91.107901}
  {\path{doi:10.1103/PhysRevLett.91.107901}}.

\bibitem{cohen}
Oliver Cohen.
\newblock Unlocking hidden entanglement with classical information.
\newblock {\em Phys. Rev. Lett.}, 80:2493--2496, Mar 1998.
\newblock URL: \url{https://link.aps.org/doi/10.1103/PhysRevLett.80.2493},
  \href {https://doi.org/10.1103/PhysRevLett.80.2493}
  {\path{doi:10.1103/PhysRevLett.80.2493}}.

\bibitem{Pye:2026siw}
Jason Pye, Atharva Hingane, and Robert~H. Jonsson.
\newblock {The negativity core of a 1+1D massless scalar quantum field}.
\newblock 5 2026.
\newblock \href {https://arxiv.org/abs/2605.23824} {\path{arXiv:2605.23824}}.

\bibitem{Pye:2026ehc}
Jason Pye, Atharva Hingane, and Robert~H. Jonsson.
\newblock {Exact calculation of entanglement negativity for a 1+1D massless
  scalar field using phase space methods}.
\newblock 6 2026.
\newblock \href {https://arxiv.org/abs/2606.30231} {\path{arXiv:2606.30231}}.

\bibitem{Mollabashi:2024gik}
Ali Mollabashi and Saleh Rahimi-Keshari.
\newblock {Information scrambling in bosonic Gaussian dynamics}.
\newblock {\em Phys. Rev. E}, 112(3):034213, 2025.
\newblock \href {https://arxiv.org/abs/2408.12089} {\path{arXiv:2408.12089}},
  \href {https://doi.org/10.1103/wvpl-jdgk} {\path{doi:10.1103/wvpl-jdgk}}.

\bibitem{Botero:2004twx}
Alonso Botero and Benni Reznik.
\newblock {BCS-like Modewise Entanglement of Fermion Gaussian States}.
\newblock {\em Phys. Lett. A}, 331:39--44, 2004.
\newblock \href {https://arxiv.org/abs/quant-ph/0404176}
  {\path{arXiv:quant-ph/0404176}}, \href
  {https://doi.org/10.1016/j.physleta.2004.08.037}
  {\path{doi:10.1016/j.physleta.2004.08.037}}.

\bibitem{Eisler:2015tgq}
Viktor Eisler and Zolt{\'a}n Zimbor{\'a}s.
\newblock {On the partial transpose of fermionic Gaussian states}.
\newblock {\em New J. Phys.}, 17(5):053048, 2015.
\newblock \href {https://arxiv.org/abs/1502.01369} {\path{arXiv:1502.01369}},
  \href {https://doi.org/10.1088/1367-2630/17/5/053048}
  {\path{doi:10.1088/1367-2630/17/5/053048}}.

\bibitem{Shapourian:2016cqu}
Hassan Shapourian, Ken Shiozaki, and Shinsei Ryu.
\newblock {Partial time-reversal transformation and entanglement negativity in
  fermionic systems}.
\newblock {\em Phys. Rev. B}, 95:165101, 2017.
\newblock \href {https://arxiv.org/abs/1611.07536} {\path{arXiv:1611.07536}},
  \href {https://doi.org/10.1103/PhysRevB.95.165101}
  {\path{doi:10.1103/PhysRevB.95.165101}}.

\bibitem{Shapourian:2018ozl}
Hassan Shapourian and Shinsei Ryu.
\newblock {Entanglement negativity of fermions: monotonicity, separability
  criterion, and classification of few-mode states}.
\newblock {\em Phys. Rev. A}, 99(2):022310, 2019.
\newblock \href {https://arxiv.org/abs/1804.08637} {\path{arXiv:1804.08637}},
  \href {https://doi.org/10.1103/PhysRevA.99.022310}
  {\path{doi:10.1103/PhysRevA.99.022310}}.

\bibitem{Herzog:2016ohd}
Christopher~P. Herzog and Yihong Wang.
\newblock {Estimation for Entanglement Negativity of Free Fermions}.
\newblock {\em J. Stat. Mech.}, 1607(7):073102, 2016.
\newblock \href {https://arxiv.org/abs/1601.00678} {\path{arXiv:1601.00678}},
  \href {https://doi.org/10.1088/1742-5468/2016/07/073102}
  {\path{doi:10.1088/1742-5468/2016/07/073102}}.

\bibitem{Coser:2015eba}
Andrea Coser, Erik Tonni, and Pasquale Calabrese.
\newblock {Towards the entanglement negativity of two disjoint intervals for a
  one dimensional free fermion}.
\newblock {\em J. Stat. Mech.}, 1603(3):033116, 2016.
\newblock \href {https://arxiv.org/abs/1508.00811} {\path{arXiv:1508.00811}},
  \href {https://doi.org/10.1088/1742-5468/2016/03/033116}
  {\path{doi:10.1088/1742-5468/2016/03/033116}}.

\end{thebibliography}

\newpage

\appendix

\section{Uniform boundedness of logarithmic negativity
in the massless limit}
\label{app:LNbound}

We establish a mass-independent upper bound on the
logarithmic negativity between fixed finite regions
$A$ and $B$ of the infinite Lifshitz harmonic chain.
The argument applies to any fixed $z>0$, with the
lattice spacing held fixed. We use lattice units and
the dispersion
\begin{equation}
\omega_m(k)
=
\sqrt{m^{2z}+\left(2\left|\sin\frac{k}{2}\right|\right)^{2z}},
\end{equation}
where the absolute value is included to accommodate non-integer $z$ as well. Let $R=A\cup B$ contain $n_R$ sites, with integer coordinates $x_a$. For $m>0$, the restricted vacuum
covariance matrix is
$\sigma_R(m)=G_R(m)\oplus H_R(m)$, where
\begin{align}
[G_R(m)]_{ab}
&=
\frac12\int_{-\pi}^{\pi}\frac{dk}{2\pi}\,
\frac{e^{ik(x_a-x_b)}}{\omega_m(k)},
\\
[H_R(m)]_{ab}
&=
\frac12\int_{-\pi}^{\pi}\frac{dk}{2\pi}\,
\omega_m(k)e^{ik(x_a-x_b)}.
\end{align}

\paragraph{Uniform lower bound on the field covariance.}
Fix $m_0>0$ and restrict the mass to $0<m\leq m_0$.
Throughout the Brillouin zone,
\begin{equation}
\omega_m(k)\leq
\omega_*:=\sqrt{m_0^{2z}+2^{2z}}.
\end{equation}
For any vector $f\in\mathbb C^{n_R}$, define
\begin{equation}
F_f(k)=\sum_{a=1}^{n_R}f_a e^{-ikx_a}.
\end{equation}
Since the sites are distinct, Fourier orthogonality
gives
\begin{equation}
\int_{-\pi}^{\pi}\frac{dk}{2\pi}\,
|F_f(k)|^2=\|f\|^2.
\end{equation}
Consequently,
\begin{align}
f^\dagger G_R(m)f
&=
\frac12\int_{-\pi}^{\pi}\frac{dk}{2\pi}\,
\frac{|F_f(k)|^2}{\omega_m(k)}
\geq \frac{1}{2\omega_*}\|f\|^2.
\end{align}
Thus,
\begin{equation}
G_R(m)\geq g_*\,\mathbb I,
\qquad
g_*=\frac{1}{2\omega_*}>0,
\label{eq:app_G_lower_bound}
\end{equation}
uniformly in $m$. The possible divergence of some
field-covariance eigenvalues as $m\to0^+$ does not
affect this lower bound.

\paragraph{Uniform lower bound on the momentum covariance.}
The massless dispersion
\begin{equation}
\omega_0(k)=
\left(2\left|\sin\frac{k}{2}\right|\right)^z
\end{equation}
is bounded and positive almost everywhere in the
Brillouin zone. Therefore $H_R(0)$ is finite.
Moreover, for every nonzero $f$,
\begin{equation}
f^\dagger H_R(0)f
=
\frac12\int_{-\pi}^{\pi}\frac{dk}{2\pi}\,
\omega_0(k)|F_f(k)|^2>0.
\end{equation}
Indeed, if this integral vanished, positivity of
$\omega_0(k)$ almost everywhere would imply
$F_f(k)=0$ almost everywhere. Fourier orthogonality
would then imply $\|f\|^2=0$, a contradiction.

Hence $H_R(0)$ is strictly positive definite.
Because $R$ is finite, its smallest eigenvalue obeys
\begin{equation}
h_*:=h_{\min}\bigl(H_R(0)\bigr)>0.
\end{equation}
Finally, $\omega_m(k)\geq\omega_0(k)$ implies,
as a quadratic-form inequality,
\begin{equation}
H_R(m)\geq H_R(0)\geq h_*\,\mathbb I.
\label{eq:app_H_lower_bound}
\end{equation}
This bound is also independent of $m$.

\paragraph{Lower bound on the PT symplectic spectrum.}
Let $\Gamma$ be diagonal, with entries $+1$ on $A$
and $-1$ on $B$. Partial transposition reverses
the momenta in $B$, giving
\begin{equation}
\sigma^{\Gamma}_R
=
G_R\oplus\bigl(\Gamma H_R\Gamma\bigr).
\end{equation}
Since $\Gamma^T=\Gamma$ and $\Gamma^2=\mathbb I$,
Eq.~\eqref{eq:app_H_lower_bound} yields
\begin{equation}
\Gamma H_R(m)\Gamma\geq h_*\,\mathbb I.
\end{equation}

The squared PT symplectic eigenvalues are the
eigenvalues of $G_R\Gamma H_R\Gamma$.
For every $m>0$, this matrix is similar to the
positive definite symmetric matrix
\begin{equation}
M_R(m)
=
G_R^{1/2}\Gamma H_R\Gamma G_R^{1/2}.
\end{equation}
Taking a congruence with $G_R^{1/2}$ and using
Eq.~\eqref{eq:app_G_lower_bound}, we obtain
\begin{equation}
M_R(m)\geq h_*G_R(m)
\geq g_*h_*\,\mathbb I.
\end{equation}
It follows that every PT symplectic eigenvalue satisfies
\begin{equation}
\nu_j(m)\geq \sqrt{g_*h_*}>0,
\qquad 0<m\leq m_0.
\end{equation}

\paragraph{Bound on logarithmic negativity and its scope.}
In the covariance convention used here,
\begin{equation}
\mathcal E(m)
=
\sum_{j=1}^{n_R}
\max\!\left\{0,-\log\!\bigl(2\nu_j(m)\bigr)\right\}.
\end{equation}
The uniform spectral lower bound therefore gives
\begin{equation}
\boxed{
\mathcal E(m)
\leq
n_R\max\!\left\{0,-\log(2\sqrt{g_*h_*})\right\}<\infty,
\qquad 0<m\leq m_0.
}
\end{equation}
In particular,
\begin{equation}
\limsup_{m\to0^+}\mathcal E(m)\leq n_R\max\!\left\{0,-\log(2\sqrt{g_*h_*})\right\}.
\end{equation}

The constants may depend on $z$, $m_0$, the retained
sites, and the fixed lattice spacing, but not on $m$.
The bound excludes an infrared divergence of
logarithmic negativity for fixed finite regions,
whether adjacent or separated.
It does not by itself establish convergence to a
unique limiting value or determine the leading mass
corrections. Nor is it asserted to remain uniform
when the regions grow or the lattice spacing vanishes.

The proof requires the field covariance only at
positive mass. For $z\geq1$, $G_R(0)$ has divergent
entries, even though $H_R(0)$ is finite and strictly
positive. Thus the result concerns the limit
$m\to0^+$ and does not assume a finite covariance
matrix at exactly zero mass.
For $0<z<1$, the field-covariance integral is also
infrared convergent at $m=0$, and the same inequalities
apply directly there.

This proof concerns fixed finite lattice regions and
does not establish a bound uniform in the continuum
limit. Direct use of pointwise continuum correlators
requires ultraviolet regularization, whereas an
analogous argument applies to a fixed finite collection
of suitably smeared canonical modes. Boundedness for
the complete continuum regions requires additional
control of the ultraviolet limit and of the sum over
PT symplectic eigenvalues.

\newpage

\section{Momentum-space structure of the dominant partially transposed mode}\label{appendix:SKWavefunction}

This appendix reproduces the PT-mode analysis of Ref.~\cite{Klco:2021biu} and examines its extension
to the dispersions studied here.
It supplies the canonical-wavefunction comparison used in the main text to distinguish ultraviolet
mode structure from infrared-dominated spatial participation.

As discussed in the main text, the canonical components
of weakly contributing entanglement modes can exhibit
pronounced short-wavelength oscillations.
This momentum-space structure should be distinguished
from their spatial participation, which can remain
smooth and collective.
Here we reproduce the PT-eigenvector analysis of
Ref.~\cite{Klco:2021biu} for the Lifshitz harmonic
chains considered in this work.
That study showed that, as two regions are moved
apart in the relativistic harmonic chain, fewer
PT symplectic eigenvalues contribute to logarithmic
negativity, until a single eigenvalue accounts for
the surviving negativity near the negativity sphere.
The associated eigenvector, corresponding to the
smallest eigenvalue of $GH^\Gamma$, develops
increasingly ultraviolet-dominated momentum-space
support.
Following the same procedure, we examine its Fourier
profile at different separations to illustrate this
behavior in the Lifshitz generalizations.

\begin{figure}[h!]
\begin{center}
\includegraphics[scale=0.3]{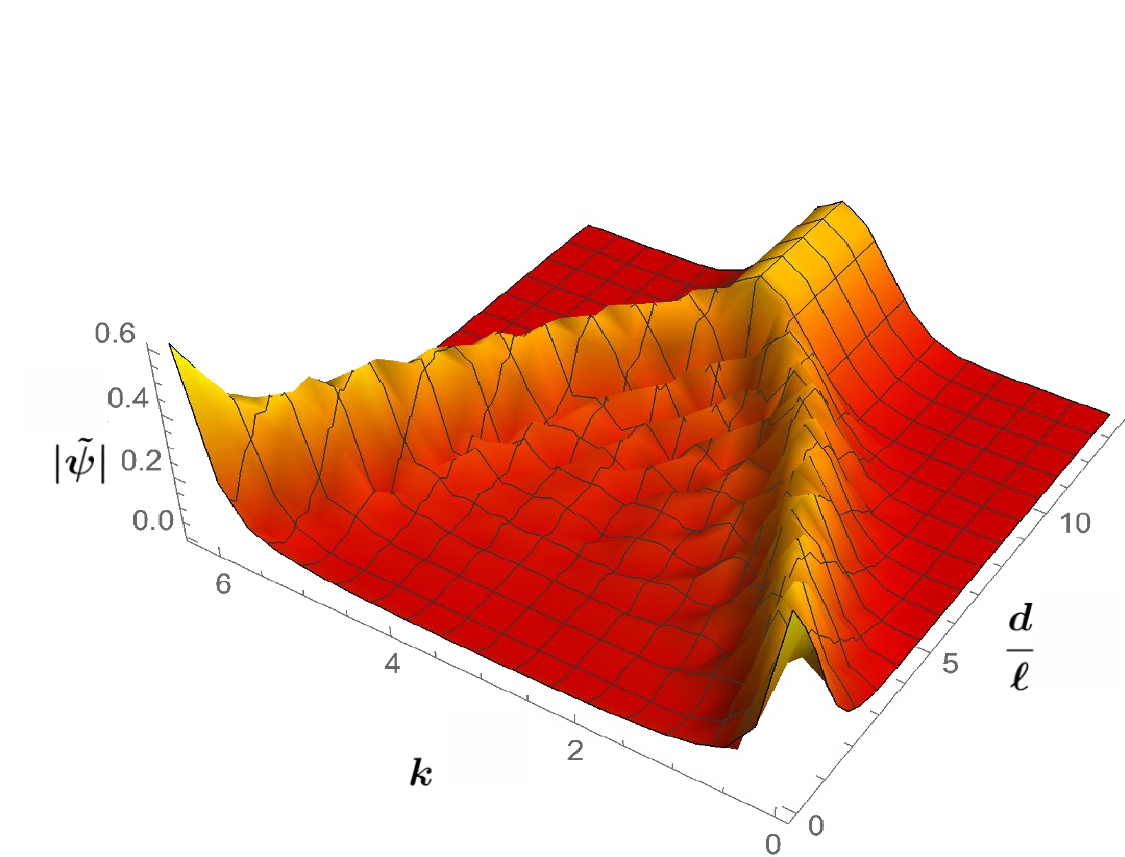}
\hspace{4mm}
\includegraphics[scale=0.3]{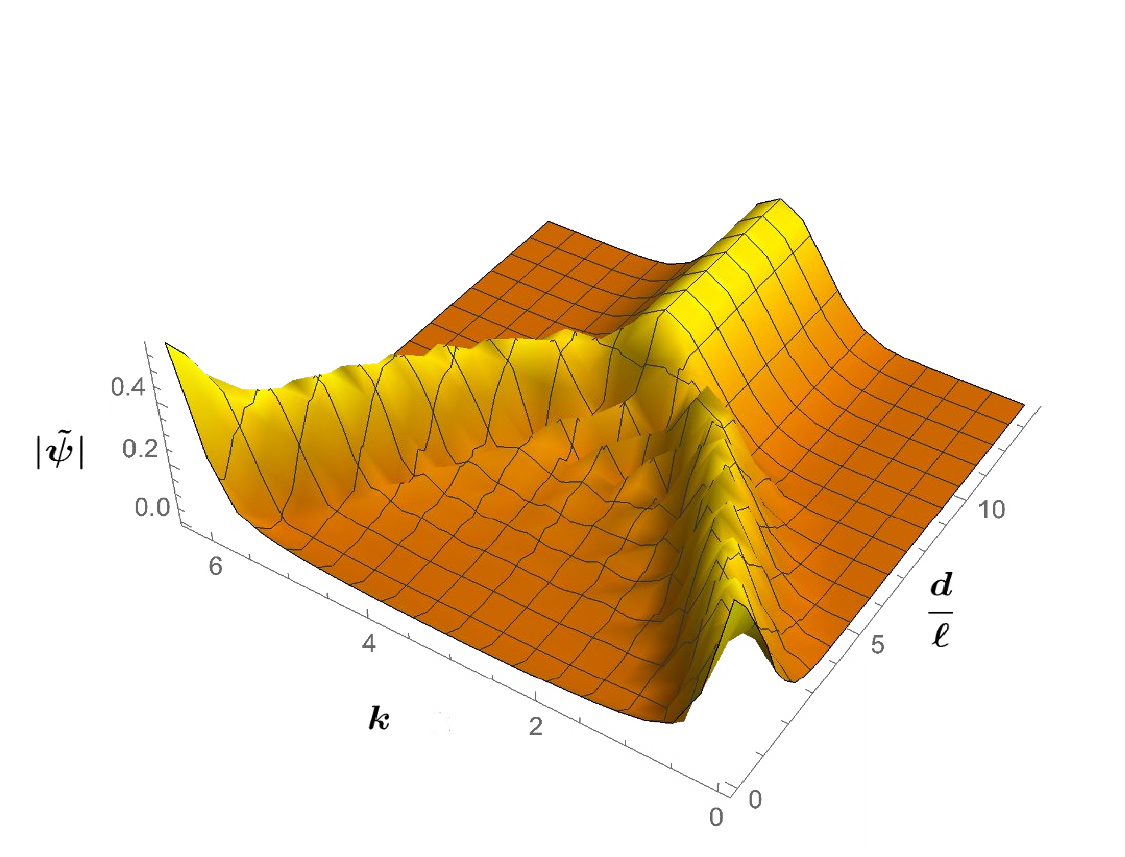}
\hspace{4mm}
\includegraphics[scale=0.3]{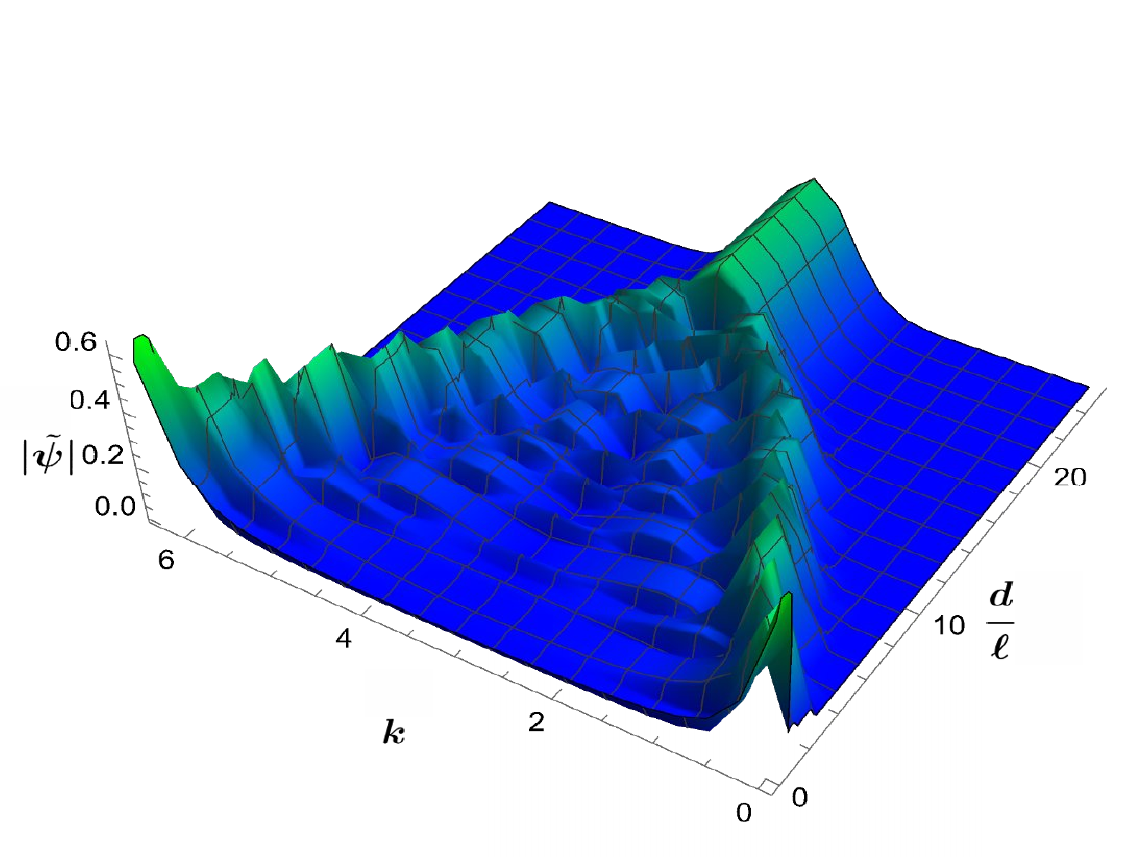}
\hspace{4mm}
\includegraphics[scale=0.3]{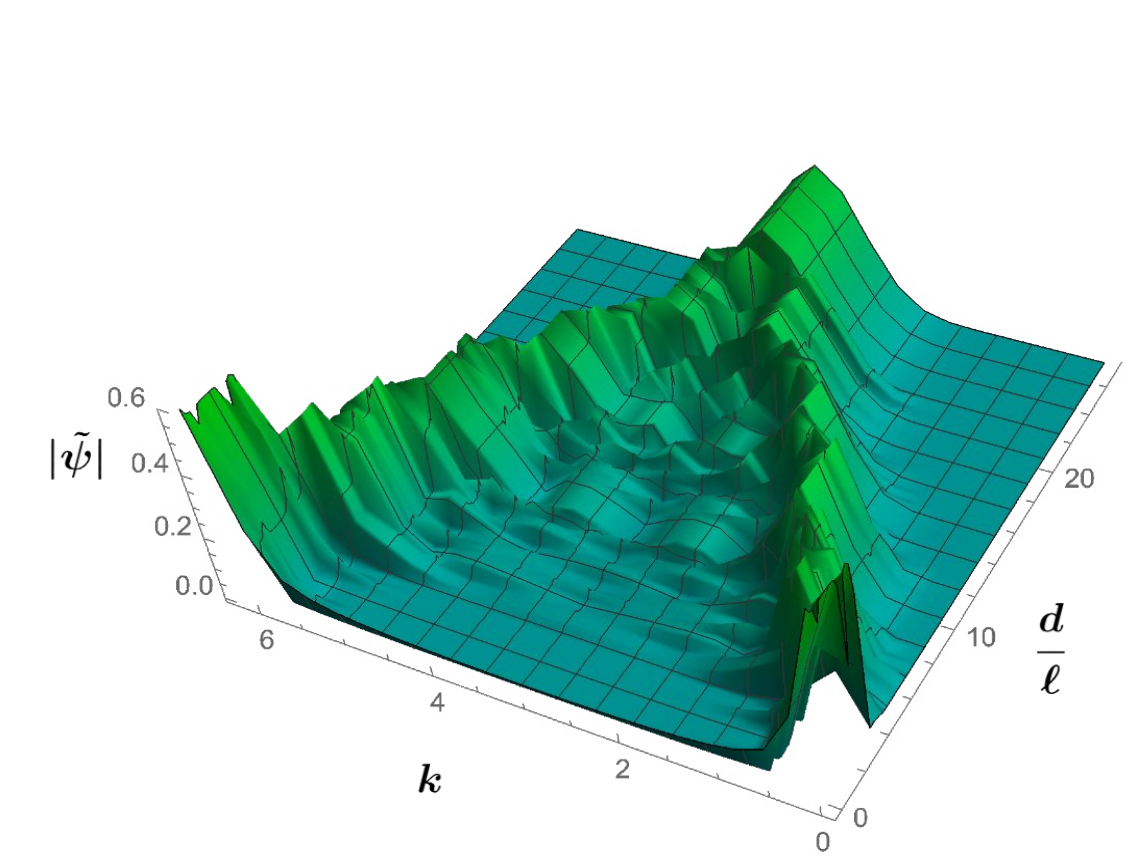}
\end{center}
\caption{The eigenvector associated with the smallest eigenvalue of $GH^{\Gamma}$ in momentum space for $\ell=22$, $m=10^{-6}$, $z=3$ (left) and $z=5$ (right) in the top panel. 
For $\ell=22$, $m=10^{-1}$, $z=2$ (left) and $z=4$ (right) in the bottom panel.
The negativity radius is $\frac{d}{\ell}=10.2727$ (left) and $\frac{d}{\ell}=8.5909$ (right) in the top panel and $\frac{d}{\ell}=18.2727$ (left) and
$\frac{d}{\ell}=22$ (right) in the bottom panel.
Beyond the negativity sphere, the momentum-space wavefunction, now corresponding to a separable state, 
no longer evolves further and effectively saturates the ultraviolet truncation.}
\label{figg1}
\end{figure}

In this appendix, we show this behavior does appear in the LHC, as presented in Figure \ref{figg1}. 
In this Figure, we display the one-dimensional wavefunction of region A, $\tilde{\psi}=\frac{1}{\sqrt{\ell}}\displaystyle\sum_{x}\psi(x)e^{ikx}$, in momentum space, plotted within the Brillouin zone, $k\in \frac{2\pi}{\ell}\mathbb{Z}_{\ell}$. By analyzing the symplectic eigenvalues of $\sqrt{GH^{\Gamma}}$ that contribute to the negativity, we observe that for long separations, the entanglement is effectively characterized by a single eigenvector associated with the smallest eigenvalue. By plotting this dominant eigenvector in phase space in Figure \ref{figg1}, we see that for small separations the wavefunction support in momentum space is localized around low momenta (near zero and the Brillouin zone edges at $2\pi$). As the separation increases, the momentum content within the region shifts towards higher-momentum modes with $k=\pi$. 

\section{Core and halo negativities in the preparation decomposition}\label{Appendix:halo}

\begin{table}[H]
\centering

\begin{minipage}[t]{0.32\textwidth}
\centering
\resizebox{0.75\textwidth}{!}{%
\begin{tabular}{|p{0.8cm}|p{2.0cm}|p{2.0cm}|}
\hline
\multicolumn{3}{|c|}{\textbf{$z=3$}} \\
\hline
$d/\ell$ & $LN(\sigma_{c})_{A|B}$ & $LN(\sigma_{h})_{A|B}$ \\
\hline\hline
0.016 & \num{7.573e-1} & 13.224 \\
0.031 & \num{3.484e-1} & 13.474 \\
0.047 & \num{2.269e-1} &  13.512\\
0.062 & \num{1.548e-1} & 13.535 \\
0.078 & \num{1.101e-1} & 13.559 \\
0.094 & \num{8.077e-2} &13.583  \\
0.109 & \num{6.113e-2} & 13.607 \\
0.125 & \num{4.783e-2} &  13.631\\
0.141 & \num{3.860e-2} & 13.654 \\
0.156 & \num{3.193e-2} &  13.676\\
0.172 & \num{2.693e-2} & 13.696 \\
0.188 & \num{2.304e-2} &  13.715\\
0.203 & \num{1.994e-2} &  13.732\\
0.219 & \num{1.740e-2} &  13.749\\
0.234 & \num{1.528e-2} & 13.764 \\
0.250 & \num{1.349e-2} & 13.778 \\
0.266 & \num{1.196e-2} & 13.791 \\
0.281 & \num{1.064e-2} &  13.804\\
0.297 & \num{9.481e-3} & 13.815 \\
0.312 & \num{8.468e-3} & 13.827 \\
0.328 & \num{7.574e-3} & 5.672 \\
0.344 & \num{6.780e-3} &  1.186\\
0.359 & \num{6.071e-3} & \num{4.593e-1} \\
0.375 & \num{5.436e-3} &  \num{9.276e-2}\\
0.391 & \num{4.864e-3} & \num{5.406e-3}  \\
0.406 & \num{4.349e-3} &  \num{6.879e-3} \\
0.422 & \num{3.883e-3} & \num{1.520e-2}  \\
0.438 & \num{3.462e-3} & \num{1.403e-3} \\
0.453 & \num{3.081e-3} &  \num{1.387e-3}\\
0.469 & \num{2.737e-3} &   \num{1.478e-3} \\
0.484 & \num{2.426e-3} &  \num{2.537e-3}\\
0.500 & \num{2.146e-3} &  \num{9.357e-4} \\
0.516 & \num{1.894e-3} &  \num{8.941e-4}\\          
0.531 & \num{1.669e-3} &  \num{8.483e-4}\\
0.547 & \num{1.468e-3} &  \num{8.410e-4}\\
0.562 & \num{1.290e-3} &  \num{1.124e-3}\\
0.578 & \num{1.133e-3} &  \num{8.500e-4}\\
0.594 & \num{9.952e-4} &  \num{8.720e-4}\\
0.609 & \num{8.751e-4} & \num{9.019e-4} \\
0.625 & \num{7.708e-4} &  \num{9.330e-4}\\
0.641 & \num{6.804e-4} & \num{9.647e-4} \\
0.656 & \num{6.022e-4} & \num{9.906e-4} \\
0.672 & \num{5.346e-4} & \num{1.013e-3} \\
0.688 & \num{4.762e-4} &  \num{1.029e-3}\\
0.703 & \num{4.254e-4} &  \num{1.039e-3}\\
0.719 & \num{3.813e-4} & \num{1.040e-3} \\
0.734 & \num{3.428e-4} &  \num{1.028e-3}\\
0.750 & \num{3.089e-4} &  \num{9.997e-4}\\
0.766 & \num{2.792e-4} &  \num{9.642e-4}\\
0.781 & \num{2.528e-4} &  \num{9.200e-4}\\
0.797 & \num{2.294e-4} &  \num{8.713e-4}\\
0.812 & \num{2.084e-4} & \num{8.122e-4} \\
0.828 & \num{1.897e-4} & \num{7.462e-4} \\
0.844 & \num{1.728e-4} & \num{6.752e-4} \\
0.859 & \num{1.576e-4} & \num{6.036e-4} \\
0.875 & \num{1.438e-4} & \num{5.337e-4} \\
0.891 & \num{1.313e-4} &  \num{4.638e-4} \\
0.906 & \num{1.199e-4} &  \num{3.960e-4}\\
0.922 & \num{1.094e-4} &  \num{3.319e-4} \\
0.938 & \num{9.988e-5} &  \num{2.720e-4}\\
0.953 & \num{9.113e-5} &  \num{2.164e-4}\\
0.969 & \num{8.310e-5} &  \num{1.658e-4} \\
0.984 & \num{7.572e-5} & \num{1.212e-4} \\
1.000 & \num{6.893e-5} &  \num{8.570e-5} \\
1.016 & \num{6.268e-5} &  \num{6.071e-5}\\
1.031 & \num{5.693e-5} & \num{4.480e-5} \\
1.047 & \num{5.163e-5} &  \num{3.545e-5}\\
1.062 & \num{4.676e-5} & \num{2.905e-5} \\
1.078 & \num{4.228e-5} &  \num{2.400e-5}\\
1.094 & \num{3.817e-5} &  \num{2.140e-5}\\
1.109 & \num{3.441e-5} &  \num{1.959e-5}\\
1.125 & \num{3.096e-5} &  \num{1.834e-5} \\
1.141 & \num{2.782e-5} &  \num{1.749e-5}\\
1.156 & \num{2.496e-5} &  \num{1.698e-5}\\
1.172 & \num{2.237e-5} & \num{1.656e-5} \\
1.188 & \num{2.003e-5} & \num{1.641e-5} \\
1.203 & \num{1.792e-5} & \num{1.641e-5}  \\
1.219 & \num{1.603e-5} & \num{1.656e-5} \\
1.234 & \num{1.434e-5} & \num{1.661e-5} \\
1.250 & \num{1.284e-5} &  \num{1.677e-5}\\
\hline
\end{tabular}%
}
\end{minipage}
\hfill
\begin{minipage}[t]{0.32\textwidth}
\centering
\resizebox{0.75\textwidth}{!}{%
\begin{tabular}{|p{0.8cm}|p{2.0cm}|p{2.0cm}|}
\hline
\multicolumn{3}{|c|}{\textbf{$z=5$}} \\
\hline
$d/\ell$ & $LN(\sigma_{c})_{A|B}$ & $LN(\sigma_{h})_{A|B}$ \\
\hline\hline
0.031 & \num{6.080e-1} & 22.490 \\
0.047 & \num{2.338e-1} & 23.208 \\
0.062 & \num{9.480e-2} & 23.678 \\
0.078 & \num{5.165e-2} & 24.123 \\
0.094 & \num{3.678e-2} & 24.523 \\
0.109 & \num{2.854e-2} & 24.877 \\
0.125 & \num{2.286e-2} & 25.194 \\
0.141 & \num{1.857e-2} & 25.481 \\
0.156 & \num{1.515e-2} & 25.743 \\
0.172 & \num{1.239e-2} & 25.985 \\
0.188 & \num{1.011e-2} & 26.208 \\
0.203 & \num{8.199e-3} & 26.416 \\
0.219 & \num{6.586e-3} & 26.611 \\
0.234 & \num{5.239e-3} & 26.794 \\
0.250 & \num{4.141e-3} & 26.967 \\
0.266 & \num{3.275e-3} & 27.130 \\
0.281 & \num{2.613e-3} & 27.285 \\
0.297 & \num{2.114e-3} & 27.433 \\
0.312 & \num{1.737e-3} & 27.574 \\
0.328 & \num{1.448e-3} & \num{6.477e+0} \\
0.344 & \num{1.223e-3} & \num{6.112e+0} \\
0.359 & \num{1.044e-3} & \num{2.071e-1} \\
0.375 & \num{8.978e-4} & \num{3.003e-1} \\
0.391 & \num{7.777e-4} & \num{4.603e-3} \\
0.406 & \num{6.773e-4} & \num{3.881e-3} \\
0.422 & \num{5.925e-4} & \num{4.470e-3} \\
0.438 & \num{5.203e-4} & \num{2.213e-3} \\
0.453 & \num{4.583e-4} & \num{2.662e-3} \\
0.469 & \num{4.048e-4} & \num{2.343e-3} \\
0.484 & \num{3.584e-4} & \num{1.956e-3} \\
0.500 & \num{3.178e-4} & \num{1.032e-3} \\
0.516 & \num{2.823e-4} & \num{1.009e-3} \\
0.531 & \num{2.510e-4} & \num{1.381e-3} \\
0.547 & \num{2.233e-4} & \num{2.793e-4} \\
0.562 & \num{1.988e-4} & \num{2.151e-4} \\
0.578 & \num{1.769e-4} & \num{1.471e-4} \\
0.594 & \num{1.574e-4} & \num{9.173e-5} \\
0.609 & \num{1.399e-4} & \num{5.127e-5} \\
0.625 & \num{1.243e-4} & \num{3.169e-5} \\
0.641 & \num{1.102e-4} & \num{3.157e-5} \\
0.656 & \num{9.749e-5} & \num{1.868e-5} \\
0.672 & \num{8.606e-5} & \num{1.651e-5} \\
0.688 & \num{7.576e-5} & \num{1.532e-5} \\
0.703 & \num{6.649e-5} & \num{1.480e-5} \\
0.719 & \num{5.816e-5} & \num{1.452e-5} \\
0.734 & \num{5.069e-5} & \num{1.452e-5} \\
0.750 & \num{4.404e-5} & \num{1.478e-5} \\
0.766 & \num{3.814e-5} & \num{1.520e-5} \\
0.781 & \num{3.295e-5} & \num{1.570e-5} \\
0.797 & \num{2.842e-5} & \num{1.625e-5} \\
0.812 & \num{2.450e-5} & \num{1.676e-5} \\
0.828 & \num{2.113e-5} & \num{1.716e-5} \\
0.844 & \num{1.825e-5} & \num{1.738e-5} \\
0.859 & \num{1.580e-5} & \num{1.736e-5} \\
0.875 & \num{1.373e-5} & \num{1.706e-5} \\
0.891 & \num{1.196e-5} & \num{1.646e-5} \\
0.906 & \num{1.046e-5} & \num{1.561e-5} \\
0.922 & \num{9.182e-6} & \num{1.459e-5} \\
0.938 & \num{8.086e-6} & \num{1.350e-5} \\
0.953 & \num{7.143e-6} & \num{1.201e-5} \\
0.969 & \num{6.329e-6} & \num{1.072e-5} \\
0.984 & \num{5.622e-6} & \num{9.505e-6} \\
1.000 & \num{5.007e-6} & \num{8.389e-6} \\
1.016 & \num{4.468e-6} & \num{7.399e-6} \\
1.031 & \num{3.994e-6} & \num{6.564e-6} \\
1.047 & \num{3.577e-6} & \num{5.935e-6} \\
1.062 & \num{3.208e-6} & \num{4.944e-6} \\
1.078 & \num{2.880e-6} & \num{4.287e-6} \\
1.094 & \num{2.588e-6} & \num{3.685e-6} \\
1.109 & \num{2.328e-6} & \num{3.132e-6} \\
1.125 & \num{2.095e-6} & \num{2.619e-6} \\
1.141 & \num{1.886e-6} & \num{2.142e-6} \\
1.156 & \num{1.698e-6} & \num{1.701e-6} \\
1.172 & \num{1.528e-6} & \num{1.301e-6} \\
1.188 & \num{1.375e-6} & \num{9.547e-7} \\
1.203 & \num{1.236e-6} & \num{6.793e-7} \\
1.219 & \num{1.111e-6} & \num{4.857e-7} \\
1.234 & \num{9.968e-7} & \num{3.621e-7} \\
1.250 & \num{8.932e-7} & \num{2.863e-7} \\
\hline
\end{tabular}%
}
\end{minipage}
\hfill
\begin{minipage}[t]{0.32\textwidth}
\centering
\resizebox{0.75\textwidth}{!}{%
\begin{tabular}{|p{0.8cm}|p{2.0cm}|p{2.0cm}|}
\hline
\multicolumn{3}{|c|}{\textbf{$z=7$}} \\
\hline
$d/\ell$ & $LN(\sigma_{c})_{A|B}$ & $LN(\sigma_{h})_{A|B}$ \\
\hline\hline
0.047 & \num{2.795e-1} & 57.453 \\
0.062 & \num{1.114e-1} & 58.071 \\
0.078 & \num{3.673e-2} & 58.442 \\
0.094 & \num{1.323e-2} & 58.789 \\
0.109 & \num{8.488e-3} & 59.109 \\
0.125 & \num{6.272e-3} & 59.398 \\
0.141 & \num{4.832e-3} & 59.661 \\
0.156 & \num{3.775e-3} & 59.902 \\
0.172 & \num{2.952e-3} & 60.126 \\
0.188 & \num{2.293e-3} & 60.334 \\
0.203 & \num{1.754e-3} & 60.528 \\
0.219 & \num{1.305e-3} & 60.711 \\
0.234 & \num{9.350e-4} & 60.883 \\
0.250 & \num{6.464e-4} & 61.045 \\
0.266 & \num{4.431e-4} & 61.200 \\
0.281 & \num{3.127e-4} & 61.346 \\
0.297 & \num{2.307e-4} & 19.151 \\
0.312 & \num{1.772e-4} & 20.118 \\
0.328 & \num{1.401e-4} & 5.294 \\
0.344 & \num{1.131e-4} & 5.378 \\
0.359 & \num{9.272e-5} & \num{7.291e-2} \\
0.375 & \num{7.681e-5} & \num{7.847e-1} \\
0.391 & \num{6.414e-5} & \num{2.355e-1} \\
0.406 & \num{5.389e-5} & \num{6.352e-4} \\
0.422 & \num{4.549e-5} & \num{6.137e-4} \\
0.438 & \num{3.854e-5} & \num{2.782e-3} \\
0.453 & \num{3.276e-5} & \num{2.848e-3} \\
0.469 & \num{2.790e-5} & \num{3.285e-4} \\
0.484 & \num{2.380e-5} & \num{3.018e-4} \\
0.500 & \num{2.030e-5} & \num{2.046e-4} \\
0.516 & \num{1.730e-5} & \num{6.640e-5} \\
0.531 & \num{1.472e-5} & \num{8.288e-6} \\
0.547 & \num{1.248e-5} & \num{2.929e-6} \\
0.562 & \num{1.054e-5} & \num{1.761e-6} \\
0.578 & \num{8.848e-6} & \num{1.452e-6} \\
0.594 & \num{7.373e-6} & \num{1.340e-6} \\
0.609 & \num{6.089e-6} & \num{1.311e-6} \\
0.625 & \num{4.976e-6} & \num{1.353e-6} \\
0.641 & \num{4.022e-6} & \num{1.454e-6} \\
0.656 & \num{3.216e-6} & \num{1.612e-6} \\
0.672 & \num{2.551e-6} & \num{1.811e-6} \\
0.688 & \num{2.017e-6} & \num{2.042e-6} \\
0.703 & \num{1.597e-6} & \num{2.285e-6} \\
0.719 & \num{1.273e-6} & \num{2.517e-6} \\
0.734 & \num{1.025e-6} & \num{2.717e-6} \\
0.750 & \num{8.342e-7} & \num{2.851e-6} \\
0.766 & \num{6.861e-7} & \num{2.921e-6} \\
0.781 & \num{5.696e-7} & \num{2.921e-6} \\
0.797 & \num{4.769e-7} & \num{2.856e-6} \\
0.812 & \num{4.021e-7} & \num{2.736e-6} \\
0.828 & \num{3.410e-7} & \num{2.567e-6} \\
0.844 & \num{2.907e-7} & \num{2.367e-6} \\
0.859 & \num{2.488e-7} & \num{2.117e-6} \\
0.875 & \num{2.137e-7} & \num{1.858e-6} \\
0.891 & \num{1.840e-7} & \num{1.571e-6} \\
0.906 & \num{1.587e-7} & \num{1.283e-6} \\
0.922 & \num{1.372e-7} & \num{9.928e-7} \\
0.938 & \num{1.186e-7} & \num{7.105e-7} \\
0.953 & \num{1.026e-7} & \num{4.389e-7} \\
0.969 & \num{8.865e-8} & \num{2.081e-7} \\
0.984 & \num{7.652e-8} & \num{7.109e-8} \\
1.000 & \num{6.590e-8} & \num{3.152e-8} \\
1.016 & \num{5.659e-8} & \num{2.038e-8} \\
1.031 & \num{4.840e-8} & \num{1.628e-8} \\
1.047 & \num{4.119e-8} & \num{1.429e-8} \\
1.062 & \num{3.485e-8} & \num{1.341e-8} \\
1.078 & \num{2.929e-8} & \num{1.286e-8} \\
1.094 & \num{2.444e-8} & \num{1.273e-8} \\
1.109 & \num{2.024e-8} & \num{1.283e-8} \\
1.125 & \num{1.665e-8} & \num{1.330e-8} \\
1.141 & \num{1.363e-8} & \num{1.363e-8} \\
1.156 & \num{1.112e-8} & \num{1.426e-8} \\
1.172 & \num{9.076e-9} & \num{1.493e-8} \\
1.188 & \num{7.428e-9} & \num{1.555e-8} \\
1.203 & \num{6.108e-9} & \num{1.602e-8} \\
1.219 & \num{5.053e-9} & \num{1.622e-8} \\
1.234 & \num{4.208e-9} & \num{1.618e-8} \\
1.250 & \num{3.526e-9} & \num{1.575e-8} \\
\hline
\end{tabular}
}
\end{minipage}
\caption{Numerical values of logarithmic negativity $LN(\sigma_{c})_{A|B}$ and the inaccessible negativity in the halo $LN(\sigma_{h})_{A|B}$ 
for $m=5\times10^{-3}$ and $\ell=64$. The consolidation condition is held for $d>\frac{z-1}{2}$.}
\label{tableL64}
\end{table}

\begin{table}[H]
\centering

\begin{minipage}[t]{0.3\textwidth}
\centering
\resizebox{0.75\textwidth}{!}{%
\begin{tabular}{|p{0.8cm}|p{1.6cm}|p{1.6cm}|}
\hline
\multicolumn{3}{|c|}{\textbf{$z=5/10$}} \\
\hline
$d/\ell$ & $LN(\sigma_{c})_{A|B}$ & $LN(\sigma_{h})_{A|B}$ \\
\hline\hline
0.000 & 0.90742 & 0.14113 \\
0.016 & 0.56812 & 0.14137 \\
0.031 & 0.44045 & 0.14200 \\
0.047 & 0.36794 & 0.14632 \\
0.063 & 0.32036 & 0.15272 \\
0.078 & 0.28569 & 0.15925 \\
0.094 & 0.25881 & 0.16544 \\
0.109 & 0.23707 & 0.17116 \\
0.125 & 0.21891 & 0.17638 \\
0.141 & 0.20339 & 0.18114 \\
0.156 & 0.18990 & 0.18548 \\
0.172 & 0.17800 & 0.18945 \\
0.188 & 0.16739 & 0.19312 \\
0.203 & 0.15785 & 0.19651 \\
0.219 & 0.14922 & 0.19967 \\
0.234 & 0.14135 & 0.20263 \\
0.250 & 0.13415 & 0.20542 \\
0.266 & 0.12753 & 0.20804 \\
0.281 & 0.12141 & 0.21053 \\
0.297 & 0.11574 & 0.21290 \\
0.313 & 0.11047 & 0.21515 \\
0.328 & 0.10555 & 0.21730 \\
0.344 & 0.10096 & 0.21936 \\
0.359 & 0.09666 & 0.07396 \\
0.375 & 0.09262 & 0.07189 \\
0.391 & 0.08883 & 0.07924 \\
0.406 & 0.08525 & 0.13477 \\
0.422 & 0.08189 & 0.05670 \\
0.438 & 0.07871 & 0.05851 \\
0.453 & 0.07570 & 0.04336 \\
0.469 & 0.07285 & 0.04155 \\
0.484 & 0.07016 & 0.03946 \\
0.500 & 0.06760 & 0.03774 \\
0.516 & 0.06516 & 0.03621 \\
0.531 & 0.06285 & 0.03503 \\
0.547 & 0.06064 & 0.03349 \\
0.563 & 0.05854 & 0.03180 \\
0.578 & 0.05654 & 0.03041 \\
0.594 & 0.05462 & 0.02915 \\
0.609 & 0.05279 & 0.02765 \\
0.625 & 0.05103 & 0.02602 \\
0.641 & 0.04935 & 0.02446 \\
0.656 & 0.04774 & 0.02287 \\
0.672 & 0.04620 & 0.02127 \\
0.688 & 0.04472 & 0.01969 \\
0.703 & 0.04329 & 0.01816 \\
0.719 & 0.04192 & 0.01670 \\
0.734 & 0.04061 & 0.01534 \\
0.750 & 0.03934 & 0.01409 \\
0.766 & 0.03812 & 0.01295 \\
0.781 & 0.03695 & 0.01194 \\
0.797 & 0.03581 & 0.01104 \\
0.813 & 0.03472 & 0.01025 \\
0.828 & 0.03367 & 0.00955 \\
0.844 & 0.03265 & 0.00895 \\
0.859 & 0.03167 & 0.00843 \\
0.875 & 0.03072 & 0.00798 \\
0.891 & 0.02980 & 0.00759 \\
0.906 & 0.02891 & 0.00724 \\
0.922 & 0.02806 & 0.00693 \\
0.938 & 0.02723 & 0.00665 \\
0.953 & 0.02642 & 0.00639 \\
0.969 & 0.02565 & 0.00614 \\
0.984 & 0.02489 & 0.00592 \\
1.000 & 0.02417 & 0.00571 \\
1.016 & 0.02346 & 0.00551 \\
1.031 & 0.02278 & 0.00532 \\
1.047 & 0.02211 & 0.00514 \\
1.063 & 0.02147 & 0.00497 \\
1.078 & 0.02084 & 0.00481 \\
1.094 & 0.02024 & 0.00467 \\
1.109 & 0.01965 & 0.00453 \\
1.125 & 0.01908 & 0.00440 \\
1.141 & 0.01853 & 0.00429 \\
1.156 & 0.01799 & 0.00418 \\
1.172 & 0.01747 & 0.00408 \\
1.188 & 0.01696 & 0.00400 \\
1.203 & 0.01647 & 0.00392 \\
1.219 & 0.01599 & 0.00385 \\
1.234 & 0.01552 & 0.00379 \\
1.250 & 0.01507 & 0.00373 \\
\hline
\end{tabular}%
}
\end{minipage}
\hfill
\begin{minipage}[t]{0.3\textwidth}
\centering
\resizebox{0.75\textwidth}{!}{%
\begin{tabular}{|p{0.8cm}|p{1.6cm}|p{1.6cm}|}
\hline
\multicolumn{3}{|c|}{\textbf{$z=7/10$}} \\
\hline
$d/\ell$ & $LN(\sigma_{c})_{A|B}$ & $LN(\sigma_{h})_{A|B}$ \\
\hline\hline
0.000 & 1.29196 & 0.30522 \\
0.016 & 0.72574 & 0.30575 \\
0.031 & 0.53908 & 0.30857 \\
0.047 & 0.44031 & 0.32170 \\
0.063 & 0.37768 & 0.33745 \\
0.078 & 0.33293 & 0.35251 \\
0.094 & 0.29852 & 0.36614 \\
0.109 & 0.27075 & 0.37832 \\
0.125 & 0.24762 & 0.38924 \\
0.141 & 0.22793 & 0.29629 \\
0.156 & 0.21087 & 0.40810 \\
0.172 & 0.19591 & 0.41638 \\
0.188 & 0.18266 & 0.42404 \\
0.203 & 0.17083 & 0.43118 \\
0.219 & 0.16018 & 0.43787 \\
0.234 & 0.15055 & 0.44417 \\
0.250 & 0.14178 & 0.45012 \\
0.266 & 0.13377 & 0.45577 \\
0.281 & 0.12642 & 0.46114 \\
0.297 & 0.11964 & 0.46626 \\
0.313 & 0.11339 & 0.47116 \\
0.328 & 0.10759 & 0.47586 \\
0.344 & 0.10221 & 0.48038 \\
0.359 & 0.09721 & 0.18300 \\
0.375 & 0.09254 & 0.17522 \\
0.391 & 0.08818 & 0.15290 \\
0.406 & 0.08410 & 0.19080 \\
0.422 & 0.08027 & 0.13333 \\
0.438 & 0.07667 & 0.05135 \\
0.453 & 0.07328 & 0.04785 \\
0.469 & 0.07009 & 0.05148 \\
0.484 & 0.06707 & 0.03866 \\
0.500 & 0.06421 & 0.03549 \\
0.516 & 0.06151 & 0.03264 \\
0.531 & 0.05894 & 0.02945 \\
0.547 & 0.05650 & 0.02653 \\
0.563 & 0.05418 & 0.02401 \\
0.578 & 0.05197 & 0.02198 \\
0.594 & 0.04987 & 0.01988 \\
0.609 & 0.04786 & 0.01902 \\
0.625 & 0.04594 & 0.01606 \\
0.641 & 0.04411 & 0.01462 \\
0.656 & 0.04236 & 0.01337 \\
0.672 & 0.04069 & 0.01232 \\
0.688 & 0.03908 & 0.01142 \\
0.703 & 0.03755 & 0.01066 \\
0.719 & 0.03607 & 0.01002 \\
0.734 & 0.03466 & 0.00948 \\
0.750 & 0.03331 & 0.00902 \\
0.766 & 0.03201 & 0.00863 \\
0.781 & 0.03076 & 0.00830 \\
0.797 & 0.02956 & 0.00802 \\
0.813 & 0.02840 & 0.00778 \\
0.828 & 0.02729 & 0.00758 \\
0.844 & 0.02623 & 0.00739 \\
0.859 & 0.02520 & 0.00723 \\
0.875 & 0.02421 & 0.00708 \\
0.891 & 0.02326 & 0.00694 \\
0.906 & 0.02235 & 0.00681 \\
0.922 & 0.02147 & 0.00669 \\
0.938 & 0.02062 & 0.00658 \\
0.953 & 0.01981 & 0.00647 \\
0.969 & 0.01902 & 0.00638 \\
0.984 & 0.01827 & 0.00629 \\
1.000 & 0.01754 & 0.00621 \\
1.016 & 0.01684 & 0.00613 \\
1.031 & 0.01616 & 0.00607 \\
1.047 & 0.01552 & 0.00601 \\
1.063 & 0.01489 & 0.00596 \\
1.078 & 0.01429 & 0.00593 \\
1.094 & 0.01371 & 0.00589 \\
1.109 & 0.01316 & 0.00587 \\
1.125 & 0.01262 & 0.00585 \\
1.141 & 0.01211 & 0.00584 \\
1.156 & 0.01162 & 0.00584 \\
1.172 & 0.01114 & 0.00584 \\
1.188 & 0.01069 & 0.00584 \\
1.203 & 0.01025 & 0.00584 \\
1.219 & 0.00983 & 0.00585 \\
1.234 & 0.00943 & 0.00585 \\
1.250 & 0.00904 & 0.00584 \\
\hline
\end{tabular}%
}
\end{minipage}
\hfill
\begin{minipage}[t]{0.3\textwidth}
\centering
\resizebox{0.75\textwidth}{!}{%
\begin{tabular}{|p{0.8cm}|p{1.6cm}|p{1.6cm}|}
\hline
\multicolumn{3}{|c|}{\textbf{$z=9/10$}} \\
\hline
$d/\ell$ & $LN(\sigma_{c})_{A|B}$ & $LN(\sigma_{h})_{A|B}$ \\
\hline\hline
0.000 & 1.66317 & 0.53201 \\
0.016 & 0.79894 & 0.53293 \\
0.031 & 0.56452 & 0.54155 \\
0.047 & 0.45089 & 0.56981 \\
0.063 & 0.38152 & 0.59974 \\
0.078 & 0.33259 & 0.62665 \\
0.094 & 0.29521 & 0.65032 \\
0.109 & 0.26529 & 0.67125 \\
0.125 & 0.24059 & 0.68998 \\
0.141 & 0.21974 & 0.70622 \\
0.156 & 0.20184 & 0.72239 \\
0.172 & 0.18627 & 0.73663 \\
0.188 & 0.17259 & 0.74983 \\
0.203 & 0.16046 & 0.76214 \\
0.219 & 0.14961 & 0.00000 \\
0.234 & 0.13985 & 0.78452 \\
0.250 & 0.13101 & 0.79476 \\
0.266 & 0.12298 & 0.80445 \\
0.281 & 0.11564 & 0.81366 \\
0.297 & 0.10891 & 0.82243 \\
0.313 & 0.10273 & 0.83081 \\
0.328 & 0.09703 & 0.83883 \\
0.344 & 0.09175 & 0.84651 \\
0.359 & 0.08686 & 0.25150 \\
0.375 & 0.08232 & 0.25150 \\
0.391 & 0.07808 & 0.26677 \\
0.406 & 0.07412 & 0.18402 \\
0.422 & 0.07040 & 0.03956 \\
0.438 & 0.06692 & 0.03412 \\
0.453 & 0.06364 & 0.03232 \\
0.469 & 0.06055 & 0.03176 \\
0.484 & 0.05763 & 0.02375 \\
0.500 & 0.05488 & 0.02161 \\
0.516 & 0.05227 & 0.01957 \\
0.531 & 0.04979 & 0.01782 \\
0.547 & 0.04745 & 0.01595 \\
0.563 & 0.04522 & 0.01462 \\
0.578 & 0.04310 & 0.01377 \\
0.594 & 0.04109 & 0.01347 \\
0.609 & 0.03917 & 0.01164 \\
0.625 & 0.03734 & 0.01097 \\
0.641 & 0.03560 & 0.01041 \\
0.656 & 0.03394 & 0.00994 \\
0.672 & 0.03235 & 0.00955 \\
0.688 & 0.03084 & 0.00922 \\
0.703 & 0.02939 & 0.00895 \\
0.719 & 0.02801 & 0.00873 \\
0.734 & 0.02669 & 0.00855 \\
0.750 & 0.02543 & 0.00841 \\
0.766 & 0.02423 & 0.00829 \\
0.781 & 0.02308 & 0.00820 \\
0.797 & 0.02198 & 0.00812 \\
0.813 & 0.02093 & 0.00804 \\
0.828 & 0.01993 & 0.00798 \\
0.844 & 0.01898 & 0.00792 \\
0.859 & 0.01806 & 0.00787 \\
0.875 & 0.01719 & 0.00784 \\
0.891 & 0.01636 & 0.00777 \\
0.906 & 0.01556 & 0.00772 \\
0.922 & 0.01481 & 0.00768 \\
0.938 & 0.01409 & 0.00764 \\
0.953 & 0.01340 & 0.00761 \\
0.969 & 0.01274 & 0.00757 \\
0.984 & 0.01212 & 0.00754 \\
1.000 & 0.01153 & 0.00751 \\
1.016 & 0.01096 & 0.00748 \\
1.031 & 0.01043 & 0.00745 \\
1.047 & 0.00992 & 0.00743 \\
1.063 & 0.00944 & 0.00740 \\
1.078 & 0.00898 & 0.00736 \\
1.094 & 0.00854 & 0.00733 \\
1.109 & 0.00813 & 0.00728 \\
1.125 & 0.00774 & 0.00723 \\
1.141 & 0.00736 & 0.00716 \\
1.156 & 0.00701 & 0.00707 \\
1.172 & 0.00668 & 0.00698 \\
1.188 & 0.00636 & 0.00686 \\
1.203 & 0.00606 & 0.00672 \\
1.219 & 0.00578 & 0.00657 \\
1.234 & 0.00551 & 0.00639 \\
1.250 & 0.00525 & 0.00620 \\
\hline
\end{tabular}%
}
\end{minipage}
\caption{Numerical values of logarithmic negativity $LN(\sigma_{c})_{A|B}$ and the inaccessible negativity in the halo $LN(\sigma_{h})_{A|B}$ for $z=5/10$, $z=7/10$, and $z=9/10$, with $m=5\times10^{-3}$ and $\ell=64$.}
\label{tableL64fractionalz}
\end{table}

\clearpage
\section{Adjacent-core and adjacent-halo weights versus separation}\label{Appendix:SeparationFlow}
\begin{figure}[h!]
    \centering
    \setlength{\tabcolsep}{1pt}
    \renewcommand{\arraystretch}{0.2}
    \begin{minipage}[t]{0.32\textwidth}
        \vspace{0pt}
        \centering
        \resizebox{\linewidth}{!}{%
        \begin{tabular}{c|c|c}
            \hline
            \multicolumn{3}{c}{$\ell=200\quad m=0.2\quad z=1$} \\
            \hline
            $\rho$
            & $W_{C}$
            & $W_{H}$\\
            \hline
            0.0000 & 1.000 & $8.529\times10^{-1998}$ \\
            0.0166 & 1.000 & $3.538\times10^{-33}$ \\
            0.0342 & 1.000 & $8.079\times10^{-29}$ \\
            0.0508 & 1.000 & $3.972\times10^{-26}$ \\
            0.0674 & 1.000 & $3.070\times10^{-23}$ \\
            0.0849 & 1.000 & $2.027\times10^{-20}$ \\
            0.1016 & 1.000 & $3.106\times10^{-18}$ \\
            0.1182 & 1.000 & $1.307\times10^{-15}$ \\
            0.1357 & 1.000 & $1.097\times10^{-13}$ \\
            0.1524 & 1.000 & $1.239\times10^{-11}$ \\
            0.1699 & 1.000 & $1.295\times10^{-9}$ \\
            0.1865 & 1.000 & $3.820\times10^{-8}$ \\
            0.2031 & 1.000 & $1.922\times10^{-6}$ \\
            0.2207 & 1.000 & $1.484\times10^{-5}$ \\
            0.2373 & 0.9999 & $8.052\times10^{-5}$ \\
            0.2539 & 0.9968 & $3.202\times10^{-3}$ \\
            0.2715 & 0.9918 & $8.202\times10^{-3}$ \\
            0.2881 & 0.9836 & $1.640\times10^{-2}$ \\
            0.3047 & 0.9405 & $5.955\times10^{-2}$ \\
            0.3223 & 0.9519 & $4.814\times10^{-2}$ \\
            0.3389 & 0.9094 & $9.057\times10^{-2}$ \\
            0.3555 & 0.8924 & $1.076\times10^{-1}$ \\
            0.3730 & 0.9026 & $9.738\times10^{-2}$ \\
            0.3897 & 0.8310 & $1.690\times10^{-1}$ \\
            0.4072 & 0.8920 & $1.080\times10^{-1}$ \\
            0.4238 & 0.7341 & $2.659\times10^{-1}$ \\
            0.4404 & 0.8851 & $1.149\times10^{-1}$ \\
            0.4580 & 0.7787 & $2.213\times10^{-1}$ \\
            0.4746 & 0.8329 & $1.671\times10^{-1}$ \\
            0.4912 & 0.8249 & $1.751\times10^{-1}$ \\
            0.5088 & 0.8215 & $1.785\times10^{-1}$ \\
            0.5254 & 0.6656 & $3.344\times10^{-1}$ \\
            0.5420 & 0.8786 & $1.214\times10^{-1}$ \\
            0.5596 & 0.7523 & $2.477\times10^{-1}$ \\
            0.5762 & 0.7919 & $2.081\times10^{-1}$ \\
            0.5928 & 0.7501 & $2.499\times10^{-1}$ \\
            0.6103 & 0.8191 & $1.809\times10^{-1}$ \\
            0.6270 & 0.8291 & $1.709\times10^{-1}$ \\
            0.6445 & 0.7439 & $2.561\times10^{-1}$ \\
            0.6611 & 0.7450 & $2.550\times10^{-1}$ \\
            0.6777 & 0.7551 & $2.449\times10^{-1}$ \\
            0.6953 & 0.7184 & $2.816\times10^{-1}$ \\
            0.7119 & 0.8394 & $1.606\times10^{-1}$ \\
            0.7285 & 0.8351 & $1.649\times10^{-1}$ \\
            0.7461 & 0.7357 & $2.643\times10^{-1}$ \\
            0.7627 & 0.7244 & $2.756\times10^{-1}$ \\
            0.7793 & 0.7634 & $2.366\times10^{-1}$ \\
            0.7969 & 0.7751 & $2.249\times10^{-1}$ \\
            0.8135 & 0.7987 & $2.013\times10^{-1}$ \\
            0.8301 & 0.7078 & $2.922\times10^{-1}$ \\
            0.8476 & 0.7203 & $2.797\times10^{-1}$ \\
            0.8643 & 0.7380 & $2.620\times10^{-1}$ \\
            0.8818 & 0.6656 & $3.344\times10^{-1}$ \\
            0.8984 & 0.8133 & $1.867\times10^{-1}$ \\
            0.9151 & 0.7922 & $2.078\times10^{-1}$ \\
            0.9326 & 0.7775 & $2.225\times10^{-1}$ \\
            0.9492 & 0.8243 & $1.757\times10^{-1}$ \\
            0.9658 & 0.8618 & $1.382\times10^{-1}$ \\
            0.9834 & 0.6314 & $3.686\times10^{-1}$ \\
            1.0000 & 0.7670 & $2.330\times10^{-1}$ \\
            \hline
        \end{tabular}%
        }
    \end{minipage}
    \hfill
    \begin{minipage}[t]{0.32\textwidth}
        \vspace{0pt}
        \centering
        \resizebox{\linewidth}{!}{%
        \begin{tabular}{c|c|c}
            \hline
            \multicolumn{3}{c}{$\ell=200\quad m=0.4\quad z=1$} \\
            \hline
            $\rho$
            & $W_{C}$
            & $W_{H}$\\
            \hline
            0.0000 & 1.000 & $-6.853\times10^{-588}$ \\
            0.0013 & 1.000 & $3.786\times10^{-118}$ \\
            0.0027 & 1.000 & $7.168\times10^{-112}$ \\
            0.0040 & 1.000 & $2.548\times10^{-107}$ \\
            0.0054 & 1.000 & $4.141\times10^{-104}$ \\
            0.0081 & 1.000 & $5.888\times10^{-99}$ \\
            0.0108 & 1.000 & $1.038\times10^{-95}$ \\
            0.0161 & 1.000 & $5.322\times10^{-88}$ \\
            0.0228 & 1.000 & $5.145\times10^{-85}$ \\
            0.0309 & 1.000 & $1.710\times10^{-80}$ \\
            0.0444 & 1.000 & $1.607\times10^{-71}$ \\
            0.0632 & 1.000 & $5.046\times10^{-61}$ \\
            0.0901 & 1.000 & $1.079\times10^{-52}$ \\
            0.1277 & 1.000 & $9.525\times10^{-39}$ \\
            0.1815 & 1.000 & $1.369\times10^{-29}$ \\
            0.2581 & 1.000 & $3.559\times10^{-17}$ \\
            0.3669 & 1.000 & $1.657\times10^{-5}$ \\
            0.3683 & 1.000 & $1.708\times10^{-5}$ \\
            0.3763 & 0.9998 & $1.952\times10^{-4}$ \\
            0.3831 & 0.9994 & $6.479\times10^{-4}$ \\
            0.3911 & 0.9994 & $5.884\times10^{-4}$ \\
            0.3992 & 0.9984 & $1.640\times10^{-3}$ \\
            0.4059 & 0.9922 & $7.850\times10^{-3}$ \\
            0.4140 & 0.9927 & $7.288\times10^{-3}$ \\
            0.4207 & 0.9839 & $1.608\times10^{-2}$ \\
            0.4288 & 0.9787 & $2.133\times10^{-2}$ \\
            0.4368 & 0.9838 & $1.619\times10^{-2}$ \\
            0.4435 & 0.9653 & $3.468\times10^{-2}$ \\
            0.4516 & 0.9454 & $5.461\times10^{-2}$ \\
            0.4597 & 0.9547 & $4.525\times10^{-2}$ \\
            0.4664 & 0.9694 & $3.058\times10^{-2}$ \\
            0.4745 & 0.9272 & $7.276\times10^{-2}$ \\
            0.4987 & 0.9487 & $5.130\times10^{-2}$ \\
            0.5215 & 0.9346 & $6.539\times10^{-2}$ \\
            0.5269 & 0.9078 & $9.218\times10^{-2}$ \\
            0.5470 & 0.8995 & $1.005\times10^{-1}$ \\
            0.5726 & 0.8923 & $1.077\times10^{-1}$ \\
            0.5995 & 0.8897 & $1.103\times10^{-1}$ \\
            0.6290 & 0.8975 & $1.025\times10^{-1}$ \\
            0.6586 & 0.8931 & $1.069\times10^{-1}$ \\
            0.6895 & 0.8363 & $1.637\times10^{-1}$ \\
            0.7231 & 0.8288 & $1.712\times10^{-1}$ \\
            0.7567 & 0.6623 & $3.377\times10^{-1}$ \\
            0.7930 & 0.7816 & $2.184\times10^{-1}$ \\
            0.8306 & 0.6308 & $3.692\times10^{-1}$ \\
            0.8696 & 0.7174 & $2.826\times10^{-1}$ \\
            0.9113 & 0.8346 & $1.654\times10^{-1}$ \\
            0.9543 & 0.8032 & $1.968\times10^{-1}$ \\
            1.0000 & 0.7833 & $2.167\times10^{-1}$ \\
            \hline
        \end{tabular}%
        }
    \end{minipage}
    \hfill
    \begin{minipage}[t]{0.32\textwidth}
        \vspace{0pt}
        \centering
        \resizebox{\linewidth}{!}{%
        \begin{tabular}{c|c|c}
            \hline
            \multicolumn{3}{c}{$\ell=200\quad m=0.5\quad z=1$} \\
            \hline
            $\rho$
            & $W_{C}$
            & $W_{H}$\\
            \hline
            0.0000 & 1.000 & $1.219\times10^{-576}$ \\
            0.0076 & 1.000 & $6.728\times10^{-140}$ \\
            0.0152 & 1.000 & $3.797\times10^{-123}$ \\
            0.0242 & 1.000 & $2.192\times10^{-111}$ \\
            0.0500 & 1.000 & $2.010\times10^{-98}$ \\
            0.0758 & 1.000 & $6.640\times10^{-79}$ \\
            0.1000 & 1.000 & $2.296\times10^{-69}$ \\
            0.1242 & 1.000 & $6.950\times10^{-63}$ \\
            0.1500 & 1.000 & $5.997\times10^{-52}$ \\
            0.1758 & 1.000 & $1.869\times10^{-41}$ \\
            0.2000 & 1.000 & $1.479\times10^{-37}$ \\
            0.2242 & 1.000 & $3.677\times10^{-32}$ \\
            0.2500 & 1.000 & $6.900\times10^{-28}$ \\
            0.2758 & 1.000 & $1.530\times10^{-23}$ \\
            0.3000 & 1.000 & $4.895\times10^{-19}$ \\
            0.3076 & 1.000 & $1.634\times10^{-18}$ \\
            0.3167 & 1.000 & $3.533\times10^{-17}$ \\
            0.3242 & 1.000 & $8.472\times10^{-16}$ \\
            0.3318 & 1.000 & $1.200\times10^{-14}$ \\
            0.3394 & 1.000 & $2.096\times10^{-14}$ \\
            0.3485 & 1.000 & $2.523\times10^{-13}$ \\
            0.3561 & 1.000 & $7.753\times10^{-12}$ \\
            0.3636 & 1.000 & $8.511\times10^{-11}$ \\
            0.3727 & 1.000 & $4.884\times10^{-10}$ \\
            0.3803 & 1.000 & $3.615\times10^{-9}$ \\
            0.3879 & 1.000 & $3.312\times10^{-8}$ \\
            0.3955 & 1.000 & $7.377\times10^{-8}$ \\
            0.4045 & 1.000 & $5.108\times10^{-7}$ \\
            0.4121 & 1.000 & $5.175\times10^{-6}$ \\
            0.4197 & 1.000 & $3.328\times10^{-5}$ \\
            0.4273 & 1.000 & $2.384\times10^{-5}$ \\
            0.4364 & 0.9999 & $1.230\times10^{-4}$ \\
            0.4439 & 0.9990 & $9.822\times10^{-4}$ \\
            0.4515 & 0.9969 & $3.107\times10^{-3}$ \\
            0.4606 & 0.9971 & $2.888\times10^{-3}$ \\
            0.4682 & 0.9952 & $4.841\times10^{-3}$ \\
            0.4758 & 0.9829 & $1.713\times10^{-2}$ \\
            0.4833 & 0.9766 & $2.342\times10^{-2}$ \\
            0.4924 & 0.9835 & $1.646\times10^{-2}$ \\
            0.5000 & 0.9739 & $2.614\times10^{-2}$ \\
            0.5424 & 0.9655 & $3.448\times10^{-2}$ \\
            0.5833 & 0.9161 & $8.392\times10^{-2}$ \\
            0.6242 & 0.9312 & $6.877\times10^{-2}$ \\
            0.6667 & 0.8320 & $1.680\times10^{-1}$ \\
            0.7091 & 0.8443 & $1.557\times10^{-1}$ \\
            0.7500 & 0.8624 & $1.376\times10^{-1}$ \\
            0.7909 & 0.7941 & $2.059\times10^{-1}$ \\
            0.8333 & 0.8492 & $1.508\times10^{-1}$ \\
            0.8758 & 0.7538 & $2.462\times10^{-1}$ \\
            0.9167 & 0.8483 & $1.517\times10^{-1}$ \\
            0.9576 & 0.7774 & $2.226\times10^{-1}$ \\
            1.0000 & 0.7977 & $2.023\times10^{-1}$ \\
            \hline
        \end{tabular}%
        }
    \end{minipage}
\caption{The numerical values of $W_C$ and $W_H$ for $m=0.2,0.4,0.5$ and $\ell=200$. These data are presented in Fig.~\ref{fig:adjacent_basis_flow}.
The tiny nonzero values of $W_H$ at $\rho=0$ should be identified as numerical residuals.}
    \label{tab:core_halo_weights}
\end{figure}

\newpage

\section{Singular-value hierarchy of cross-region correlations}
\label{app:correlation_channels}

Having examined the spatial and momentum-space structure of the
dominant negativity-carrying mode, we now consider a complementary
description of the separation dependence through the singular-value
spectra of the cross-region covariance blocks. Writing
\begin{align}
	G=\begin{pmatrix}G_A&G_{AB}\\G_{BA}&G_B\end{pmatrix},
	\qquad
	H=\begin{pmatrix}H_A&H_{AB}\\H_{BA}&H_B\end{pmatrix},
\end{align}
we characterize $G_{AB}$ and $H_{AB}$ by their singular values,
\begin{align}
	C_{AB}=\sum_{j}s_j^{(C)}\,a_j^{(C)}\bigl(b_j^{(C)}\bigr)^T,
	\qquad C=G,H,
\end{align}
ordered as $s_1\geq s_2\geq\cdots$. The singular-value decomposition is the appropriate diagnostic here because a cross block need not itself be a symmetric matrix. Each term defines a spatial correlation channel between a collective direction in $A$ and one in $B$, with strength $s_j$.

\begin{figure}[h!]
	\begin{center}
		\includegraphics[scale=0.35]{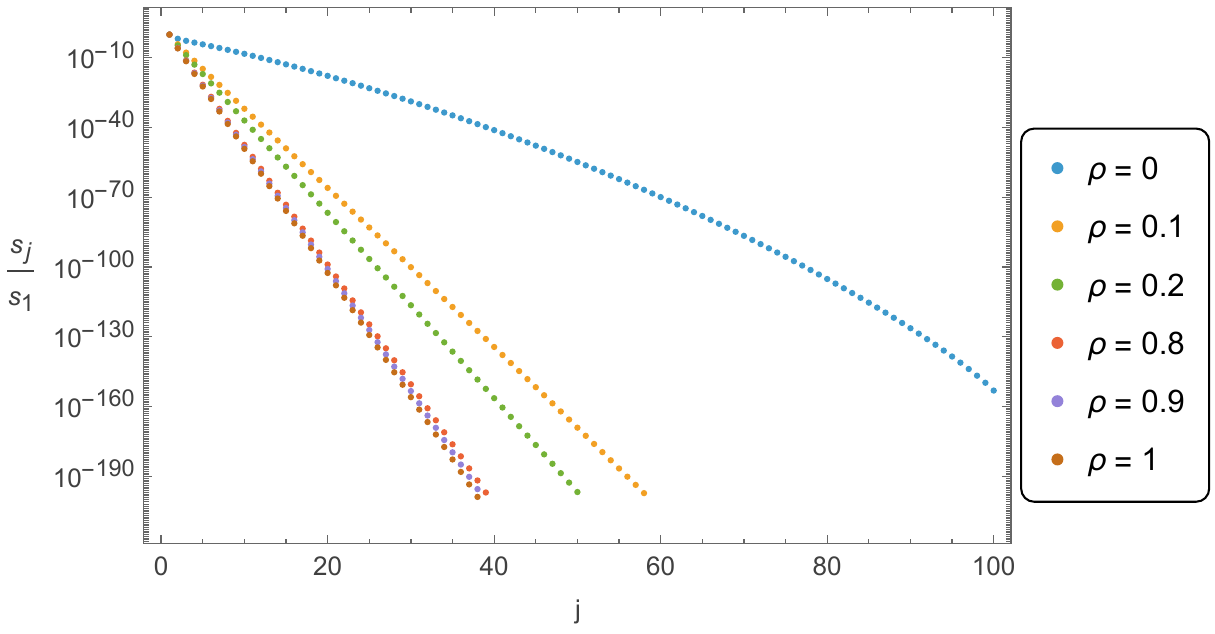}
		\hspace{3mm}
		\includegraphics[scale=0.37]{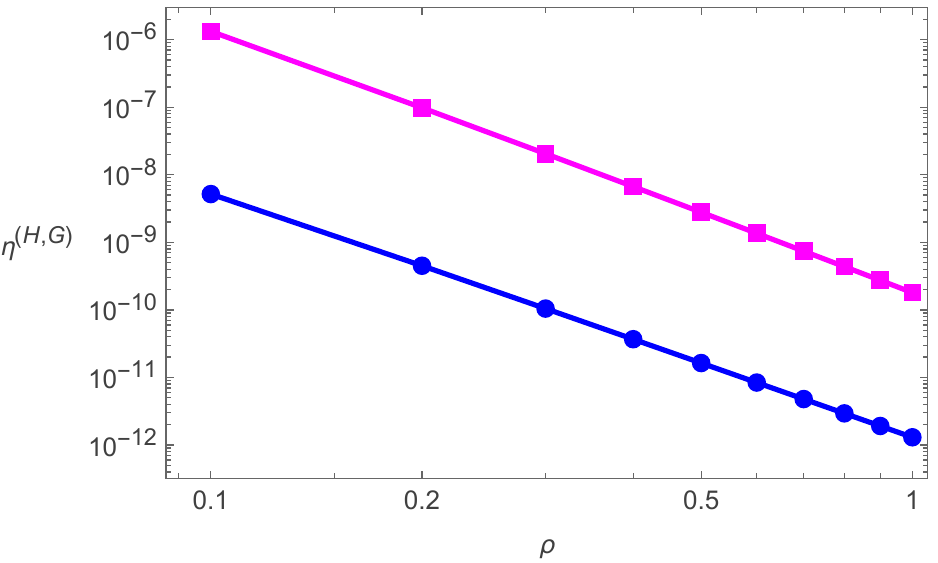}
	\end{center}
	\caption{Left: normalized singular-value spectrum $s_j/s_1$ of $G_{AB}$ for $\ell=100$ and $m=10^{-6}$ at several separations. Here $d_{{\cancel{\mathcal{E}}}}=11041$ corresponds to $\rho=1$. The approximately exponential decay with channel index steepens as separation increases. Right: fraction of the squared Frobenius norm carried by subleading channels for $G_{AB}$ and $H_{AB}$. Points are numerical results; the solid curves show the fitted power laws $\rho^{-3.6}$ for $G$ and $\rho^{-3.9}$ for $H$.}
	\label{fig:singular_spectrum_channels}
\end{figure}

The normalized spectra $s_j/s_1$ reveal a pronounced hierarchy that steepens with separation. Because the spectra are smooth rather than separated by a sharp gap, an integer ``effective rank'' defined through a tolerance $s_j/s_1>\epsilon$ would necessarily depend on the arbitrary choice of $\epsilon$. We therefore work directly with the normalized spectra and with their logarithmic decay rate. Over a broad range of the subleading spectrum the numerical data are approximately exponential,
\begin{align}
	\log\!\left(\frac{s_j}{s_1}\right)
	\simeq b(d)-\alpha(d)j,
	\label{eq:sv_exponential}
\end{align}
so that $\alpha(d)$ measures the suppression of additional correlation channels without imposing a singular-value threshold.
Figure~\ref{fig:singular_spectrum_channels} shows that the hierarchy
among the subleading channels steepens with increasing separation,
with the normalized spectrum approaching an approximately
separation-independent profile before the negativity sphere
is reached.

To quantify the relative weight of the subleading correlation
channels, we define
\begin{align}
	\eta^{(C)}(d)
	\equiv
	\frac{\|C_{AB}(d)-C_{AB}^{(1)}(d)\|_F^2}
	{\|C_{AB}(d)\|_F^2}
	=
	\frac{\sum_{j>1}[s_j^{(C)}(d)]^2}
	{\sum_{j\geq1}[s_j^{(C)}(d)]^2},
	\qquad C=G,H,
\end{align}
where $\|M\|_F^2=\mathrm{Tr}(M^\dagger M)$ and
$C_{AB}^{(1)}=s_1^{(C)}a_1^{(C)}(b_1^{(C)})^T$
is the optimal rank-one approximation obtained from the
singular-value decomposition. Thus $\eta_C$ measures the fraction
of the squared Frobenius norm carried by all channels beyond the
leading one. In Fig. \ref{fig:singular_spectrum_channels} we present the values for $\eta^{(G,H)}$ where small values indicate that the cross-region
correlation block is well approximated by a single spatial channel. 

The normalized spectra at $\rho=0.8$, $\rho=0.9$, and $\rho\to 1$ are nearly indistinguishable within numerical resolution. In this range, the relative channel weights change little while the overall cross-region correlations continue to weaken. The same qualitative suppression of subleading channels is observed in both the field and momentum blocks. These singular-value spectra describe cross-region correlations, however, and do not by themselves identify the NPT modes, which depend on the full PT symplectic problem.

The observed hierarchy is consistent with a large-distance expansion
of the cross-region correlation kernels. Let $D=d+\ell$ denote the
center-to-center distance between the intervals, and let $x$ and $y$
be coordinates measured from their respective centers. For a smooth
translation-invariant kernel,
\begin{align}
	K(D+y-x)
	=
	K(D)+(y-x)K'(D)
	+\frac{1}{2}(y-x)^2K''(D)+\cdots.
	\label{eq:largeD_kernel}
\end{align}
The leading constant term defines a rank-one matrix between the two
intervals. Its dominance requires the kernel to vary weakly across
their extent, with the relevant relative corrections
$\ell^n |K^{(n)}(D)/K(D)|$ small. For an algebraically decaying kernel
$K(D)\propto D^{-p}$, these corrections are controlled by powers of
$\ell/D$ at fixed $p$. Higher-order terms introduce additional spatial
directions with parametrically suppressed amplitudes. This provides
an analytic motivation for an increasingly hierarchical singular-value
spectrum in the slowly varying regime.

For exponentially decaying kernels, large separation alone does not
justify retaining only the constant Taylor term. Nevertheless, a
low-dimensional structure can emerge by factoring out the exponential.
For example, if $K(D)\sim D^{-\beta}e^{-\kappa D}$, then
\begin{align}
	K(D+y-x)
	\sim
	D^{-\beta}e^{-\kappa D}
	e^{\kappa x}e^{-\kappa y}
	\left[
	1-\beta\frac{y-x}{D}
	+O\!\left(\frac{\ell^2}{D^2}\right)
	\right].
\end{align}
The leading term is separable in $x$ and $y$ and therefore rank one,
even when $\kappa\ell$ is not small. An oscillatory exponential
contribution generally produces up to two real spatial channels at
leading order. Thus the asymptotic channel structure depends on the
kernel and need not always reduce to a single constant channel.

These expansions motivate the observed reduction of correlation
channels, but do not by themselves establish the fitted exponential
dependence of singular values on channel index or a distance-independent
normalized spectrum. Nor do they identify the surviving NPT mode,
which depends on the full covariance matrix, including its local blocks.

Cross-region correlations persist beyond the negativity sphere, even though the number of NPT modes has fallen to zero. Thus the threshold marks the loss of NPT entanglement, rather than the disappearance of correlations. The increasingly steep channel hierarchy characterizes this long-distance regime without equating singular channels with negativity-carrying modes.

\newpage

\section{Spatial profiles of canonical mode components}\label{Appendix:uvConfig}

In this appendix we present how the core-mode's profile is interpolated between the largest and smallest contributing modes to logarithmic negativity for different mass and dynamical exponent parameters.

\begin{figure}[h!]
    \centering

    \begin{subfigure}{0.9\textwidth}
        \centering
        \includegraphics[width=\textwidth]{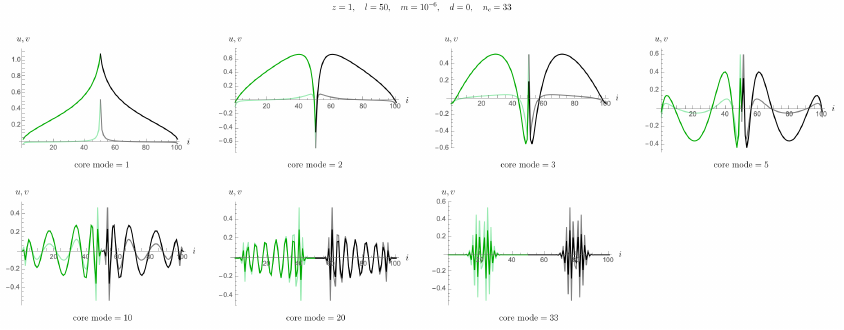}
        \label{fig:mixed_z1_m6}
    \end{subfigure}
    \hfill
    \begin{subfigure}{0.9\textwidth}
        \centering
        \includegraphics[width=\textwidth]{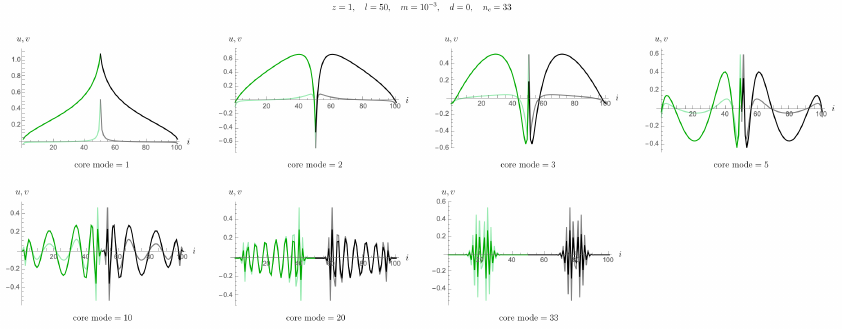}
        \label{fig:mixed_z1_m3}
    \end{subfigure}

    \vspace{6mm}

    \begin{subfigure}{0.9\textwidth}
        \centering
        \includegraphics[width=\textwidth]{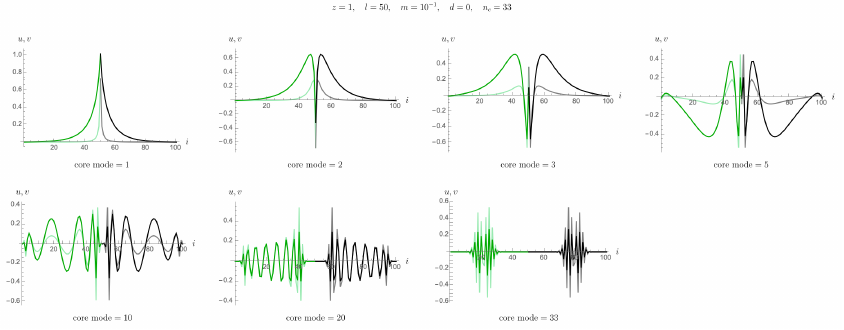}
        \label{fig:mixed_z1_m1}
    \end{subfigure}

    \caption{Core mode components $u$ and
    $v$ of subregions $A$ (green) and $B$ (black) for $z=1$, $\ell=50$ and $d=0$, shown 
    for three different values of the mass parameter $m$. From top to bottom, the masses are $10^{-6}$, $10^{-3}$, and $10^{-1}$, respectively. The $u$-modes are represented by light colors and the $v$-modes are shown in darker shades.}
    \label{fig:mixed_uv_z1_d0}
\end{figure}

\begin{figure}[h!]
    \centering

    \begin{subfigure}{0.9\textwidth}
        \centering
        \includegraphics[width=\textwidth]{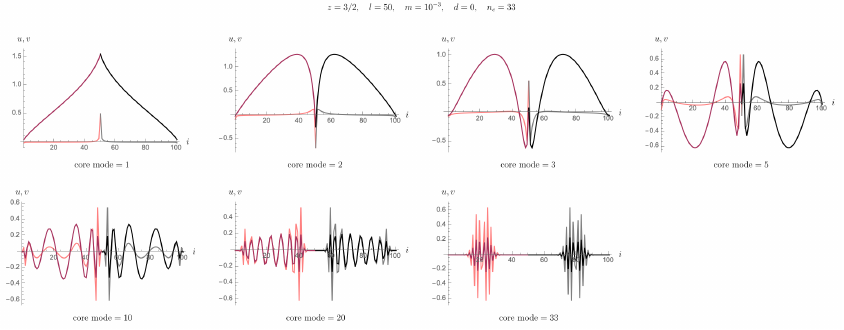}
        \label{fig:mixed_z3o2_m3_d0}
    \end{subfigure}
    \hfill
    \begin{subfigure}{0.9\textwidth}
        \centering
        \includegraphics[width=\textwidth]{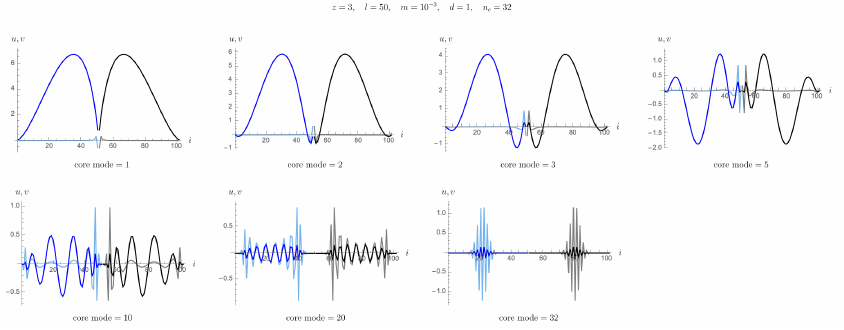}
        \label{fig:mixed_z3_m3_d1}
    \end{subfigure}

    \caption{Core mode components $u$ and
    $v$ of length-$50$ subregions $A$ (pink and blue) and $B$ (black) for $z=\frac{3}{2}$ (upper panel) and $z=3$ (lower panel). For $d=0$ (upper panel) and $d=1$ (lower panel) and $m=10^{-3}$. The $u$-modes are represented by light colors and the $v$-modes are shown in darker shades.}
    \label{fig:mixed_uv_comparison}
\end{figure}

\begin{figure}[h!]
    \centering

    \begin{subfigure}{0.9\textwidth}
        \centering
        \includegraphics[width=\textwidth]{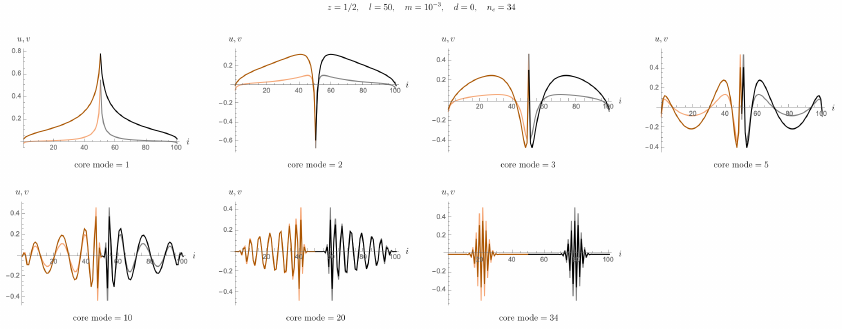}
        \label{fig:mixed_z1o2_m3_d0}
    \end{subfigure}
    \hfill
    \begin{subfigure}{0.9\textwidth}
        \centering
        \includegraphics[width=\textwidth]{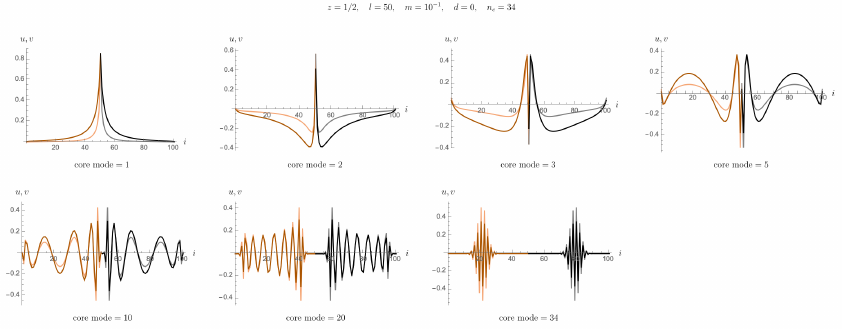}
        \label{fig:mixed_z1o2_m1_d0}
    \end{subfigure}

    \caption{Core mode components $u$ and
    $v$ of subregions $A$ (orange) and $B$ (black) for $z=\frac{1}{2}$, $\ell=50$ and $d=0$, shown 
    for two different values of the mass parameter $m$. $m=10^{-3}$ for upper panel and $m=10^{-1}$ for lower panel. The $u$-modes are represented by light colors and the $v$-modes are shown in darker shades.}
    \label{fig:mixed_uv_z1o2_d0}
\end{figure}

\begin{figure}[h!]
    \centering

    \begin{subfigure}{0.9\textwidth}
        \centering
        \includegraphics[width=\textwidth]{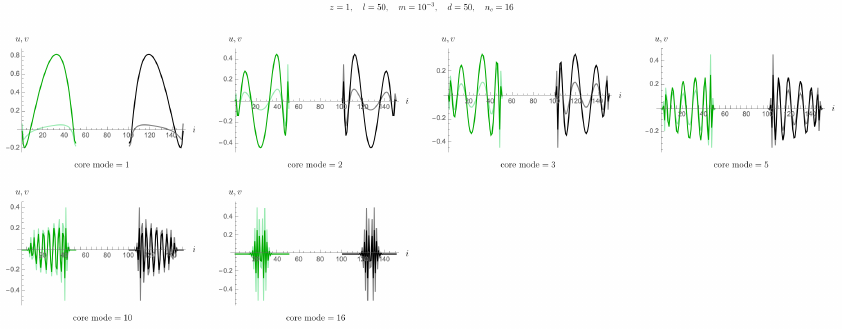}
        \caption{$z=1$, $m=10^{-3}$, $d=50$}
        \label{fig:mixed_z1_m3_d50}
    \end{subfigure}
    \hfill
    \begin{subfigure}{0.9\textwidth}
        \centering
        \includegraphics[width=\textwidth]{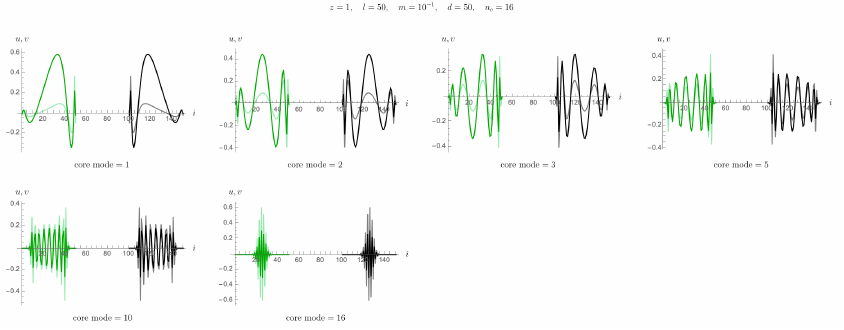}
        \caption{$z=1$, $m=10^{-1}$, $d=50$}
        \label{fig:mixed_z1_m1_d50}
    \end{subfigure}

    \caption{Core mode components $u$ and
    $v$ of subregions $A$ (green) and $B$ (black) for $z=1$, $\ell=50$ and $d=50$, shown 
    for two different values of the mass parameter $m$. $m=10^{-3}$ for upper panel and $m=10^{-1}$ for lower panel. The $u$-modes are represented by light colors and the $v$-modes are shown in darker shades.}
    \label{fig:mixed_uv_z1_d50}
\end{figure}

\begin{figure}[h!]
    \centering

    \begin{subfigure}{0.9\textwidth}
        \centering
        \includegraphics[width=\textwidth]{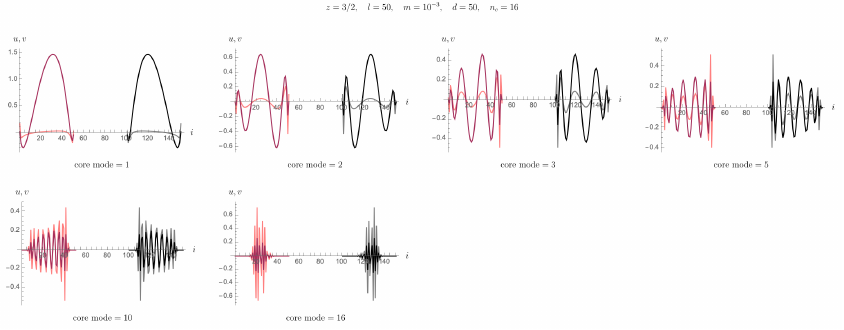}
        \label{fig:mixed_z3o2_m3_d50}
    \end{subfigure}
    \hfill
    \begin{subfigure}{0.9\textwidth}
        \centering
        \includegraphics[width=\textwidth]{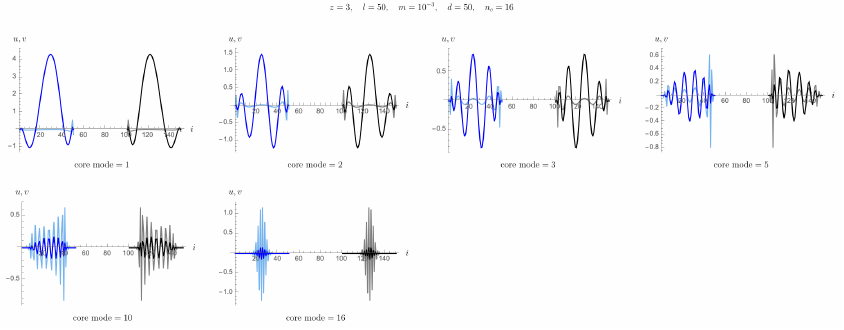}
        \label{fig:mixed_z3_m3_d50}
    \end{subfigure}

    \caption{Core mode components $u$ and
    $v$ of length-$50$ subregions $A$ (pink and blue) and $B$ (black) for $z=\frac{3}{2}$ (upper panel) and $z=3$ (lower panel). For $d=50$ and $m=10^{-3}$. The $u$-modes are represented by light colors and the $v$-modes are shown in darker shades.}
    \label{fig:mixed_uv_z32_d50}
\end{figure}

\begin{figure}[h!]
    \centering

    \begin{subfigure}{0.9\textwidth}
        \centering
        \includegraphics[width=\textwidth]{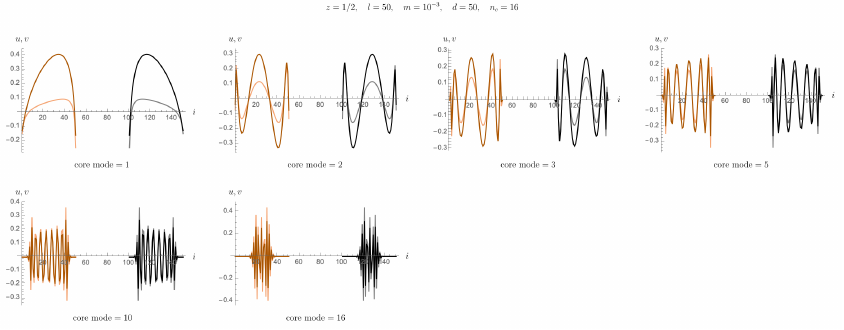}
        \label{fig:mixed_z1o2_m3_d50}
    \end{subfigure}
    \hfill
    \begin{subfigure}{0.9\textwidth}
        \centering
        \includegraphics[width=\textwidth]{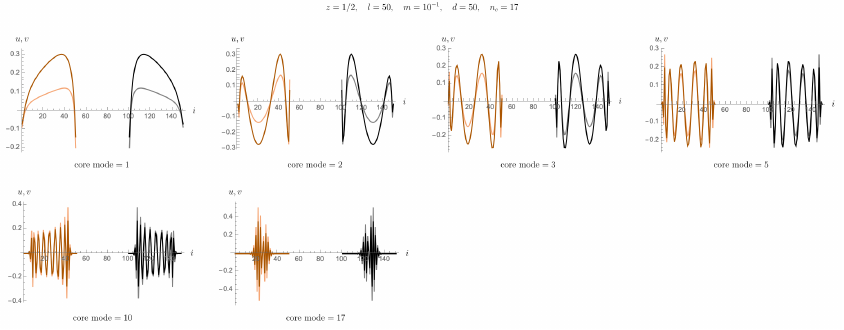}
        \label{fig:mixed_z1o2_m1_d50}
    \end{subfigure}

    \caption{Core mode components $u$ and
    $v$ of subregions $A$ (orange) and $B$ (black) for $z=\frac{1}{2}$, $\ell=50$ and $d=50$, shown 
    for two different values of the mass parameter $m$. $m=10^{-3}$ for upper panel and $m=10^{-1}$ for lower panel. The $u$-modes are represented by light colors and the $v$-modes are shown in darker shades.}
    \label{fig:mixed_uv_z1o2_d50}
\end{figure}

\end{document}